\documentclass[a4paper,11pt]{article}
\pdfoutput=1 % if your are submitting a pdflatex (i.e. if you have
\usepackage{jheppub} % for details on the use of the package, please
\usepackage[T1]{fontenc} % if needed

\usepackage{amssymb,amsmath}
\usepackage{bbm}
\usepackage{mathtools}
\usepackage{graphicx}
\usepackage{epsfig}
\usepackage{epstopdf}
\usepackage{setspace}
\usepackage{dsfont}
\usepackage{latexsym}
\usepackage{tikz}
\usepackage{psfrag}
\usepackage{booktabs}
\usepackage{bbm}
\usepackage[toc]{appendix}
\usepackage{color}
\usepackage{datetime}
\usepackage{tensor}
\usepackage{lscape}
\usepackage{lipsum}

\usepackage{tikz}
\usetikzlibrary{arrows,shapes}
\usetikzlibrary{trees}
\usetikzlibrary{patterns}
\usetikzlibrary{matrix,arrows} 				% For commutative diagram
\usetikzlibrary{positioning}				% For "above of=" commands
\usetikzlibrary{calc,through}				% For coordinates
\usetikzlibrary{decorations.pathreplacing}  % For curly braces
\usepackage{pgffor}							% For repeating patterns
\usetikzlibrary{decorations.pathmorphing}	% For Feynman Diagrams
\usetikzlibrary{decorations.markings}
\tikzset{
    vector/.style={decorate, decoration={snake}, draw},
	provector/.style={decorate, decoration={snake,amplitude=2.5pt}, draw},
	antivector/.style={decorate, decoration={snake,amplitude=-2.5pt}, draw},
        smallvector/.style={decorate, decoration={snake,amplitude=1.5pt,post length=0.5mm}, draw},
    fermion/.style={draw=black, postaction={decorate},
        decoration={markings,mark=at position .55 with {\arrow[draw=black]{>}}}},
    fermionbar/.style={draw=black, postaction={decorate},
        decoration={markings,mark=at position .55 with {\arrow[draw=black]{<}}}},
    fermionnoarrow/.style={draw=black},
    gluon/.style={decorate, draw=black,
        decoration={coil,amplitude=4pt, segment length=5pt}},
    scalar/.style={dashed,draw=black, postaction={decorate},
        decoration={markings,mark=at position .55 with {\arrow[draw=black]{>}}}},
    scalarbar/.style={dashed,draw=black, postaction={decorate},
        decoration={markings,mark=at position .55 with {\arrow[draw=black]{<}}}},
    scalarnoarrow/.style={dashed,draw=black},
    electron/.style={draw=black, postaction={decorate},
        decoration={markings,mark=at position .55 with {\arrow[draw=black]{>}}}},
    bigvector/.style={decorate, decoration={snake,amplitude=4pt}, draw},
    arrow/.style={draw=black, postaction={decorate},
        decoration={markings,mark=at position 1 with {\arrow[draw=black]{>}}}},
}

\tikzstyle{block} = [draw, rectangle, 
    minimum height=3em, minimum width=6earticlem]

\definecolor{darkblue}{rgb}{0.2, 0, 0.8}

\numberwithin{equation}{section}

\newcommand{\ang}[1]{\langle #1\rangle}

\usepackage{comment}

\newcommand{\reef}[1]{(\ref{#1})}

\def\be{\begin{equation}}
\def\ee{\end{equation}}
\def\bea{\begin{eqnarray}}
\def\eea{\end{eqnarray}}
\def\ba{\begin{array}}
\def\ea{\end{array}}
\def\bd{\begin{displaymath}}
\def\ed{\end{displaymath}}

\def\Tr{{\rm Tr}}
\def\tr{{\rm tr}}

\def\>{\rangle} %right angle
\def\<{\langle} %left angle
\def\Dsl{D \hskip-.6em \raise1pt\hbox{$ / $ } }
\def\to{\rightarrow}

\newcommand{\lra}{\leftrightarrow}

\title{Generalizations of the Double-Copy: the KLT Bootstrap}

\author[a]{Huan-Hang Chi}
\author[a]{Henriette Elvang}
\author[a]{Aidan Herderschee}
\author[a,b]{Callum R. T. Jones}
\author[a]{Shruti Paranjape}

\affiliation[a]{Leinweber Center for Theoretical Physics, Randall Laboratory of Physics\\ The University of Michigan, Ann Arbor, MI 48109-1040, USA}
\affiliation[b]{Mani L. Bhaumik Institute for Theoretical Physics, Department of Physics and Astronomy, UCLA, Los Angeles, CA 90095, USA}

\emailAdd{hhchi@umich.edu, elvang@umich.edu, aidanh@umich.edu, cjones@physics.ucla.edu, shrpar@umich.edu}

\abstract{
  We formulate a new program to generalize the double-copy of tree amplitudes. The approach exploits the link between the identity element of the ``KLT algebra'' and the KLT kernel and we demonstrate how this leads to a set of KLT bootstrap equations that the double-copy kernel has to satisfy (in addition to locality constraints). We solve the KLT bootstrap equations perturbatively to find the most general higher-derivative corrections to the 4- and 5-point field theory KLT kernel. The new kernel generalizes the string KLT kernel and its associated monodromy relations. It admits new color-structures in the effective theories it double-copies. It provides distinct generalized KK and BCJ relations for the left and right single-color theories and is in that sense a `heterotic'-type double-copy. We illustrate the generalized double-copy in detail for 4d Yang-Mills theory with higher-derivative corrections that produce dilaton-axion-gravity with local operators up order $\nabla^{10} R^4$. Finally, we initiate a search for new double-copy kernels. 
}

\begin{document} 
 \begin{flushright}
{\tt LCTP-21-16} \\
\end{flushright}
\maketitle
\flushbottom

%%%%%%%%%%%%

%%%%%%%%%%%%%%%%%%%%%%%%%%%%%%
%%%%%%%%%%%%%%%%%%%%%%%%%%%%%%
%%%%%%%%%%%%%%%%%%%%%%%%%%%%%%%%%%%%%%%%%%%%%%%%%%%%%%%%%%%%%%%%%%%%%%%%%%%%%%%%%%%%%%%%%%%%%%%%%%%%%%%%%%%%%%%%%%%%%%%%%%%%%%%%%%%%%%%%%%%%%%%%%%%%%%%%%%%%%%%%%%%%%%%%%%%%%%%%%%%%%%%%%%%%%%%%%%%%%%%%%%%%%%%%%%%%%%%%%%%%%%%%%%%%%%%%%%%%%%%%%%%%%%%%%%%%%%%%%%%%%
\section{Introduction}
\label{sec:intro}

Beginning with the pioneering discovery of Kawai-Lewellen-Tye (KLT) \cite{Kawai:1985xq}, the existence of a multiplicative structure, called the  \textit{double-copy}, on the space of relativistic field theories and string theories has become an indispensable tool for the study of scattering amplitudes and beyond. The review \cite{Bern:2019prr} provides an excellent introduction and overview. 

 In this paper, we present a generalization of the double-copy inspired by the KLT formula. Our approach is to \textit{bootstrap} the double-copy multiplication rule that takes the color-ordered tree amplitudes from a pair of (possibly distinct) models, often referred to as \textit{single copies}, and returns a non-color-ordered tree amplitude in a different model known as the \textit{double-copy}. 
The map takes the general bilinear form 
 \begin{equation}
\label{KLT00}
    \mathcal{A}_n^{\text{L}\otimes \text{R}} = \sum_{\alpha,\beta} 
    \mathcal{A}_n^{\textrm{L}}[\alpha]\,
    S_n[\alpha|\beta]\,
    \mathcal{A}_n^{\textrm{R}}[\beta]\,,
\end{equation}
where $\alpha$ and $\beta$ index the set of color-orderings and $S_n[\alpha|\beta]$ is some \textit{a priori} unknown, universal function of Mandelstam invariants called the \textit{double-copy kernel}. The well-known field theory and string theory kernels are reviewed in Sections \ref{sec:KLTAlgebra} and \ref{s:Bklt}. Generalizations of $S_n$ are non-trivial: the kernel plays a key role in canceling potential double-poles in the product of single-copy amplitudes and it also supplies ``missing'' poles. Naive modifications can have dire consequences for locality and may result in an nonsensical result on the LHS of \reef{KLT00}.

Within our proposed generalized framework, the problems we address in this paper are:
\begin{itemize}
\item  \textit{What are the rules for generalizing the kernel $S_n$? Are there multiple distinct double-copy ``multiplication rules''?}
\item \textit{For a given kernel, which models can be used as single-copies: what constraints do they need to satisfy?}
\item \textit{Which models can be produced as a double-copy?}
\end{itemize}
The double-copy can be considered a map
\be 
  \label{KLTmap0}
  \text{(theory)}_\text{L} 
  \otimes 
  \text{(theory)}_\text{R}
  ~=~ \text{(theory)}_{\text{L} \otimes \text{R}} \,,
\ee
in which the multiplication rule $\otimes$ is determined by the double-copy kernel $S_n$. 
A key feature is that both the field theory and string theory KLT double-copy maps contain an \textit{identity element}: a model whose tree amplitudes double-copy with the L/R tree amplitudes of single-copy models to give those same L/R amplitudes as output. 
This is summarized as the {\em KLT algebra}
\begin{equation}
  \label{KLTalgebra0}
     \text{L}\otimes \mathds{1} = \text{L}, \hspace{10mm} \mathds{1} \otimes  \text{R} =  \text{R}, \hspace{10mm} \mathds{1} \otimes \mathds{1} = \mathds{1}\,.
\end{equation}
We refer to the identity element as the {\em zeroth copy} \cite{Monteiro:2014cda}. Intuitively, changing the multiplication rule \eqref{KLTalgebra0} also changes the identity element, and vice versa. In fact, the tree amplitudes of the zeroth copy uniquely determine the double-copy kernel. 

To systematically explore generalizations of KLT, our fundamental assumption is that any KLT double-copy has an identity element $\mathds{1}$  that satisfies \reef{KLTalgebra0}. 
Hence, the first question above can be restated as:
\begin{itemize}
    \item \textit{What are the allowed choices of zeroth copy?}
\end{itemize}
The zeroth copy must have a certain double color-structure, however, not every such choice of local field theory defines a valid double-copy multiplication rule: important constraints arise when we demand the absence of spurious singularities in the double-copy amplitudes. In other words, non-trivial conditions come from requiring that the output of the double-copy are indeed tree amplitudes of a local field theory. 

In this paper, we develop a systematic analysis of the constraints imposed by the KLT algebra \reef{KLTalgebra0} and locality 
on generalizations of the double-copy kernel. In short, the identity $\mathds{1} \otimes \mathds{1} = \mathds{1}$ becomes a bootstrap equation for the double-copy kernel while the two other conditions $\text{L}\otimes \mathds{1} = \text{L}$ and $\mathds{1} \otimes  \text{R} =  \text{R}$ determine the generalizations of the Kleiss-Kuijf (KK) \cite{Kleiss:1988ne}  and Bern-Carrasco-Johansson (BCJ) \cite{Bern:2008qj} relations. 

Let us outline the ideas and some results before diving more into the details. 
For the field theory KLT map, the identity element is the cubic bi-adjoint scalar theory (BAS) with Lagrangian
\begin{equation} 
\label{cubicbas0}
    \mathcal{L}_\text{BAS} = -\frac{1}{2}\left(\partial_\mu \phi^{aa'}\right)^2 -\frac{g}{6}
    f^{abc}\tilde{f}^{a'b'c'}\phi^{aa'} \phi^{bb'} \phi^{cc'}\,, 
\end{equation}
in which the scalar field $\phi^{aa'}$ carries adjoint indices of two groups, say $U(N)$ and $U(N')$. 
We sometimes call this the BAS zeroth copy. 
Furthermore, the string theory zeroth copy \cite{Mizera:2016jhj} can, in the small $\alpha'$-expansion, be viewed as the BAS model with a very particular selection of higher-derivative corrections, all with coefficients completely fixed in terms of $\alpha'$. 

This motivates the study of the class of double-copy kernels that arise from 
the most general deformations of the BAS zeroth copy model with local higher-derivative (h.d.) operators,
\begin{equation}
\label{BAShd0}
    \mathcal{L}_{\text{BAS+h.d.}} =  
    \mathcal{L}_{\text{BAS}} + \sum_{i}\frac{c_i}{\Lambda^{\Delta_i-d}}\mathcal{O}_i\,.
\end{equation}
 Here $\mathcal{O}_i \sim \partial^{2k} (\phi^{aa'})^n$ are local scalar operators of dimension $\Delta_i$ and the constants $c_i$ are the Wilson coefficients.  

The KLT kernel bootstrap equation $\mathds{1} \otimes \mathds{1} = \mathds{1}$ can be understood as a rank-condition on the matrix of doubly color-ordered zeroth-copy tree amplitudes; we elaborate on this in Sections \ref{s:Bklt} and \ref{s:KLTdefs}. The rank is $(n-3)!$ for $n$-particle scattering in the BAS and string zeroth copies \cite{Mizera:2016jhj}, but for other zeroth copies it could in principle be different. We argue that in the context of BAS+h.d., the rank of doubly color-ordered amplitudes 
should be independent of $\Lambda$; in particular if the rank changed as $\Lambda \rightarrow \infty$, the double-copy procedure would be singular. Since this is the limit in which the UV physics decouples, our observation is that for finite $\Lambda$ the rank of the associated matrix of scattering amplitudes must be the same as for the BAS zeroth copy. We call rank $(n-3)!$ {\em minimal rank}. Higher-derivative operators generically increase the rank, so imposing the minimal rank condition requires delicate cancellations between contributions of higher-dimension operators with an associated infinite set of non-trivial constraints on the Wilson coefficients $c_i$.

Our analysis shows that the leading higher-dimension operators compatible KLT bootstrap equations are
\begin{equation}
\label{BAShd4pt0}
\begin{split}
\mathcal{L}&\supset 
  -\frac{a_\text{L}+a_\text{R}}{2\Lambda^{4}}
  f^{abx}f^{cdx}f^{a'b'x'}f^{c'd'x'}
  (\partial_\mu\phi^{aa'})
  (\partial^\mu\phi^{bb'})\phi^{cc'}\phi^{dd'} \\ 
  &\hspace{5mm}+\frac{a_\text{L}}{\Lambda^{4}}
  f^{abx}f^{cdx}d^{a'b'x'}d^{c'd'x'}
  (\partial_\mu\phi^{aa'})\phi^{bb'}
  (\partial^\mu\phi^{cc'})\phi^{dd'} \\
  &\hspace{5mm}+\frac{a_\text{R}}{\Lambda^{4}}
  d^{abx}d^{cdx}f^{a'b'x'}f^{c'd'x'}
  (\partial_\mu\phi^{aa'})\phi^{bb'}
  (\partial^\mu\phi^{cc'})\phi^{dd'} 
  + \ldots\ .
\end{split}
\end{equation}
The presence of the symmetric tensors $d^{abc} = \text{Tr}\big[T^a\{T^b,T^c\}\big]$ modify the $U(1)$ decoupling identity; a feature familiar from the strings zeroth copy which has $a_\text{L}/\Lambda^4=a_\text{R}/\Lambda^4 =-1/(6\pi\alpha')$. The generalized double-copy kernel, based on \reef{BAShd4pt0}, has arbitrary coefficients for the L and R sectors and it therefore treats the L and R sector single-copy amplitudes differently; in this sense, it can be thought of as a ``heterotic''-type double-copy. We have solved for the double-copy kernel up to and including 16-derivative order at 4-point and 6-derivative order at 5-point. These results are valid in general $d \ge 4$ dimensions. 

We apply the generalized double-copy to several examples, specifically 4d Yang-Mills theory with higher-derivative terms (YM+h.d.) up to 10-derivative order.
Specifically, up to 4-derivative order, we find that the most general L sector single-copy model of non-abelian gauge bosons, compatible with the generalized KK and BCJ relations, takes the form
\begin{equation}
\label{YMHDLag}
    \mathcal{L}_{\text{L}} = -\frac{1}{4}F^{a}_{\mu\nu}F^{a\mu\nu} + \frac{g^\text{L}_{F^3}}{\Lambda^2}f^{abc}{F^a_{\mu}}^{\;\nu}{F^b_{\nu}}^{\;\rho}{F^c_{\rho}}^{\;\mu} + \frac{2a_\text{L}g_{\text{YM}}^\text{L}}{g^2\Lambda^4}d^{abcd}(F^4_{\text{ss}})^{abcd} +  \ldots,
\end{equation}
where the ellipsis indicates that the complete effective action contains higher-derivative operators of the form $D^k F^n$ with $k>0$ and $n \ge4$. In (\ref{YMHDLag}), $F^4_{\text{ss}}$ denotes the linear combination 
\begin{align}
    d^{abcd}(F^4_{\text{ss}})^{abcd} &
    \equiv \text{Tr}\left[{F_\mu}^\nu {F_\nu}^\rho {F_\rho}^\sigma {F_\sigma}^\mu\right]
    +2\text{Tr}\left[{F_\mu}^\nu {F_\rho}^\sigma{F_\nu}^\rho  {F_\sigma}^\mu\right] \nonumber\\
    &\hspace{10mm}-\frac{1}{2}\text{Tr}\left[F_{\mu\nu}F^{\mu\nu} F_{\rho\sigma}F^{\rho\sigma}\right]-\frac{1}{4} \text{Tr}\left[F_{\mu\nu}F_{\rho\sigma}F^{\mu\nu} F^{\rho\sigma}\right],
\end{align}
which also appears in the open superstring effective action \cite{Tseytlin:1997csa}, but here with the \textit{free} coefficient $a_\text{L}$. The result is a generalization of the $\alpha'$-expansion of the open-string gluon amplitude and of previous work to extend the double-copy to higher-derivative operators \cite{Broedel:2012rc}. (See also \cite{Carrasco:2019yyn,Carrasco:2021ptp}.) In particular, since the above Yang-Mills deformation contains operators with more general color-tensor structures, it is not restricted to contractions of $f^{abc}$. 

The double-copy of 4d YM+h.d.~gives dilaton-axion (NS-NS) gravity with higher-derivative interactions. Due to the $F^3$-term in \reef{YMHDLag}, the graviton amplitudes have contributions from exchanges of both dilatons and axions. 

Separately, we study the double-copy bootstrap equations with rank greater than $(n-3)!$, including zeroth-copy candidates that are not simply BAS+h.d.
In each of the higher-rank examples, we find potential obstructions from spurious poles at higher points. This leads us to suspect that a well-defined local double-copy requires a zeroth copy whose matrix of doubly color-ordered tree amplitudes must have rank $(n-3)!$.

In the following subsections, we review the KLT formula and the KLT algebra in more detail and then we describe how a naive choice of zeroth copy results in spurious singularities. The Introduction ends with an overview of results and an outline of the rest of the paper.

\subsection{Field Theory KLT}
\label{sec:KLTAlgebra}

At 4-point, the KLT formula \reef{KLT00} has only one term and can be written in various equivalent forms, for example\footnote{We use the Mandelstam conventions $s \equiv (p_1+p_2)^2$, $t \equiv (p_1+p_3)^2$ and $u \equiv (p_1+p_4)^2$.}
\be
  \label{4ptKLTex}
  \begin{split}
  \mathcal{A}_4^{\text{L}\otimes \text{R}} =
   \mathcal{A}_4^{\textrm{L}}[1234]\,
    S_4[1234|1234]\,
    \mathcal{A}_4^{\textrm{R}}[1234]\,,
    ~~~~~~ &S_4[1234|1234] =-\frac{su}{g^2t} \,,\\
  \mathcal{A}_4^{\text{L}\otimes \text{R}} =
   \mathcal{A}_4^{\textrm{L}}[1234]\,
    S_4[1234|1243]\,
    \mathcal{A}_4^{\textrm{R}}[1243]\,,
    ~~~~~~ &S_4[1234|1243] =-\frac{s}{g^2} \,,
    \end{split}
\ee
where $g$ denotes a scale which soon will be identified more precisely. The fact that both versions produce the same result $\mathcal{A}_4^{\text{L}\otimes \text{R}}$ for the double-copy relies on the identity
\be\label{4ptBCJ}
  \text{BCJ:}
  ~~~~~
  \mathcal{A}_4[1234] -\frac{t}{u} \mathcal{A}_4[1243] =0 \,,
\ee 
as can be seen from subtracting the two equations in \reef{4ptKLTex}.
The condition \reef{4ptBCJ} is an example of a {\em BCJ relation}. Similarly, the independence of ``basis choice'', i.e.~choice of which of the possible color-orderings of amplitudes participate in the double-copy, is ensured by the L and R single-copy amplitudes satisfying, in addition to \reef{4ptBCJ}, the following KK relations 
\be
  \label{4ptKKtr}
  \text{Trace-reversal:}
  ~~
  \mathcal{A}_4[1432] = \mathcal{A}_4[1234]\,,
  ~~~
  \mathcal{A}_4[1342] = \mathcal{A}_4[1243]\,,
  ~~~
  \mathcal{A}_4[1423] = \mathcal{A}_4[1324]\,,
\ee  
and
\be
  \label{4ptU1}
  \text{$U(1)$-decoupling:}
  ~~~~
  \mathcal{A}_4[1234] + \mathcal{A}_4[1243] 
  + \mathcal{A}_4[1423] = 0 \,.\\[1mm]
\ee
These 5 requirements \reef{4ptBCJ}-\reef{4ptU1} reduce the 6 independent color-ordered tree amplitudes to just 1.

At $n$-point there are similar requirements, jointly called {\em KKBCJ relations} in the following, which reduce the number of independent color-ordered amplitudes from $(n-1)!$ (due to the cyclicity of the single trace in the color-decomposition) to $(n-3)!$. This is exactly the number of color-orderings summed over in the $n$-point field theory KLT relations \reef{KLT00} and the KKBCJ relations ensure that the result of the double-copy is independent of the choice of $(n-3)!$ color-orderings used in the double-copy. 

The tree amplitudes of Yang-Mills (YM) theory and super Yang-Mills (SYM) theory solve the field theory KKBCJ relations. The same is true for the tree amplitudes of chiral perturbation theory ($\chi$PT), which is the nonlinear sigma model (NLSM) based on the coset $U(N) \times U(N)/U(N)$. Finally, the doubly color-ordered tree amplitudes of the 
bi-adjoint scalar model defined by the Lagrangian \reef{cubicbas0} also satisfy the field theory KKBCJ relations. The result of using each of these models as the L and R inputs of the double-copy is shown in Table \ref{table:KLT}.

\begin{table}[t!]
\be
 \nonumber
 \begin{array}{c|cccc}
     \text{L / R} & \text{BAS} &~~\chi\text{PT} & \text{YM} & \mathcal{N}=4~\text{SYM}  \\[1mm]
     \hline
     \text{BAS} & \text{BAS} &~~\chi\text{PT} & \text{YM} & \mathcal{N}=4~\text{SYM} \\[1mm]
     \chi\text{PT} & \chi\text{PT} &~~ \text{sGal} & \text{BI} & \mathcal{N}=4~\text{sDBI} \\[1mm]
     \text{YM} & \text{YM}  &~~  \text{BI} & ~\text{NS-NS gravity} & \mathcal{N}=4~\text{SG}\\[1mm]
    \mathcal{N}=4~\text{SYM}  ~&~ \mathcal{N}=4~\text{SYM} ~&~ \mathcal{N}=4~\text{sDBI} ~&~ \mathcal{N}=4~\text{SG} ~&~ \mathcal{N}=8~\text{SG}
 \end{array}
\ee
\caption{\label{table:KLT}The table shows the tree-level  double-copy $\text{L}\otimes \text{R}$ for a selection of different choices of L and R single-color models. BAS is the cubic bi-adjoint scalar model which is described more in the main text. The single-color models are $\chi$PT = chiral perturbation theory,  YM = Yang-Mills theory, and $\mathcal{N}=4$  super Yang-Mills theory (SYM). These theories double-copy as shown to: the special Galileon (sGal), Born-Infeld theory (BI), $\mathcal{N}=4$ supersymmetric Dirac-Born-Infeld theory (sDBI), and 
and NS-NS gravity which is the $\alpha'\rightarrow 0$ limit of the NS-NS sector of superstring theory describing Einstein gravity coupled to a dilaton and a 2-form gauge field (in 4d the latter is dualized to an axion). Finally, SG stands for supergravity. The important point we want to make here is that BAS acts as the identity under the field theory KLT map.}
\end{table}

Interestingly, the table shows that the BAS model acts as the identity element for the double-copy multiplication rule, as advertised. The tree amplitudes $m_n[\alpha|\beta]$ in the BAS model are doubly color-ordered with single-trace structures under each of the two color-group factors. Diagrams that contribute to a given  doubly color-ordered  amplitude must be compatible with the color-orderings of both group factors; for example
\be
  \label{4ptBASex}
  m_4[1234|1234] = \frac{g^{2}}{s} + \frac{g^{2}}{u}
  \,,~~~~
  m_4[1234|1243] = -\frac{g^{2}}{s} \,.
\ee
When BAS amplitudes are used as input in the KLT formula, one color-group is actively taking part in the double-copy while the other is inert. The output amplitude inherits the inert color-orderings. Specifically, with $\gamma$ and $\delta$ the inert color-orderings, we have
\begin{equation}
\label{preBCJ}
    \mathcal{A}_n^{\textrm{R}}[\gamma] = \sum_{\alpha,\beta} 
    m_n[\gamma|\alpha]\,
    S_n[\alpha|\beta]\,
    \mathcal{A}_n^{\textrm{R}}[\beta]\,,
    ~~~~~
        \mathcal{A}_n^{\textrm{L}}[\delta] = \sum_{\alpha,\beta} 
        \mathcal{A}_n^{\textrm{L}}[\alpha]\,
        S_n[\alpha|\beta] \,
        m_n[\beta|\delta]\,,
\end{equation}
and 
\begin{equation}
\label{KLTmn}
    m_n[\gamma|\delta] = \sum_{\alpha,\beta} 
    m_n[\gamma|\alpha]\,
    S_n[\alpha|\beta]\,
    m_n[\beta|\delta]\,,
\end{equation}
where the sums on $\alpha$ and $\beta$ are over a choice of two (possibly distinct) subsets of $(n-3)!$ color-orderings. 
The formulas (\ref{preBCJ}) and (\ref{KLTmn}) are the mathematical manifestations of the KLT algebra \reef{KLTalgebra0}.

We can view \reef{KLTmn} as matrix multiplication of the $(n-3)! \times (n-3)!$ 
KLT kernel $S_n$ with two $(n-3)! \times (n-3)!$ submatrices of BAS tree amplitudes $m_n$. These submatrices are invertible, hence if \reef{KLTmn} is multiplied from both the left and right by  $(m_n)^{-1}$,  we find that
\be
\label{naiveSinv}
S_n = \big(m_n\big)^{-1}\,.
\ee 
This relationship was first noted for the BAS zeroth copy in  \cite{Cachazo:2013iea}. For the 4-point case, the relation is particularly simple. For example, using \reef{4ptBASex}, we find
\be
\begin{split}
  S_4[1234|1234] &= \big(m_4[1234|1234]\big)^{-1} = -\frac{s u}{t g^2}\,, \\
  ~~~~~
  S_4[1234|1243] &= \big(m_4[1243|1234]\big)^{-1} = -\frac{s}{g^2}\,.
  \end{split}
\ee
Indeed, this result matches the kernel in \reef{4ptKLTex} and we have now identified $g$ as the BAS cubic coupling constant. Using the expressions for $S_4$ and $m_4$, one can reproduce the 4-point KKBCJ relations \reef{4ptBCJ}-\reef{4ptU1} from \reef{preBCJ}. 
This discussion illustrates the statement we made previously that  $\mathds{1} \otimes \mathds{1} = \mathds{1}$ links the field theory kernel to the tree amplitudes of BAS \reef{naiveSinv} while $\text{L}\otimes \mathds{1} = \text{L}$ and $\mathds{1} \otimes  \text{R} =  \text{R}$ encode the KKBCJ conditions. 

As described above, the KKBCJ conditions \reef{preBCJ} allow some single-copy models (such as YM, SYM, $\chi$PT) to be double-copied but not others. For example, the amplitudes of a model with an adjoint scalar $\varphi$ and a fully symmetric interaction $\tr\big(\varphi\{\varphi,\varphi\}\big)$ do not obey KKBCJ. Similarly, it was shown in \cite{Broedel:2012rc} that adding an operator Tr$[F^3]$ to YM theory is KKBCJ-compatible, 
but adding any form of Tr$[F^4]$ is not. A central motivation for our work is to understand and expand the space of theories that can be double-copied. 

\subsection{Beyond Field Theory KLT}
\label{s:Bklt}
Perhaps the reader is surprised that YM with a Tr$[F^4]$ fails the KKBCJ constraints; after all, the low-energy $\alpha'$-expansion of the open string  gluon amplitude does indeed produce an operator of the form Tr$[F^4]$, and the KLT formula originated as a map from open string amplitudes to closed string amplitudes \cite{Kawai:1985xq}. These statements appear to be in conflict with Tr$[F^4]$ not passing the KKBCJ constraints. The resolution is simple: the {\em string} KLT formula uses a {\em string} KLT kernel $S_n^{(\alpha')}$ in place of the field theory KLT kernel $S_n$. In the  $\alpha' \to 0$ limit, the strings KLT kernel reduces at the leading order to the field theory KLT kernel $S_n$ described above. If we replace $S_n$ in the KKBCJ relations \reef{preBCJ} by the $\alpha'$-expansion of $S_n^{(\alpha')}$, one can indeed show that the Tr$[F^4]$ operator {\em is} allowed. Moreover, its Wilson coefficient is fixed to be exactly its value in the open string gluon amplitude. 
This example indicates that in order to systematically examine the double-copy in the context of higher-derivative operators with  general Wilson coefficients, we need to  consider generalizations of the KLT kernel. 

As discussed around  \reef{naiveSinv}, the field theory KLT kernel, $S_n$ is the inverse of a submatrix of  BAS tree-amplitudes $m_n$ \cite{Cachazo:2013iea}. Mizera \cite{Mizera:2016jhj} showed that  the string theory KLT kernel $S_n^{(\alpha')}$ similarly is the inverse of a $(n-3)! \times (n-3)!$ submatrix of amplitudes $m_n^{(\alpha')}$ that are obtained from the 
BAS tree-amplitudes $m_n$ by replacing $g^2/p^2$ by $1/\sin(\pi \alpha' p^2)$ or $1/\tan(\pi \alpha' p^2)$. For example,
\be
  \label{exm4str}
  m_4[1234|1243] = - \frac{g^2}{s}
   ~~~\longrightarrow~~~
  m_4^{(\alpha')}[1234|1243] =-\frac{1}{\sin(\pi \alpha' s)} \,.
\ee 
When one replaces 
\be
  m_n \to m_n^{(\alpha')}
  ~~~~\text{and}~~~~
  S_n \to S_n^{(\alpha')} =  (m_n^{(\alpha')})^{-1}\,,
\ee
the KKBCJ relations \reef{preBCJ} become the open string monodromy relations \cite{BjerrumBohr:2009rd, Plahte:1970wy, Feng:2010my, Stieberger:2009hq, BjerrumBohr:2010zs}. Moreover, $m_n^{(\alpha')}$ solves 
\reef{KLTmn}. 
Thus, the model associated with the amplitudes $m_n^{(\alpha')}$ is indeed the identity for the  KLT algebra \reef{KLTalgebra0} with multiplication kernel $S_n^{(\alpha')}$.

From a conventional field theory point of view, it may be surprising that the $m_n^{(\alpha')}$ are called amplitudes. However, in the $\alpha'$-expansion, it makes perfect sense. For example, consider the low-energy expansion $s \ll (\alpha')^{-1}$ of \reef{exm4str}:
\be
  m_4^{(\alpha')}[1234|1243] 
  = -\frac{1}{\pi \alpha' s} - \frac{1}{6} \pi \alpha' s 
  -\frac{7}{360} (\alpha' \pi s)^3 
  +\dots\,.
\ee
The leading term matches that in $m_4[1234|1243]$ in \reef{exm4str} with $g^2=1/(\pi \alpha')$ and the subleading terms arise from higher-derivative corrections to the BAS model \reef{cubicbas0}. Schematically, we can write
\be 
  \label{LBASstr}
  \mathcal{L}_{\alpha'} 
  = 
  \mathcal{L}_\text{BAS}
  + \alpha' \,\partial^2 \phi^4
  + \alpha'^3 \,\partial^6 \phi^4
  + \ldots\,.
\ee
Hence, the amplitudes $m_n^{(\alpha')}$ are the re-summed tree amplitudes of an effective field theory consisting of the cubic bi-adjoint scalar model with higher-derivative corrections. One way to view this is that the string KLT kernel provides an explicit one-parameter deformation of the field theory KLT kernel with an associated KLT algebra \reef{KLTalgebra0} and it generates healthy double-copy amplitudes. 

It is clear even from the schematic form \reef{LBASstr} that $ \mathcal{L}_{\alpha'}$ involves a particular selection of higher-derivative terms; notably $\phi^4$ or $\partial^4 \phi^4$ are absent. One goal of our analysis is to  determined which local operators are allowed in a generalized zeroth copy of the form BAS+higher-derivative terms.  

To illustrate how this is done, consider the 4-point case with $(n-3)!=1$. In that case, equation \reef{naiveSinv} simply gives 
$S_{4}[\alpha|\beta]=1/m_{4}[\beta|\alpha]$ and the sum in \reef{KLTmn} is over just a single choice of color-orderings $\alpha$ and $\beta$, so the double-copy bootstrap equation $\mathds{1} \otimes \mathds{1} = \mathds{1}$ becomes
\begin{equation}
\label{KLTmn2}
    m_4[\gamma|\delta] =
    m_4[\gamma|\alpha]\,
    \frac{1}{m_4[\beta|\alpha]}\,
    m_4[\beta|\delta]\,.
\end{equation}
Rearranging \reef{KLTmn2}, it says
\be
  \label{KLTmn3}
  \det \left(
  \begin{array}{cc}
  m_4[\beta|\alpha] & m_4[\beta|\delta]
  \\
  m_4[\gamma|\alpha] & m_4[\gamma|\delta]
  \end{array}
  \right)
  =0\,.
\ee
Since this holds for all choices of the $(n-1)!=6$ color-orderings $\alpha,\beta,\gamma,\delta$, this says that all such $2 \times 2$ minors vanish. In other words, 
the $6 \times 6$ matrix of doubly color-ordered zeroth-copy tree amplitudes $m_4$ must have rank 1. The determinant in \reef{KLTmn3} trivially vanishes for amplitudes of the cubic BAS model, but once higher-derivative operators are included, \reef{KLTmn3} becomes a non-trivial constraint. This is an example of how $\mathds{1} \otimes \mathds{1} = \mathds{1}$ becomes a bootstrap equation for the zeroth copy and hence for the double-copy kernel.

More generally, at $n$-point, the equation $\mathds{1} \otimes \mathds{1} = \mathds{1}$ links the number of color-orderings summed over in the double-copy \reef{KLT00} to the rank $R_n$ of the $(n-1)! \times (n-1)!$ matrix of tree amplitudes of the zeroth copy.  For the BAS and string theory zeroth-copy models, the rank is $R_n = (n-3)!$, but in a general setting, $R_n$ is not predetermined and different values can be explored. 

Thus, to examine the possible generalizations of the double-copy kernel, 
one can consider {\em any} local model of a bi-adjoint scalar field $\phi^{aa'}$, compute its color-ordered tree amplitudes $m_n[\alpha|\beta]$, and subject them to the {\em double-copy bootstrap equation} $\mathds{1} \otimes \mathds{1} = \mathds{1}$ with some choice of rank $R_n$. Inverting the resulting $R_n$-rank matrix gives the generalized double-copy kernel. Amplitudes of L and R single-copy models can then be subjected to the generalized KKBCJ relations $\text{L}\otimes \mathds{1} = \text{L}$ and $\mathds{1} \otimes  \text{R} =  \text{R}$ and finally double-copied.

Returning to the 4-point example, the $6\times 6$ matrix of amplitudes $m_4$ can in full generality be parameterized using just 3 functions after using cyclic symmetry and momentum relabelings. The rank $R_4 = 1$ double-copy bootstrap equations  \reef{KLTmn3} then fix two of those functions in terms of the third one which must solve a self-consistency condition. This latter equation is then solved perturbatively in the momentum expansion, and subject to additional constraints of locality, we find the most general allowed higher-derivative corrections to the BAS model. This is presented in detail in Section \ref{sec:bootstrap}. 

\subsection{Locality Constraints}
\label{sec:nonlocal}
 
The requirements of locality impose additional constraints on zeroth-copy solutions to the bootstrap equation $\mathds{1} \otimes \mathds{1} = \mathds{1}$. Moreover, for a given kernel, one must make sure that the result of the double-copy represents the tree amplitude of some local theory; it must have the correct simple poles and factorize correctly on each of them, and it must be free of spurious poles.

Even in the familiar case of the field theory KLT formula, the statement that the double-copy amplitudes contain the expected physical factorization singularities, and only those singularities, is not manifest. 
The problem is three-fold: first, expressions such as (\ref{KLT00}) naively contain double-poles from the product of single-copy amplitudes with singularities in the same channel. Second,  zeroes of $\text{det}(m_n)$ may naively produce nonphysical or spurious singularities in the double-copy via poles in the kernel. Third, since we sum over a restricted set of $(n-3)!$ orderings, the combined set of singularities in the L and R single-copies could, in general, not contain all the expected singularities of the double-copy amplitude. 

It is instructive to see explicitly how these problems are resolved in the field theory KLT formula. As a simple example, consider the product of YM tree amplitudes
\begin{equation}
   \label{A4LA4R}
    \mathcal{A}_4^\text{L}[1234]\, \mathcal{A}_4^\text{R}[1234]\,.
\end{equation}
%%%
It has double poles in the $s$- and $u$-channels, but is missing the $t$-channel singularity which must be present if the double-copy is a model of gravity with graviton exchange in every channel. However, in the actual field theory KLT formula \reef{4ptKLTex}, the product \reef{A4LA4R} is multiplied by the kernel 
\begin{equation}
  \label{polesS4}
   S_4[1234|1234] = \left(m_4[1234|1234]\right)^{-1} = -\frac{su}{g^2t}\,,
\end{equation}
which cures all the problems: it reduces the $s$- and $u$-channel double-poles to simple poles, and while $\text{det}(m_4) = -g^2t/(su)$ does have a zero at $t=0$,  rather than generating a spurious singularity, it provides precisely the ``missing'' $t$-channel pole in the double-copy! 

In general, to avoid spurious or higher-order singularities, the correspondence between zeros of determinants and ``missing'' poles must persist at multiplicity $n$ and for all basis choices of $(n-3)!$ color-orderings. The fact that there exists \textit{any} model with this property is something of a miracle!   It is incredible that such an intricate structure can be provided by the mundane-looking BAS model.\footnote{As a cubic scalar model, its potential is unbounded from below and there is no sign in the Lagrangian that this model should play as fundamental a role as it does for the double-copy. The BAS model does however arise naturally from the BCJ or CHY formulations of the double copy by replacing the kinematic numerator factors in a single-copy amplitude in color-kinematic form by a set of color-factors.}

To illustrate the delicate nature of the double copy, let us analyze the pole structure again after making an innocent-looking deformation to the BAS Lagrangian by adding to it
\begin{equation}
\label{deformation}
    \Delta\mathcal{L} = \lambda\, d^{abcd}\tilde{d}^{a'b'c'd'}\phi^{aa'}\phi^{bb'}\phi^{cc'}\phi^{dd'},
\end{equation}
where $d^{abcd} \equiv \text{Tr}\left[T^a T^{(b} T^c T^{d)}\right]$. Turning on this deformation modifies the amplitudes as
\begin{align}
    m_4\left[1234|1234\right] =  m_4\left[1234|1432\right]  = -\frac{g^2t}{su}+ \lambda, \hspace{5mm} m_4\left[1234|1243\right] &=-\frac{g^2}{s} + \lambda\,,
\end{align}
with all other orderings given by relabelling. The rank of the full $6\times 6$ matrix, at a generic point in kinematic space, is found to be 2. Therefore, this deformed BAS theory solves the bootstrap equations associated with a $2\times 2$ kernel at 4-point. Repeating the analysis of potential spurious poles, we calculate the determinant of one of these full-rank submatrices
\begin{equation}
\label{sp2x2}
    \text{det}\left(\begin{array}{cc}
 m_4\left[1234|1234\right] & m_4\left[1234|1243\right] \\
 m_4\left[1243|1234\right] & m_4\left[1243|1243\right] \\
\end{array}\right) = -\frac{g^2 \lambda (s+2 t)^2}{s t u}.
\end{equation}
This expression has a zero at $s+2t=0$ which does \textit{not} correspond to a physical singularity in the double-copy. Furthermore, since this pole can occur at arbitrarily low energies, it cannot be discarded as an artifact of the EFT description. Clearly, turning on the deformation (\ref{deformation}) has broken whatever special property of cubic BAS (\ref{cubicbas0}) was responsible for the formula (\ref{KLT00}) producing a physical scattering amplitude.

Based on this example, one could take a very pessimistic view and conclude that the cubic BAS model is the \textit{unique} model with the  special properties that ensure locality.  As discussed in the previous section, we can immediately dismiss this possibility as we know the string kernel produces a healthy KLT formula free of spurious poles. At 4-point the determinant that potentially generates spurious singularities is given by
\begin{equation}
\begin{split}
    \text{det}\left(m_4^{\text{string}}[1234|1234]\right) &= -\frac{\sin(\pi \alpha' t)}{\sin(\pi \alpha' s) \sin(\pi \alpha' u)} \\
    &= \frac{-t}{\pi \alpha' s u}+\frac{\pi\alpha'}{3}t-\frac{1}{45}\pi^{3}\alpha'^{3}(s^{3}+u^{3})+O(\alpha'^{4}).
\end{split}
\end{equation}
In the EFT context, the singularity at $t=0$ in the $E\ll (\alpha')^{-1/2}$ expansion corresponds to the otherwise missing $t$-channel singularity, just as in discussion around\reef{polesS4}.\footnote{Of course the string kernel also produces a healthy KLT formula free of spurious poles even when $E\gtrsim (\alpha')^{-1/2}$, outside the EFT regime. In that case, the infinite tower of zeros in (\ref{exm4str}) with $t<0$ are ``missing'' physical poles corresponding to \textit{massive} string states exchanged in the $t$-channel. For the double-copy to make sense the physical spectrum must  contain the corresponding massive states. The other infinite tower in (\ref{exm4str}) with $t>0$ are tachyonic and non-physical. In this case, these tachyonic poles must cancel against corresponding zeros in the single-copy amplitudes for the double-copy to make sense when $E\ll (\alpha')^{-1/2}$. This is precisely what happens in the double-copy of open string theory, where the tachyonic poles precisely cancel against an infinite tower of zeros of the Veneziano amplitude. The connection between the zeros of the Veneziano amplitude and the double-copy was previously identified in \cite{Boels:2014dka}.

This example may inspire one to consider the logical possibility that the spurious pole at $s+2t=0$ in \reef{sp2x2} cancels against a zero arising from the sum in the KLT formula. This requires additional  cancellations that go beyond those of the usual field theory KLT formula and in general they may not occur.}

In both of our \textit{healthy} examples, the cubic BAS and string theory kernels, the rank of the zeroth-copy matrix is $(n-3)!$. In the string theory case, this requires intricate cancellations between the Wilson coefficients of higher-derivative operators at different mass dimensions. In the \textit{unhealthy} case \reef{deformation}, the rank was increased to 2 at 4-point. This distinction is representative of a more general (empirically observed) pattern. If the generalized zeroth copy is cubic BAS deformed by higher-derivative operators suppressed by an EFT scale $\Lambda$, then spurious determinant zeroes do not appear if the rank of the zeroth-copy matrix is continuous as $\Lambda\rightarrow \infty$. In this case, just as in the string theory example, the higher-derivative operators must be tuned to ensure that the rank remains $(n-3)!$. For such generalized kernels, physical scattering amplitudes, without spurious low-energy singularities, are observed to be produced by the formula (\ref{KLT00}), for all choices of single-copy amplitudes satisfying the associated generalized KKBCJ conditions \textit{without any additional constraints}. 

In much of the Introduction we have focused on generalizations of the double-copy that arise from adding higher-derivative operators to the BAS model, but one can also look for other types of solutions to the KLT bootstrap equations. 
We initiate such a search for zeroth copies at various ranks. The locality constraints are again essential and indicate that minimal rank $(n-3)!$ may play a special role for generalizations of the double-copy.

\subsection{Previous Work}

Let us note previous work toward generalizing the double-copy of tree amplitudes in the context of field theory. Extending the double-copy to include higher-derivative local operators with arbitrary Wilson coefficients was first studied in \cite{Broedel:2012rc} and further explored in \cite{Elvang:2018dco,Elvang:2020kuj,Bern:2017tuc,Bern:2017rjw,Carrillo-Gonzalez:2019aao}. Using the BCJ formulation of the double-copy to include higher-derivative operators, including some with different color-structures was recently studied in \cite{Carrasco:2019yyn,Carrasco:2021ptp}; we comment further on this in Section \ref{sec:discussion}.

The double-copy has also been studied in the context of massive theories. Much work on the massive double-copy focuses on massive string modes \cite{Gomis:2021hxa,Guillen:2021mwp} or the dimensional reduction of massless theories \cite{Chiodaroli:2015rdg,Chiodaroli:2017ehv,Chiodaroli:2018dbu, Bautista:2019evw, Bern:2019nnu,Brandhuber:2021kpo,Momeni:2020hmc}. A sharp break from this trend was the bottom-up approach taken in \cite{Momeni:2020vvr,Johnson:2020pny}. In particular, \cite{Johnson:2020pny} showed that spurious singularities appear in the KLT kernel at 5-point for generic mass deformations unless the minimal rank condition was imposed. The work \cite{Johnson:2020pny} was an early inspiration for the approach in this paper.

Finally, significant work has focused on generalizing the double-copy to states in more general representations, such as the pseudo-real \cite{Chiodaroli:2015wal} and (anti-)fundamental representations \cite{Johansson:2014zca,Johansson:2015oia,delaCruz:2015dpa,Brown:2018wss,Brown:2016hck,Naculich:2015coa,Anastasiou:2016csv,delaCruz:2016wbr}.
Furthermore, the double copy can be extended to quiver gauge theories, which contain states in a variety of representations \cite{Chiodaroli:2013upa}. Finally, in three dimensions, the double-copy is applicable to gauge theories with states transforming in Lie 3-algebras \cite{Huang:2012wr, Huang:2013kca}.

\subsection{Outline of this Paper}

In Section \ref{sec:introklt}, we give a precise technical description of the generalized KLT double-copy program. In particular, we detail the relation between the zeroth-copy scattering amplitudes, the generalized KLT kernel, and the generalized KKBCJ conditions. We also show how the KLT bootstrap equation is a rank condition on the matrix of doubly color-ordered zeroth-copy amplitudes. 

In Section \ref{sec:bootstrap3pt}, we study the KLT bootstrap at 3-point as an informative warm-up for the higher-point analyses. 

In Section \ref{sec:bootstrap}, we show how the rank 1 KLT bootstrap conditions at 4-point reduce to a simple non-linear constraint on a single function of Mandelstam invariants and this function determines all the zeroth-copy amplitudes. For higher-derivative corrections to BAS, this condition (together with a set of additional locality constraints) can be solved order-by-order in the derivative expansion. The most general solution is presented explicitly up to and including operators of the form $\partial^6 \phi^4$.\footnote{It is straightforward to go to much higher orders and we have done so.} 

In Section \ref{sec:HDYM}, the perturbative generalized KLT kernel is used to calculate the most general higher-derivative corrected Yang-Mills amplitudes consistent with the generalized KKBCJ conditions up to and including operators of the form $\text{Tr}[D^6 F^4]$. The resulting single-copy amplitudes are observed to contain operators that do not appear when either the cubic BAS or the string theory KLT kernel is used. The resulting generalized double-copy is examined up to order $\nabla^{10} R^4$. We examine all three helicity sectors: MHV, self-dual (all-plus), and next-to-self-dual (NSD). 

In Section \ref{sec:5pt}, we study the 5-point rank 2 double-copy bootstrap equations and calculate  
the perturbative generalized KLT kernel at 5-point up to $\partial^6 \phi^5$. 
Importantly, this imposes no constraints on the Wilson coefficients of the 4-point zeroth-copy amplitudes. This is a crucial check because if the higher-point double-copy bootstrap imposed conditions on parameters in the lower-point kernel, then we would not be able to solve for the generalized double-copy kernel individually at each $n$ since we would not know all the constraints from higher-point. 
 We solve the generalized KKBCJ conditions for Yang-Mills with higher-derivative corrections in the self-dual (all-plus) helicity sector up to operators $\text{Tr}[D^6 F^5]$. 

In Section \ref{sec:QBAS}, we explore the possibility of constructing a generalized KLT kernel based on zeroth-copy models that are \textit{not} higher-derivative corrections to cubic BAS. Specifically, we consider the addition of more general $\phi^3$ interactions including totally symmetric, $d^{abc}$, color tensors, as well as a quartic bi-adjoint theory with a leading $\phi^4$ interaction. In both cases, the matrix rank is found to be larger than the rank of the original cubic BAS theory, and at sufficient multiplicity, the kernel contains would-be spurious singularities.

In the Discussion and Outlook (Section \ref{sec:discussion}), we outline the possibility of moving terms between the kernel and the L and R single-copy amplitudes using similarity transformations. At 4-point this elucidates some of our results from the generalized double-copy with higher-derivative terms. We further discuss potential constraints from Swampland positivity constraints and how that may provide some understanding of the particular form of the $\alpha'$-expansion of the string KLT kernel. We then discuss the possible connections to the BCJ-form of generalized color-kinematics of \cite{Carrasco:2019yyn,Carrasco:2021ptp} and make some initial contact with that formalism using an exact solution to the bootstrap equation. We discuss various forms of exact solutions, both those that truncate the higher-derivative expansion and also a completely different one from Z-theory. Finally, we mention some other future directions. 

Various technical details are relegated to the Appendices. Appendix \ref{sec:nullvectors} gives the null vectors of the $(n-1)! \times (n-1)!$ matrix of zeroth copy amplitudes and shows that they imply generalized KKBCJ relations for the L and R sector single-copy amplitudes. Appendix \ref{app:NLSM} gives the results for the generalized double-copy of $\chi$PT with itself and with YM including the leading orders of compatible higher-derivative operators. Appendix \ref{app:5pt} lists analytic solutions to the KLT bootstrap at 5-point.

%%%%%%%%%%%%%%%%%%%%%%%%%%%%%%%%%%%%
%%%%%%%%%%%%%%%%%%%%%%%%%%%%%%%%%%%%
%%%%%%%%%%%%%%%%%%%%%%%%%%%%%%%%%%%%
%%%%%%%%%%%%%%%%%%%%%%%%%%%%%%%%%%%%
\section{Double-Copy Bootstrap}\label{sec:introklt}

In the Introduction, we have outlined the ideas of the double-copy bootstrap. The purpose of this section is to make each step of the procedure precise. 

\subsection{Double-Copy Kernel and Zeroth-Copy Models: Bootstrap \texorpdfstring{$\mathds{1} \otimes \mathds{1} = \mathds{1}$}{1x1=1}}
\label{s:KLTdefs}

In the KLT double-copy formula, 
 \begin{equation}
\label{KLTsec2}
    \mathcal{A}_n^{\text{L}\otimes \text{R}} = \sum_{\alpha,\beta} 
    \mathcal{A}_n^{\textrm{L}}[\alpha]\,
    S_n[\alpha|\beta]\,
    \mathcal{A}_n^{\textrm{R}}[\beta],
\end{equation}
the L (R) sector refers to field theories with all states in the adjoint representation of color groups $G_L$ ($G_R$), which could for example be $SU(N)$ ($SU(\tilde{N})$). The single-copy amplitudes  $\mathcal{A}_n^{\textrm{L}}[\alpha]$ ($\mathcal{A}_n^{\textrm{R}}[\beta]$) are color-ordered with respect to a single-trace of $n$ generators of $G_L$ ($G_R$). The structure of the formula \reef{KLTsec2} shows that the double-copy kernel $S_n$ has a color-structure associated with the product $G_L \times G_R$. 
As we indicated in the Introduction, the kernel is the inverse of a submatrix of doubly color-ordered $m_n$ amplitudes, so they must have a color-structure $G_R \times G_L$. The candidates for the zeroth-copy models are local field theories with a single scalar field $\phi^{aa'}$ that transforms in the adjoint of each group factors.

At $n$-point there are $n!$ possible color-orderings for each color-group factor, but only $(n-1)!$ are independent under the cyclicity of each color-trace. We use $\mathbf{m}_n$ to denote the $(n-1)!\times (n-1)!$ matrix of color-ordered tree amplitudes of the zeroth copy. For example, at 4-point we choose the ordering 
$\{ 1234,1243,1324,1342,1423,1432\}$ and the $6\times6$ matrix of zeroth-copy tree amplitudes is then
\begin{equation}
\label{fullm4matrix}
\begin{split}
    \mathbf{m}_4 &= 
    \begin{pmatrix}
    m_4[1234|1234] & m_4[1234|1243] & m_4[1234|1324] & \cdots & m_4[1234|1432] \\
    m_4[1243|1234] & m_4[1243|1243] & m_4[1243|1324] & \cdots & m_4[1243|1432] \\
    \vdots & \vdots & \vdots
    & & \vdots \\
    m_4[1432|1234] & m_4[1432|1243] & m_4[1432|1324] & \cdots & m_4[1432|1432] 
    \end{pmatrix}.
    \end{split}
    \end{equation}
We do not make any assumptions a priori about the properties of the color-ordered amplitudes $m_n$. For example, we do {\em not} assume trace-reversal, so in general 
\be
  m[\beta|\alpha^T] \ne m[\beta|\alpha]   
  ~~~\text{or}~~~
  m[\beta^T|\alpha] \ne m[\beta^T|\alpha],   
\ee  
where for example $\{1234\}^T = \{4321\} = \{1432\}$. Also, we do {\em not} assume that the $\mathbf{m}_n$ is symmetric, i.e.~in general we have
\be
   m_4[\beta|\alpha] \ne m_4[\alpha|\beta]\,.
\ee
For the BAS or string zeroth-copy models, the rank of the matrix $\mathbf{m}_n$ is $(n-3)!$; this is what we call {\em minimal rank}. Moreover, in those two cases, trace-reversal does hold and $\mathbf{m}_n$ is symmetric. Allowing for generalizations makes it possible to incorporate more `heterotic' double-copies in which the L and R constraints are genuinely distinct. 

Suppose more generally that, for some integer $R_n$, there are invertible $R_n \times R_n$  submatrices of $\mathbf{m}_n$. We label such submatrices by a specification of a choice of a subset of $R_n$ orderings for the rows and columns in $\mathbf{m}_n$. We denote the row or \textit{R basis} as $B_R = \{\beta_1,...,\beta_{R_n}\}$ and the column or \textit{L basis} as $B_L = \{\alpha_1,...,\alpha_{R_n}\}$. 
 In matrix notation, we then have
\begin{equation}
    \mathbf{m}_{n}(B_R,B_L) \equiv \begin{bmatrix}
m_n[\beta_1|\alpha_1] & \cdots & m_n[\beta_1|\alpha_{R_n}] \\
\vdots & \ddots & \vdots \\
m_n[\beta_{R_n}|\alpha_1] & \cdots & m_n[\beta_{R_n}|\alpha_{R_n}] 
\end{bmatrix}\,.
\end{equation}
The condition that the zeroth copy is the identity element under the double-copy multiplication rule, $\mathds{1} \otimes \mathds{1} = \mathds{1}$, is that the tree amplitudes copy to themselves using the double-copy kernel; in matrix notation, this is the requirement  
\begin{equation}
\label{1x1=1}
   \mathds{1} \otimes \mathds{1} = \mathds{1}\!:~~~~~
    \mathbf{m}_n\left(B'_R,B_L\right)
    \mathbf{S}_n\left(B_L,B_R\right)
    \mathbf{m}_n\left(B_R,B'_L\right) = \mathbf{m}_n\left(B'_R,B'_L\right).
\end{equation}
Now set $B_L' = B_L$ and $B_R' = B_R$ in \reef{1x1=1} and multiply on both the L and R by $\big(\mathbf{m}_n\left(B_R,B_L\right)\big)^{-1}$. It then follows that 
\begin{equation}
  \label{SnInvmn}
    \mathbf{S}_{n}(B_L,B_R) \equiv \big(\mathbf{m}_{n}(B_R,B_L)\big)^{-1} 
\,.
\end{equation}
Thus, requiring the zeroth copy to be an identity element under the double-copy inevitably
links its tree amplitudes to the double-copy kernel.

This, however, does not exhaust the contents of \reef{1x1=1}. 
Using \reef{SnInvmn}, we have
\begin{equation}
\label{BootstrapEq}
    \text{bootstrap eq:}~~~~~
    \mathbf{m}_n\left(B'_R,B_L\right)
    \big( \mathbf{m}_n\left(B_R,B_L\right) \big)^{-1}
    \mathbf{m}_n\left(B_R,B'_L\right) = \mathbf{m}_n\left(B'_R,B'_L\right),
\end{equation}
This equation is non-trivial for elements of $B_L'$ that are not in $B_L$ and elements of  $B_R'$ not in $B_R$. As such, it constrains the zeroth-copy amplitudes $m_n$: thus \reef{BootstrapEq} is our double-copy bootstrap equation. 

To interpret the constraint \reef{BootstrapEq}, consider for a given basis choice, $B_L$ and $B_R$, the extension of $\mathbf{m}_{n}(B_R,B_L)$ to the $(R_n+1) \times (R_n+1)$ submatrix 
\be
\label{eq:vanishsubmatricesPRE}
M=
\left[
\begin{array}{ccc|c}
 m_n[\beta_1|\alpha_1]&\cdots& m_n[\beta_1|\alpha_{R_n}] & m_{n}[\beta_1|\delta] \\ 
 \vdots & \ddots & \vdots & \vdots \\
 m_n[\beta_{R_n}|\alpha_1]&\cdots& m_n[\beta_{R_n}|\alpha_{R_n}] & m_{n}[\beta_{R_n}|\delta] \\ 
 \hline
 m_n[\gamma|\alpha_1]&\cdots& m_n[\gamma|\alpha_{R_n}] & m_{n}[\gamma|\delta] 
\end{array}
\right]\,,
\ee
where $\alpha_i\in B_L$ and $\beta_i\in B_R$ while 
$\delta \notin B_L$ and $\gamma \notin B_R$. Using that the determinant of a block matrix with $\det{A} \ne 0$ can be expressed as
\be
  \det \left[ 
  \begin{array}{c|c}
  A & B \\
  \hline
  C & D
  \end{array}
  \right]
  = \det (A)
  \, 
  \det \big(D - C A^{-1}B\big),
\ee
we can write the determinant of \reef{eq:vanishsubmatricesPRE} as
\begin{equation}
\label{eq:vanishsubmatrices}
  \det (M)
  =
  \det\big(\mathbf{m}_n\left(B_R,B_L\right)\big) \times
  \bigg(
 m_{n}[\gamma|\delta] - \mathbf{m}_n[\gamma|B_L]
  \,\big(\mathbf{m}_n\left(B_R,B_L\right)\big)^{-1}
  \,\mathbf{m}_n[B_R|\delta] 
\bigg)
,
\end{equation}
By \reef{SnInvmn}, the (negative of the) second factor can be written 
\be
  \sum_{\alpha \in B_L,\, \beta \in B_R}
  {m}_n[\gamma|\alpha]
  \,\mathbf{S}_n\left(B_L,B_R\right)[\alpha|\beta]
  \,{m}_n[\beta|\delta] - {m}_n[\gamma|\delta]\,.
\ee
The vanishing of this condition is exactly the same as the written-out matrix multiplication of \reef{BootstrapEq}. 
Thus we learn that {\bf \em the ``self-copy'' $\mathds{1} \otimes \mathds{1} = \mathds{1}$ condition \reef{BootstrapEq} is equivalent to the requirement that the full $(n-1)! \times (n-1)!$ matrix $\mathbf{m}_n$ of zeroth-copy amplitudes has rank $\mathbf{R_n}$.} This means that {\em the rank $R_n$ of the double-copy kernel must be equal to the rank of the full matrix $\mathbf{m}_n$.} 

\subsection{Single-Copy Models: Generalized KKBCJ from \texorpdfstring{$\mathds{1} \otimes \text{R} = \text{R}$}{1xR=R} and \texorpdfstring{$\text{L} \otimes \mathds{1} = \text{L}$}{Lx1=L}}
\label{s:KLTdefsSing}

It is convenient also to use a matrix notation to represent the single-copy amplitudes with orderings restricted to a given choice of basis
\begin{equation}
    \mathbf{A}^\text{L}_n\left(B_L\right) \equiv 
    \begin{pmatrix}
    \mathcal{A}_n^\text{L}[\alpha_1] & \cdots &\mathcal{A}_n^\text{L}[\alpha_{R_n}]
    \end{pmatrix}^\top, \hspace{10mm} \mathbf{A}^\text{R}_n\left(B_R\right) \equiv 
    \begin{pmatrix}
    \mathcal{A}_n^\text{R}[\beta_1] & \cdots &\mathcal{A}_n^\text{R}[\beta_{R_n}]
    \end{pmatrix}^\top\,,
\end{equation}
where ${}^\top$ denotes transpose. 
In this notation, the double-copy formula \reef{KLTsec2} can be written as a simple matrix product
\begin{equation}
\label{KLTmatrixnot}
    \mathcal{A}_n^{\text{L}\otimes \text{R}} = \left(\mathbf{A}^{\text{L}}_n\left(B_L\right)\right)^\top \mathbf{S}_{n}(B_L,B_R)\,
    \mathbf{A}^\text{R}_n\left(B_R\right).
\end{equation}
and the generalized KKBCJ relations arise from the KLT algebra:
\bea
\label{RBCJmatrixnot}
   \mathds{1} \otimes \text{R} = \text{R}\!:&~~~&~~~~
    \mathbf{m}_n\left(B'_R,B_L\right)\mathbf{S}_n\left(B_L,B_R\right)\mathbf{A}_n^\text{R}\left(B_R\right) = \mathbf{A}_n^\text{R}\left(B'_R\right)\,,\\[2mm]
\label{LBCJmatrixnot}
   \text{L} \otimes \mathds{1} = \text{L}\!:&~~~&
    \left(\mathbf{A}_n^\text{L}\left(B_L\right)\right)^\top\mathbf{S}_n\left(B_L,B_R\right)\mathbf{m}_n\left(B_R,B'_L\right) = \left(\mathbf{A}_n^\text{L}\left(B'_L\right)\right)^\top.
\eea
These conditions are non-trivial only for elements in $B'_R$ ($B'_L$) that are not in $B_R$ ($B_L$).

The relations \reef{RBCJmatrixnot} and \reef{LBCJmatrixnot} are the \textit{generalized KKBCJ conditions}. They ensure that the result of the double-copy is independent of the choice of bases $B_L$ and $B_R$. To see this, rewrite \reef{RBCJmatrixnot} as
\begin{equation}
\label{RBCJmatrixnot2}
    \mathbf{S}_n\left(B_L,B_R\right)\mathbf{A}_n^\text{R}\left(B_R\right) - \mathbf{S}_n\left(B_L,B'_R\right)\mathbf{A}_n^\text{R}\left(B'_R\right) = 0\,.
\end{equation}
When multiplied from the left by $\left(\mathbf{A}_n^\text{L}\left(B_L\right)\right)^\top$, equation \reef{RBCJmatrixnot2} states that the double copy resulting from the two basis choices of $B_R$ and $B_R'$ are the same. Similarly, \reef{LBCJmatrixnot} ensures that the double-copy is independent of the choice of L-basis. Without basis independence, we cannot think of the double-copy as a map between field theories. 

When the $m_n$ are the amplitudes of the BAS model, the generalized KKBCJ conditions become the standard KKBCJ relations discussed in the Introduction. Likewise, they are equivalent to the string monodromy relations when the $m_n$ are the amplitudes of the string zeroth copy. 

When the full matrix $(n-1)! \times (n-1)!$ of zeroth-copy amplitudes $\mathbf{m}_n$ has non-maximal rank $R_n$, it must have $(n-1)!-R_n$ null vectors $n^\text{L}_i$ and $n^\text{R}_i$ for multiplication from the left and right, respectively. We show in Appendix \ref{sec:nullvectors} that these null vectors precisely encode the generalized KKBCJ relations as 
\be
  \label{nullvecKKBCJ}
  \sum_{\text{all}\,\beta} n^\text{R}_i[\beta] \, \mathcal{A}_n^\text{R}[\beta]= 0\,
  ~~~~\text{and}~~~~
  \sum_{\text{all}\,\alpha}   \mathcal{A}_n^\text{L}[\alpha]\, n^\text{L}_i[\alpha]= 0\,
  \ee
for each $i = 1,2, \ldots , (n-1)!-R_n$. The relation between null vectors and BCJ conditions was introduced previously in the context of a massive double-copy formalism in  \cite{Johnson:2020pny}. They are useful for understanding how the generalized KKBCJ relations modify the regular field theory KKBCJ relations.

\subsection{Roadmap for the Generalized Double-Copy}

Let us summarize how the double-copy bootstrap proceeds: 
\begin{enumerate}
\item Choose a candidate for a zeroth-copy model, i.e.~a local field theory with a bi-adjoint scalar field $\phi^{aa'}$ and some choice of interactions. Compute its color-ordered tree amplitudes $m_n[\alpha|\beta]$. 
\item Subject the matrix of these amplitudes to the double-copy bootstrap equation $\mathds{1} \otimes \mathds{1} = \mathds{1}$ in the form \reef{BootstrapEq} with some choice of rank $R_n$, possibly restricting the couplings in the model. Inverting the resulting rank $R_n$ matrices gives the generalized double-copy kernel $S_n$ via \reef{SnInvmn}.
\item  Tree amplitudes of L and R single-copy local models are then subjected to the generalized KKBCJ relations $\text{L}\otimes \mathds{1} = \text{L}$ and $\mathds{1} \otimes  \text{R} =  \text{R}$ in the form \reef{RBCJmatrixnot} and \reef{LBCJmatrixnot}.
\item Double-copy using \reef{KLTmatrixnot}.
\end{enumerate}
Along the way, locality constraints must be imposed. In particular, we have pointed out that zeroes of $\det(\mathbf{m}_{n}(B_R,B_L))$ may signal issues with spurious poles in the double-copy. 

A particularly prominent example is the bootstrap of a zeroth-copy model based on BAS with higher-derivative corrections, so let us comment more on this.

\subsection{Perturbative KLT Bootstrap}
\label{sec:pertKLT}

To study a double-copy kernel based on BAS + higher-derivative (h.d.) operators, let $m_{n}^{(0)}$ denote the BAS amplitudes and $\mathcal{A}_{n}$ the single-copy amplitudes that obey the regular field theory KKBCJ relations associated with the BAS zeroth copy. We can then write the BAS + h.d.~amplitudes and single-copy amplitudes (suppressing L and R superscripts) as 
\begin{equation}\label{eq:generalphiansatz}
\begin{split}
m_{n}[\beta|\alpha]&=m_{n}^{(0)}[\beta|\alpha]+m_{n}^{(1)}[\beta|\alpha]+\ldots \,, \\
\mathcal{A}_{n}[\alpha]&=\mathcal{A}_{n}^{(0)}[\alpha]+\mathcal{A}_{n}^{(1)}[\alpha]+\ldots \,,
\end{split}
\end{equation}
where $m^{(i)}[\beta|\alpha]$ and $\mathcal{A}_{n}^{(i)}[\alpha]$ with $i>0$ are the contributions from higher-dimension operators. They are systematically organized by increasing powers in $1/\Lambda$ of the UV scale of  BAS+h.d.~EFT such that the limits
\begin{equation}
    \lim_{\Lambda\rightarrow \infty}m_{n}[\beta|\alpha]=m_{n}^{(0)}[\beta|\alpha], \quad \lim_{\Lambda\rightarrow \infty}\mathcal{A}_{n}[\alpha]=\mathcal{A}_{n}^{(0)}[\alpha] \, 
\end{equation}
are smooth.  
This is the expected behavior in a physical EFT where the $\Lambda$ corresponds to the scale of some, perhaps unknown, UV physics which decouples from the IR dynamics in an appropriate limit.\footnote{We are not considering mass-deformations of the BAS model. See \cite{Johnson:2020pny} for a discussion of double-copy construction with masses.}

Taking the double-copy to be perturbative in $1/\Lambda$ means that the double-copy amplitude also has an expansion in $1/\Lambda$. For example, the double copy of YM+h.d.~with itself should give  NS-NS gravity plus higher dimension operators and the contributions from these operators should go smoothly to zero as $\Lambda \to \infty$. 
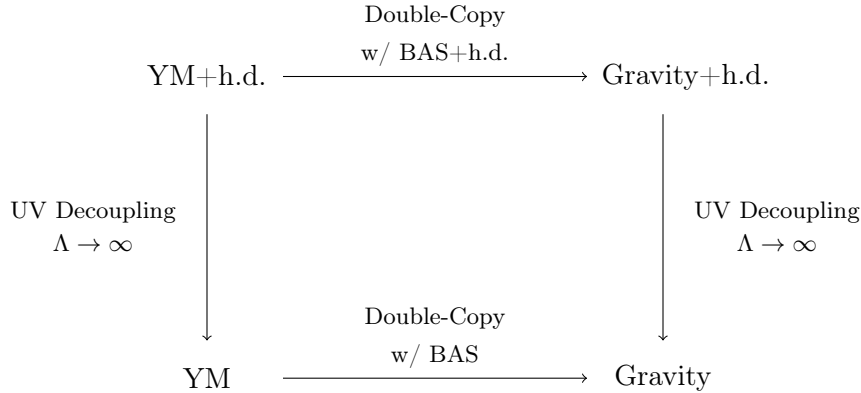
\begin{figure}
    \centering
    \begin{tikzpicture}
        \draw[->] (1,0)--(5,0);
        \draw[->] (1,-4)--(5,-4);
        \draw[->] (0,-0.5)--(0,-3.5);
        \draw[->] (6,-0.5)--(6,-3.5);
        \node at (0,0) {YM+h.d.};
        \node at (0,-4) {YM};
        \node at (6.3,0) {Gravity+h.d.};
        \node at (6,-4) {Gravity};
        \node at (3,0.8) {\footnotesize Double-Copy};
        \node at (3,0.3) {\footnotesize w/ BAS+h.d.};
        \node at (3,-3.2) {\footnotesize Double-Copy};
        \node at (3,-3.7) {\footnotesize w/ BAS};
        \node at (-1.5,-1.8) {\footnotesize UV Decoupling};
        \node at (-1.5,-2.2) {\footnotesize $\Lambda\rightarrow \infty$};
        \node at (7.5,-1.8) {\footnotesize UV Decoupling};
        \node at (7.5,-2.2) {\footnotesize $\Lambda\rightarrow \infty$};
    \end{tikzpicture}
    \caption{Illustration of the physical meaning of the perturbative double-copy. Physics at the UV scale $\Lambda$ decouples in both the single- and double-copies as $\Lambda \rightarrow \infty$ (i.e. this diagram commutes) only if the rank of the higher-derivative corrected BAS is the same as the rank of the uncorrected BAS.}
    \label{fig:my_label}
\end{figure}
This physically sensible requirement has implications for the rank of the double-copy kernel. As we saw in Section \ref{sec:nonlocal}, generic deformations of cubic BAS will increase the rank from $(n-3)!$ of BAS. Since the double-copy kernel is the inverse of a rank $R_n$ matrix of BAS+h.d.~amplitudes, the kernel would be divergent in the limit $\Lambda \to \infty$ where the cubic BAS amplitudes are recovered. This would imply that the double-copy amplitudes do not have sensible $\Lambda \to \infty$ limits and then we can no longer identify $\Lambda\rightarrow \infty$ as the limit of decoupling UV physics. 

To avoid a situation of unnatural UV-IR mixing in the double-copy, we must therefore require that the rank of the zeroth copy  \textit{does not change} as a function of $\Lambda$.\footnote{Of course, it is logically possible that the UV dynamics may not fully decouple, and we are free to entertain the possibility of a discontinuity in the rank of the zeroth copy. Curiously, for all examples studied in this paper, relaxing this naturalness assumption also leads to spurious poles in the double-copy.} This means that to study 
the most general double-copy kernel based on BAS+h.d.~we must work with the rank $R_n = (n-3)!$. 

The double-copy bootstrap with rank $(n-3)!$ is studied  at 3-, 4-, and 5-point in Sections \ref{sec:bootstrap3pt}-\ref{sec:5pt}. In Section \ref{sec:QBAS} we consider examples of models with rank $R_n > (n-3)!$ that are not UV deformations of the BAS model.

\section{KLT Bootstrap at 3-Point}
\label{sec:bootstrap3pt}

Let us begin at 3-point as an informative warm-up for the higher-point analysis. In 3-particle kinematics, all Mandelstam variables vanish on-shell, so it is impossible for an on-shell 3-point scalar amplitude to have momentum dependence. At the level of the Lagrangian, this means that any higher-derivative corrections at 3-point can be moved into higher-point by a field redefinition. Thus we only need to consider {\em constant} 3-point scalar amplitudes. 

By cyclic symmetry, there are two independent options for the double color ordered bi-adjoint scalar amplitudes and we parameterize them using couplings $g$ and $\lambda_3$  as
\begin{equation}
    m_3[123|123] = g + \lambda_3, 
    \hspace{10mm} 
    m_3[123|132] =  - g + \lambda_3\,.
\end{equation}
These amplitudes arise from Lagrangian interactions of the form\footnote{A mixed term ,
$f^{abc}\tilde{d}^{a'b'c'}\phi^{aa'}\phi^{bb'}\phi^{cc'} = 0$, vanishes
due to the symmetric-antisymmetric index contractions.}
\begin{equation}
   \label{fanddat3pt}
    \mathcal{L}_{\text{3}} = -\frac{g}{6}\, f^{abc}\tilde{f}^{a'b'c'}\phi^{aa'}\phi^{bb'}\phi^{cc'} + \frac{\lambda_3}{6}\, d^{abc}\tilde{d}^{a'b'c'}\phi^{aa'}\phi^{bb'}\phi^{cc'}.
\end{equation}
The first term is the cubic interaction from the cubic BAS model \reef{cubicbas0} and the second one is its fully symmetric counterpart. In terms of generators, we have 
\begin{equation}
    i\,f^{abc} = \text{Tr}\left[T^a[T^b,T^c]\right], \hspace{10mm} d^{abc} = \text{Tr}\left[T^a\{T^b,T^c\}\right].
\end{equation}
The invariant $d^{abc}$ is sometimes called the anomaly coefficient and it is non-zero for generic representations of $SU(N)$ groups with $N>2$.

The general $2\times 2$ bi-adjoint scalar matrix labeled by the $(n-1)!=2$ independent color-orderings $\{123,132\}$ is then
\begin{equation} 
    \mathbf{m}_3
    =
    \begin{pmatrix}
    m_3[123|123] & m_3[123|132]\\
    m_3[132|123] & m_3[132|132]
    \end{pmatrix}
    =
    \begin{pmatrix}
    g + \lambda_3 & -g + \lambda_3\\
    -g + \lambda_3 & g + \lambda_3
    \end{pmatrix},
\end{equation}
and its determinant is
\be
   \det(\mathbf{m}_3) = 4 g \lambda_3\,.
\ee
Thus, for non-zero values of $g$ and $\lambda_3$ the matrix $\mathbf{m}_3$ has rank 2, however, whenever one of the two couplings vanishes, the rank is reduced to 1. In Section \ref{sec:QBAS}, we show that the model with $\lambda_3 \ne 0$ does not satisfy the minimal-rank condition at 4-point and it leads to a generalized KLT kernel with spurious poles at 5-point. For this reason, we set 
\be
\lambda_3=0 \,
\ee
in our studies of generalizations of the KLT double-copy. This in particular means that the null vectors of $\mathbf{m}_3$ are $\{1,1\}$, which via \reef{nullvecKKBCJ} imply the usual $3$-point KK relation
\be
  \label{3ptKK}
  \mathcal{A}_3[132] + \mathcal{A}_3[123] = 0
\ee
on the L and R sector 3-point amplitudes.

%%%%%%%%%%%%%%%%%%%%%%%%%%%%%%%%%%%%%%%%%%
%%%%%%%%%%%%%%%%%%%%%%%%%%%%%%%%%%%%%%%%%%
%%%%%%%%%%%%%%%%%%%%%%%%%%%%%%%%%%%%%%%%%%

\section{KLT Bootstrap at 4-Point}
\label{sec:bootstrap}

In this section, we solve the KLT bootstrap equation at 4-point for minimal rank $(n-3)!=1$. The general solution can be written in terms of a single function and we derive from it the generalized KKBCJ relations. 
Next, we use these results to find the most general higher-derivative corrections to the BAS model at 4-points subject to constraints of locality and minimal rank. This gives a generalized  KLT formula for double-copying single-color EFTs at 4-point. We compare the result to the string KLT kernel and comment in general properties of the result.

\subsection{4-Point Bootstrap Equations}
\label{s:boot4pt}

The KLT bootstrap for minimal rank $(n-3)!=1$ imposes that all 2-by-2 minors of the matrix $\mathbf{m}_4$ in \reef{fullm4matrix} must vanish, 
\be
  \label{boot}
   m_4[\alpha|\beta] m_4[\delta|\gamma] 
    = m_4[\alpha|\gamma] m_4[\delta|\beta],
\ee
for any choice of $\alpha, \beta, \gamma, \delta \in 
\{ 1234, 1243, 1324, 1342, 1423, 1432\}$.

Using cyclic symmetry and momentum relabeling, the six different doubly color-ordered 4-point amplitudes $m_4$ can be expressed in terms of three functions, $f_1$, $f_2$, and $f_6$, as follows:
\begingroup
\allowdisplaybreaks
\be
\label{m4s}
\begin{split}
m_{4}[1234|1234] &= f_1(s,t)    ~~~~\text{with}~~~~    f_1(s,t)=f_1(u,t)\,,\\
m_{4}[1234|1243] &= f_2(s,t)\,,\\
m_{4}[1234|1324] &= f_3(s,t) = f_2(u,t)\,,\\
m_{4}[1234|1342] &= f_4(s,t)= f_2(s,t)\,,\\
m_{4}[1234|1423] &= f_5(s,t)= f_2(u,t)\,,\\
m_{4}[1234|1432] &= f_6(s,t)   ~~~~\text{with}~~~~    f_6(s,t)=f_6(u,t)\,.
\end{split}
\ee
\endgroup
where here and in the following it is always understood that $s+t+u=0$. For example, to obtain the 3rd line, we use that 
\be
  \begin{split}
  f_3(s,t) &= m_{4}[1234|1324] = m_{4}[4123|4132] = m_{4}[1234|1243] \Big|_{1\to 4 \to 3 \to 2 \to 1}\\
  &= f_2(s,t)\Big|_{1\to 4 \to 3 \to 2 \to 1}
   = f_2(u,t) \,.
  \end{split}
\ee
We allow for the possibility that $m_{n}[\beta|\alpha] \ne m_{n}[\alpha|\beta]$ and we do not assume trace reversal symmetry, e.g.~$m_{n}[\alpha|\beta^T]$ is not necessarily related to $m_{n}[\alpha|\beta]$. 
In terms of $f_1$, $f_2$, and $f_6$, the matrix $\mathbf{m}_4$ of \reef{fullm4matrix} then takes the form
\be
\label{m4f1f2f6}
    \mathbf{m}_4 = 
    \begin{pmatrix}
    f_1(s, t) &  f_2(s, t) &  f_2(u, t) &  f_2(s, t) &  f_2(u, t) & 
  f_6(s, t) \\ 
  f_2(s, u)  &  f_1(s, u) &  f_2(t, u)  & f_6(s, u)  &  f_2(t, u)  & 
  f_2(s, u) \\
  f_2(u, s)  &  f_2(t, s) &  f_1(t, s) &  f_2(t, s) &  f_6(t, s)  &  
  f_2(u, s) \\
  f_2(s, u)  &  f_6(t, u) &  f_2(t, u) &  f_1(t, u) &  f_2(t, u)  &  
  f_2(s, u) \\
  f_2(u, s)  &  f_2(t, s)  &  f_6(u, s)  &  f_2(t, s)  &  f_1(u, s)  &  
  f_2(u, s) \\
  f_6(u, t)  &  f_2(s, t) &  f_2(u, t) &  f_2(s, t) &  f_6(u, t)  & 
  f_2(u, t)
    \end{pmatrix}
    .
\ee

Generically this matrix has rank 6, so we must impose the rank 1 bootstrap condition by setting all 2-by-2 minor to zero. This can be done very simply. Consider the vanishing of the 2-by-2 minor of \reef{m4f1f2f6} with rows 1 and 2  and columns 1 and 6:

\be
     \big(f_1(s,t) -f_6(s,t)\big) f_2(s,u) = 0\,.
\ee
This implies\footnote{When $f_{2}=0$, the other rank 1 conditions set $f_{1}=f_{6}=0$. So we assume $f_{2}$ to be non-zero.}
\be 
  \label{f6sol}
   \boxed{
   f_6(s, t) = f_1(s, t).  
   }
\ee
Next, the 2-by-2 minor of \reef{m4f1f2f6} with rows 1 and 3 and columns 1 and 2 vanishes when 
\be
  \label{f1f2A}
  0=f_1(s, t) f_2(t, s)-f_2(s, t) f_2(u, s)\,,
\ee
while the vanishing of the minor with rows 1 and 2 and columns 1 and 3 requires
\be
  \label{f1f2B}
  0=f_1(s, t) f_2(t,u)-f_2(s,u) f_2(u, t) \ .
\ee
It follows from \reef{f1f2A} that $f_1$ is fixed in terms of $f_2$ as
\be
  \label{f1sol}
  \boxed{f_1(s, t) = \frac{f_2(s, t) f_2(u, s)}{f_2(t, s)}\,,
  }
\ee
and combining \reef{f1f2A}  and \reef{f1f2B} gives a final self-consistency condition for $f_2$,
\be
  \label{f2cond}
  \boxed{
  f_2(s, t) f_2(u, s) f_2(t,u) = f_2(t, s)  f_2(u, t) f_2(s, u)
  \,.}
\ee
When the three equations \reef{f6sol}, \reef{f1sol}, and \reef{f2cond} are imposed, the matrix \reef{m4f1f2f6} has rank 1, as desired; thus, these three conditions  are the {\bf 4-point KLT bootstrap equations}. 

The cubic bi-adjoint scalar amplitudes \reef{4ptBASex} have
\be
  \label{f1f2BAS}
  f^\text{BAS}_1(s,t) = f^\text{BAS}_6(s,t) =\frac{g^2}{s}+\frac{g^2}{u}\,,
  ~~~~
  f^\text{BAS}_2(s,t) = -\frac{g^2}{s}\,,  
\ee
and it is easy to see that they solve the three 4-point bootstrap equations \reef{f6sol}, \reef{f1sol}, and \reef{f2cond}. They are likewise solved by the string theory 4-point amplitudes of Mizera which have \cite{Mizera:2016jhj}   
\be
  \label{f1f2str}
   f_1^\text{string}(s,t) 
   =f_6^\text{string}(s,t) = \frac{1}{\tan( \alpha' \pi s)} + \frac{1}{\tan(\alpha' \pi u)} \,,
   ~~~~
   f_2^\text{string}(s,t)
   = - \frac{1}{\sin(\alpha' \pi s)} \, .
\ee

As we shall see, the bootstrap equations   are not quite sufficient to guarantee that \reef{m4s} correspond to doubly color-ordered amplitudes of a local theory, i.e.~that the only singularities in the amplitudes correspond to physical poles. This is a nontrivial constraint. For example,  $f_2(s,t)=s$ solves \reef{f2cond}, but gives $f_1(s,t) = su/t$ which has a $t$-pole not permitted by the color structure of $m_{4}[1234|1234] = f_1(s,t)$. 
Therefore, locality constraints on $f_1$ and $f_2$ generally further restrict the solution. 

%%%%%%%%%%%%%%%%%%%%%%%%%%%%%%%%%%%%%%%%%%%%%%%%
%%%%%%%%%%%%%%%%%%%%%%%%%%%%%%%%%%%%%%%%%%%%%%%%
%%%%%%%%%%%%%%%%%%%%%%%%%%%%%%%%%%%%%%%%%%%%%%%%

\subsection{Generalized KKBCJ Conditions}
\label{s:genKKBCJ4pt}
Equations \reef{f6sol}, \reef{f1sol}, and \reef{f2cond} ensure that the $6\times6$ matrix has rank 1. Hence it must have five null vectors under left and right multiplication. One can directly verify that 
\be
  \label{nullKK}
  (1,0,0,0,0,-1)\,,~~~~
  (0, 1, 0,-1, 0, 0)\,,~~~~
  (0, 0, 1, 0, -1, 0),
\ee
are null vectors under both left and right multiplication for any solution $f_2$. Via \reef{nullvecKKBCJ}, the null vectors imply the following relations among both L and R sector single-copy amplitudes: 
\be
\label{simpleKK4pt}
\text{Trace reversal:}~~
\mathcal{A}_{4}[1234]= \mathcal{A}_{4}[1432]\,,~~~~
\mathcal{A}_{4}[1243]= \mathcal{A}_{4}[1342]\,,~~~~
\mathcal{A}_{4}[1324]= \mathcal{A}_{4}[1423]\,.
\ee
These are exactly the three 4-point KK relations \reef{4ptKKtr} that are not the $U(1)$-decoupling relation. This means that any L or R sector amplitudes must satisfy the trace-reversal identity
\begin{equation}\label{eq:KKnonU1gen}
\mathcal{A}_{4}[\alpha]=\mathcal{A}_{4}[\alpha^{T}]  \,,
\end{equation}
where the $\alpha^{T}$ denotes the color-ordering that has the reverse ordering of $\alpha$, for example~$(1234)^T = (4321)$. 

Under right multiplication, the two remaining null vectors can be written as
\be
  \Big(1,1,0,0,
  -\frac{f_2(s,t)}{f_2(u,t)}-\frac{f_2(s,u)}{f_2(t,u)},0\Big)\,,
  ~~~~~\Big(1, -\frac{f_2(u,s)}{f_2(t,s)}, 0, 0,0, 0\Big).
\ee
By \reef{nullvecKKBCJ}, they imply
\bea
  \label{4ptgenU1L}
  \text{L generalized $U(1)$:}&&
   \mathcal{A}^\text{L}_{4}[1234] + \mathcal{A}^\text{L}_{4}[1243]
   - \bigg(\frac{f_2(s,t)}{f_2(u,t)}+\frac{f_2(s,u)}{f_2(t,u)}\bigg)
   \mathcal{A}^\text{L}_{4}[1423] = 0\,,~~~~\\[3mm]
  \label{4ptgenBCJL}
  \text{L generalized BCJ:}  
   &&
   \mathcal{A}^\text{L}_{4}[1234]-\frac{f_2(u,s)}{f_2(t,s)}\mathcal{A}^\text{L}_{4}[1243] = 0 \,.
\eea
 When $f_2(s,t) = -g^2/s$, these relations reduce to the familiar $U(1)$-decoupling relation \reef{4ptU1} and BCJ conditions \reef{4ptBCJ}. Note that if one {\em insists} that the usual BCJ relation \reef{4ptBCJ} hold, i.e.~if we impose 
\be
    \frac{f_2(u,s)}{f_2(t,s)} = \frac{t}{u}\,,
\ee
then \reef{4ptgenU1L} reduces to the usual $U(1)$ decoupling identity \reef{4ptU1} after using $s+t+u=0$. This is natural; the BCJ relation can be derived from color-kinematics duality \cite{Bern:2008qj} in which the color-structures are all generated by the structure constants $f^{abc}$ and therefore $U(1)$-decoupling must hold. Our generalized KKBCJ
relations \reef{simpleKK4pt} allow for more general color-structures, such as $d^{abc}$ and $d^{abcd}$ in the higher-derivative operators and therefore they modify the $U(1)$ decoupling identity and the BCJ relations. 

For the R sector, if follows from the left-multiplication null vectors that 
\bea
  \label{4ptgenU1R}
  \text{R generalized $U(1)$:}&&
   \mathcal{A}^\text{R}_{4}[1234] + \mathcal{A}^\text{R}_{4}[1243]
   - \bigg(\frac{f_2(s,t)}{f_2(t,s)}+\frac{f_2(s,u)}{f_2(u,s)}\bigg)
   \mathcal{A}^\text{R}_{4}[1423] = 0\,,~~~~\phantom{s}\\[3mm]
  \label{4ptgenBCJR}
  \text{R generalized BCJ:}  
   &&
   \mathcal{A}^\text{R}_{4}[1234]-\frac{f_2(u,t)}{f_2(t,u)}\mathcal{A}^\text{R}_{4}[1243] = 0 \,.
\eea
The generalized R and L sector KKBCJ relations are generally distinct. This differs from the usual field theory KKBCJ relations or the string monodromies.\footnote{Note that the modified $U(1)$ decoupling relations above are not exactly the same as the string theory monodromy relations for the choice \reef{f1f2str}; rather the string monodromy relations arise from combinations of our general KKBCJ relations with $f_2$ chosen as in \reef{f1f2str}.} In particular, this means that the criteria for being a valid input for the KLT relations are different for the R and L sector models. For example, the higher-derivative corrections allowed for YM may be different for the R and L sectors. 
We discuss the generalized KKBCJ further in  Section \ref{s:4ptcomments}.

%%%%%%%%%%%%%%%%%%%%%%%%%%%%%%%%%%%%%%%%%%%%%%%%
%%%%%%%%%%%%%%%%%%%%%%%%%%%%%%%%%%%%%%%%%%%%%%%%
%%%%%%%%%%%%%%%%%%%%%%%%%%%%%%%%%%%%%%%%%%%%%%%%

%%%%%%%%%%%%%

%%%%%%%%%%%%%%%%%%%%%%%%%%%%%%%%%%%%%%%%%%%%%%%%
%%%%%%%%%%%%%%%%%%%%%%%%%%%%%%%%%%%%%%%%%%%%%%%%
%%%%%%%%%%%%%%%%%%%%%%%%%%%%%%%%%%%%%%%%%%%%%%%%

%%%%%%%%%%%%%%%%%%%%%%%%%%%%%%%%%%%%%%%%%%%%%%%%
%%%%%%%%%%%%%%%%%%%%%%%%%%%%%%%%%%%%%%%%%%%%%%%%
%%%%%%%%%%%%%%%%%%%%%%%%%%%%%%%%%%%%%%%%%%%%%%%%

\subsection{Perturbative Solution}
\label{s:pert4pt}

We now determine the most general local higher-derivative  corrections to the BAS model that are compatible with the minimal rank condition at 4-point. 
We begin with an ansatz for $f_2$ whose leading term is the usual bi-adjoint $s$-pole. Since there can be no higher-derivative corrections to the 3-point amplitudes, there can be no other pole terms and hence any higher-derivative corrections at 4-point must be a power-expansion in $s$ and $t$. The most general ansatz for $f_2$ is, therefore,\footnote{The analysis is valid in $d$-dimensions and as such we leave $g$ to have mass-dimension $3-d/2$. To keep $\Lambda$ mass-dimension $1$ dimension-counting parameter then implies that the coefficients $a_{k,r}$ have mass-dimension ${6-d}$.} 
\be
  \label{f2ansatz}
  f_2(s,t) = -\frac{g^2 }{s} + \sum_{k=0}^N \sum_{r=0}^k
  \frac{a_{k,r}}{\Lambda^{2(k+1)}} \,s^r \,t^{k-r} \ ,
\ee
where $N$ is the highest power in Mandelstams used in the expansion and $a_{k,r}$ are coefficients that encode the higher-derivative corrections. We use $\Lambda$ as a mass-dimension parameter that keeps track of the power-expansion such that we reduce to the BAS model in the limit $\Lambda \rightarrow\infty$.  
 
Using the ansatz for $f_2$ in \reef{f1sol}, we find 
\be
  f_1(s,t) = \frac{g^2 }{s} - \frac{g^2 }{s+t}
  +\frac{2a_{0,0}}{\Lambda^2}\frac{t^2}{s (s + t)} + \ldots \ , 
\ee
The $t$-dependent residue of the $s$ and $u$ poles is not possible in a pure scalar theory, so we must set $a_{0,0}=0$. Going to higher orders in the power-expansion, one finds unphysical poles with coefficients $a_{2r,2r}$, so we must take 
\be
  \label{f1loc}
  a_{2r,2r} = 0 ~~~~\text{for any}~r=0,1,2,3,\ldots
\ee
With this choice, $f_1$ only has physical poles. 

Imposing the bootstrap condition \reef{f2cond} on the ansatz \reef{f2ansatz} with \reef{f1loc}, we find that 
\be
a_{2,1} =  a_{2, 0}\,, 
~~~~~~
a_{4,3} = a_{4, 2}- a_{4, 1} + a_{4, 0}+ \ldots \,,
\ee
so that the result for $f_2$ is
\be   
 \label{f2res}
  \begin{split}
  f_2(s,t) =& -\frac{g^2 }{s} +  \frac{1}{\Lambda^4}( a_{1, 0} t +a_{1, 1}s) +
  \frac{a_{2, 0}}{\Lambda^6}  t (s + t) +
  \\
  &+
  \frac{1}{\Lambda^8}  \Big[ a_{3, 0}t^3  + a_{3, 1}s t^2 + a_{3, 2} s^2 t + a_{3, 3}s^3  \Big]+ \\
  &+
  \frac{1}{\Lambda^{10}}   \Big[ a_{4, 0} t^4  +a_{4, 1} s t^3  + a_{4, 2}  s^2 t^2 + 
 (a_{4, 0} - a_{4, 1} + a_{4, 2}) s^3 t \Big]\\
 & +
  \frac{1}{\Lambda^{12}}   \bigg[
  a_{5, 0} t^5  + a_{5, 1} s t^4  + a_{5, 2} s^2 t^3  + a_{5, 3} s^3 t^2 \\
  &\hspace{0.5cm}+
  \bigg(
 \frac{a_{1, 0} a_{1, 1} (a_{1, 0} - a_{1, 1}) }{g^4}  
 + \frac{a_{1, 1} (a_{3, 1} - a_{3, 2}) - 
  a_{1, 0} (a_{3, 0} - a_{3, 2} + a_{3, 3})}{g^2} 
  \\
  & \hspace{2cm} + a_{5, 0} - a_{5, 1}+ 
 a_{5, 3}\bigg)
 s^4 t
  +a_{5, 5} s^5 
  \bigg]
   + O\bigg(\frac{1}{\Lambda^{14}} \bigg) \ ,
   \end{split}
\ee
and for $f_1$ we then have,
\be   
 \label{f1res}
  \begin{split}
   f_1(s,t) =&\, g^2 \bigg( \frac{1}{s} - \frac{1}{s+t}\bigg)
   +
   \frac{a_{1, 0} - 2 a_{1, 1}}{\Lambda^4} t - 
  \frac{a_{2, 0}}{\Lambda^6}  t^2 \\
    &\, -\frac{1}{\Lambda^8}
       \bigg[
          \Big(  2 a_{3, 3} - a_{3, 2} + a_{3, 1}-a_{3, 0} 
            -g^{-2} a_{1, 1}(a_{1, 1} -a_{1, 0})
          \Big) \, s t (s + t) 
          \\
          &\,
          \phantom{r}\hspace{0.6cm}+
    \Big(  (4 a_{3, 3} - 2 a_{3, 2} + a_{3, 0}) 
    +2 g^{-2} a_{1, 1}(a_{1, 1} -a_{1, 0})
    \Big) \,t^3
       \bigg]
 + O\bigg(\frac{1}{\Lambda^{10}} \bigg) \ .
   \end{split}
\ee
Equations (\ref{f2res}) and (\ref{f1res}) correspond to the most general solution to the bootstrap equations assuming that the leading contribution to the partial amplitudes is cubic BAS. 

%%%%%%
\subsection{Comparison with String Theory}\label{sec:4pstringkernel}

The results  for the bi-adjoint 4-point amplitude with higher-derivative corrections are more general than the string kernel result \cite{Mizera:2016jhj}. Let us now expand the string functions \reef{f1f2str} in small $\alpha'$, 
\be
\begin{split}
   f_1^\text{string}(s,t)
   &= \frac{1}{\alpha' \pi s} 
  +\frac{1}{\alpha' \pi u} 
  +\frac{1}{3} \alpha' t
   - \frac{1}{45} \alpha'^3 \pi^3 (s^3  + u^3) 
   + O\big( \alpha'^5 \big)\\
   f_2^\text{string}(s,t)
   &= -\frac{1}{\alpha' \pi s} 
   - \frac{1}{6} \alpha' \pi s
    - \frac{7}{360} (\alpha' \pi s)^3
    + O\big( \alpha'^5 \big) \ .
    \end{split}
\ee

Setting $g^{2}=1/(\pi \alpha')$ and identifying $\Lambda^2 = 1/(\pi \alpha')$, we compare $f_2^\text{string}$ with our $f_2$ in \reef{f2res} and find the choice of $a_{i,j}$'s that reproduce the string result, namely:
\bea
\nonumber 
  &&a_{2k,i} = 0~~\text{for all}~~k,i\,,\\
\label{stringsas}
  &&a_{k,i}=0~~\text{for}~~k>i\,,
  \\ \nonumber
  &&a_{1, 1} = -\frac{1}{6\pi\alpha'} \,, ~~~
  a_{3, 3} = -\frac{7}{360\pi\alpha'}\,, ~~~
  a_{5, 5} = -\frac{31}{15120\pi\alpha'}\,,~~~
  a_{7, 7} =  -\frac{127}{604800\pi\alpha'}\,,
  \ldots.
\eea
We use this to compare our results for generalized KLT to that of string theory.

%%%%%

\subsection{Comments on Perturbative Solution}
\label{s:4ptcomments}

Consider the lowest orders of the higher-derivative bi-adjoint 4-point amplitudes
\be
   \begin{split}
   m_{4}[1234|1234] &= f_1(s,t) =
   \frac{g^2}{s}+\frac{g^2}{u}
   +
   \frac{a_{1,0}-2a_{1,1}}{\Lambda^4} \,t
   - \frac{a_{2,0}}{\Lambda^6} \,t^2 + \ldots
   ,\\[2mm]
   m_{4}[1234|1243] &= f_2(s,t) =
   - \frac{g^2}{s}
   +
   \frac{a_{1,0}}{\Lambda^4} \,t
   +\frac{a_{1,1}}{\Lambda^4} \,s
   - \frac{a_{2,0}}{\Lambda^6} \,t u
   + \ldots
   \,.
   \end{split}
\ee
These matrix elements derive from a Lagrangian of the form\footnote{The normalization is chosen such that $\sum_{a} \textrm{Tr}[XT^{a}]\textrm{Tr}[YT^{a}]=\textrm{Tr}[XY]$} 
\bea
\nonumber
\mathcal{L}&=& \mathcal{L}_\text{BAS}\\
\nonumber
&&
-\frac{a_\text{L}+a_\text{R}}{2\Lambda^{4}}
f^{abx}f^{cdx}f^{a'b'x'}f^{c'd'x'}(\partial_\mu\phi^{aa'})(\partial^\mu\phi^{bb'})\phi^{cc'}\phi^{dd'} \\[1mm]
\label{lagrangianoper}
&&+\frac{a_\text{L}}{\Lambda^{4}}
f^{abx}f^{cdx}d^{a'b'x'}d^{c'd'x'}(\partial_\mu\phi^{aa'})\phi^{bb'}(\partial^\mu\phi^{cc'})\phi^{dd'} \\[1mm]
\nonumber
&&+\frac{a_\text{R}}{\Lambda^{4}}d^{abx}d^{cdx}f^{a'b'x'}f^{c'd'x'}(\partial_\mu\phi^{aa'})\phi^{bb'}(\partial^\mu\phi^{cc'})\phi^{dd'} 
\\[1mm]
\nonumber
&&-\frac{a_{2,0}}{2\Lambda^6}f^{abx}f^{cdx}f^{a'b'x'}f^{c'd'x'}(\partial_{\nu}\partial_{\mu}\phi^{a,a'}\partial^{\nu}\partial^{\mu}\phi^{b,b'}\phi^{c,c'}\phi^{d,d'}+\partial_{\nu}\partial_{\mu}\phi^{a,a'}\partial^{\mu}\phi^{b,b'}\partial^{\nu}\phi^{c,c'}\phi^{d,d'}) \\[2mm]
\nonumber
&&+\ldots
\ ,
\eea
where
\be
 \label{aLaR}
 a_\text{L} = \frac{1}{4}(a_{1, 1} - a_{1, 0})\,,
 ~~~~
 a_\text{R} = \frac{1}{4} a_{1, 1} \,.
\ee
There are no $\phi^4$ operators; they are simply not permitted by the rank 1 bootstrap equations at 4-point. That is consistent with the example in the Introduction, where including a $\phi^4$ operator led to non-minimal rank. The lowest dimension operators allowed by the KLT bootstrap are of the form $\partial^{2}\phi^{4}$.

At 2-derivative order, the effective action has two independent couplings, $a_\text{R}$ and $a_\text{L}$, that encode different color-structures in the L and R sectors. This reflects that generalized KLT kernel built from the $m_4$-amplitudes is not symmetric: it treats the L and R sectors separately and it is in this sense ``heterotic''. If we insist on a symmetric kernel, this requires $\mathbf{m}_4$ in \reef{m4f1f2f6} to be symmetric which is achieved by taking $f_2(s,u)=f_2(s,t)$; this is one particular solution to the bootstrap equation \reef{f2cond}. It is solved perturbatively by $a_{1,0} = 0$, i.e.~$a_\text{R}=a_\text{L}$, while $a_{2,0}$ can remain non-zero. Thus we see that the string solution \reef{stringsas} is an example of a symmetric kernel, but it is not the most general one. 

It is clear from the effective action \reef{lagrangianoper} that the contribution at order $\partial^2 \phi^4$ is necesssarily linked to violation of the $U(1)$-decoupling relation due to the presence of the symmetric $d^{abc}$ color-structures. It is explicitly $a_\text{L}$ that is responsible for the leading-order modifications of the BCJ and $U(1)$-decoupling identities in the L sector; and likewise $a_\text{R}$ that is responsible for the modifications in the R sector, e.g.~
\begin{equation}
\begin{split}
0=&\mathcal{A}_{4}^{\textrm{R}}[1234]+\mathcal{A}_{4}^{\textrm{R}}[1243]
+\left ( 1-\frac{12 a_\text{R}}{g^2 \Lambda ^4}\, tu
   +O(\Lambda^{-8})\right ) \mathcal{A}_{4}^{\textrm{R}}[1423] \ , \\
0=&\mathcal{A}_{4}^{\textrm{R}}[1234]-
\bigg( \frac{t}{u}
-\frac{4 a_\text{R}}{g^2 \Lambda^4}
\frac{st (t-u)}{u} +O(\Lambda^{-8})\bigg)\mathcal{A}_{4}^{\textrm{R}}[1243] \ . 
\end{split}
\end{equation}
The $1/\Lambda^6$ term in \reef{lagrangianoper} controlled by $a_{2,0}$ does not modify the BCJ and $U(1)$-decoupling identities. If we wanted a generalized KLT kernel that preserved the $U(1)$-decoupling identity, we would set $a_\text{R}=a_\text{L}=0$ while keeping $a_{2,0}$. 
The KKBCJ constraints  generalize the string theory monodromy relations and allow the L and R sector amplitudes to be distinct.

\section{Example: Higher-Derivative YM to Gravity}
\label{sec:HDYM}

In this section we present the double-copy of YM plus higher dimension operators at 3- and 4-point as an illustrative example of our generalized double-copy. Another interesting example is chiral perturbation theory ($\chi$PT) with higher-derivative operators: we give results for that its double copy with itself and YM+h.d.~in Appendix \ref{app:NLSM}.

\subsection{3-Point}
\label{s:YM3pt}
Up to the choice of coupling constants, 3-point amplitudes are uniquely fixed by little group scaling and locality. For gluons, there are only two options: the MHV and anti-MHV helicity amplitudes arising from the usual YM 3-point vertex of $\Tr F^2$ and the all-minus and all-plus amplitudes of $\Tr F^3$. We have 
\be\label{eq:3ptYM}
		\mathcal{A}_3[1_g^- 2_g^- 3_g^+] = g_\text{YM} \frac{\<12\>^3}{\<23\>\<31\>}\,,
		~~~~~
		\mathcal{A}_3[1_g^- 2_g^- 3_g^-] =\frac{g_{F^3}}{\Lambda^{2}}\<12\>\<23\>\<13\>
		\,.
\ee
We write out only the mostly-minus amplitudes explicitly;
the conjugate amplitudes are obtained by exchanging angle brackets with square brackets. 
These amplitudes satisfy the rank-1 KKBCJ relations \reef{3ptKK}.

The double-copy of YM with itself gives gravity coupled to the dilaton $\varphi$ and the antisymmetric 2-form. In 4d, the latter can be dualized to an axion-scalar $B$ with a shift-symmetry, $B \to B \,+ \,$constant.
The precise map of the 4d on-shell states is
\be
\label{4dmap}
	\begin{array}{ccc}
		& \text{L} & \text{R}\\
		h^+  ~& + & +\\
		h^-  ~& - & -\\
		Z  ~& + & -\\
		\bar{Z}  ~& - & +\\
	\end{array}
\ee
where $\pm$ denote the L and R sector gluon helicity states, $h^\pm$ are the graviton helicity states, and 
the complex scalar is $Z = \frac{1}{\sqrt{2}}(\varphi + i B)$.

Using \reef{eq:3ptYM} and the map \reef{4dmap}, we  construct all possible 4d gravity-dilaton-axion 3-point amplitudes arising from the double-copy relation
\be
   \label{3ptDC}
   \mathcal{M}_3(123) 
   = \frac{1}{g}\mathcal{A}_3^\text{L}[123]
   \mathcal{A}_3^\text{R}[123]\,,
\ee
The pure graviton amplitudes
\begin{align}
	\mathcal{M}_3\left(h^- h^- h^+\right)&=\kappa\, \frac{\<12\>^6}{\<23\>^2\<31\>^2}\,,\\[1mm]
	\mathcal{M}_3\left(h^- h^- h^-\right)&=\frac{\kappa_{R^3}}{\Lambda^4} \<12\>^2\<23\>^2\<13\>^2\,.
\end{align}
and their conjugates 
 correspond to the 3-point interactions of $\sqrt{g}R$ and $\sqrt{g}R^3$, respectively. We have identified the couplings as
\be
   \label{kappa}
   \kappa \equiv \frac{g^\text{R}_\text{YM} g^\text{L}_\text{YM}}{g}\,,
   ~~~~~
   \kappa_{R^3} \equiv \frac{g^\text{R}_{F^3} g^\text{L}_{F^3}}{g}.
\ee
Note that we are allowing for the possibility that the YM and $F^3$ couplings are different in the L and R sectors. 

The double-copy of the leading-order YM amplitudes also includes the coupling of the dilaton to the graviton. Specifically, in the MHV sector we get
\be
 \mathcal{M}_3(-Z\bar{Z}) = \kappa\, \frac{\<12\>^2\<13\>^2}{\<23\>^2}\, ,
\ee
and its anti-MHV conjugate. These amplitudes simply represent the canonical coupling of the complex scalars coupled to gravity via the kinetic term $g^{\mu\nu}\partial_\mu Z \partial_\nu \bar{Z}$.
The corresponding non-zero axion-dilaton amplitudes are
\begin{align}
\mathcal{M}_3(-\varphi\varphi)= \mathcal{M}_3(-BB) =& 
\kappa \,
\frac{\<12\>^2\<13\>^2}{\<23\>^2}\, ,
\end{align}
and similarly with square brackets for the corresponding amplitudes of a positive helicity graviton.

Since 3-point special kinematics means that either only angle-brackets {\em or} only square brackets are nonvanishing, the amplitudes 
$\mathcal{M}_3(\pm ZZ) 
= g^{-1} \mathcal{A}^\text{L}_3[\pm ++] \mathcal{A}^\text{L}_3[\pm --] $ vanish identically. Likewise, 3-scalar amplitudes vanish.

In the double-copy of YM with itself, dilaton- and axion-parity ($\varphi \to -\varphi$ and independently $B \to -B$) emerge from the double-copy. However, when higher-derivative corrections are included, this is no longer the case.
In the generic double-copy, where the L and R sector couplings can be distinct, dilaton- and axion-parity is broken already at 3-point by the $F^3$ operator. To see this, note that the double-copy \reef{3ptDC} gives
\be  
   \label{3ptZ}
	\mathcal{M}_3(--Z) = \frac{g^\text{L}_\text{YM}g^\text{R}_{F^3} }{g\Lambda^2} \,
	\<12\>^4\,, 
	\quad
	\mathcal{M}_3(--\bar{Z}) = \frac{g^\text{L}_{F^3} g^\text{R}_\text{YM}}{g\Lambda^2} \,
	\<12\>^4\,,
\ee
and their conjugates. The single axion and dilaton amplitudes are then
\be
   \label{3ptphiB}
	\mathcal{M}_3(--\varphi) =  \frac{\kappa_\varphi}{\Lambda^2}\,
	\<12\>^4\,, ~~~~
	\mathcal{M}_3(--B) = 
	-i\frac{\kappa_B}{\Lambda^2}
    \<12\>^4\,,
\ee
where we have defined
\be
  \label{gdilB}
  \kappa_\varphi = 
  \frac{g^\text{L}_\text{YM}g^\text{R}_{F^3}
  +g^\text{L}_{F^3} g^\text{R}_\text{YM}}{\sqrt{2}g}\,,
  ~~~~~
  \kappa_B = \frac{ g^\text{L}_\text{YM}g^\text{R}_{F^3}
  -g^\text{L}_{F^3} g^\text{R}_\text{YM}}{\sqrt{2}g} \,.
\ee
The couplings $\kappa$, $\kappa_{R^3}$, $\kappa_\varphi$, and $\kappa_B$ are not independent, but satisfy
\be
  \label{kapparel}
  \kappa_\varphi^2 - \kappa_B^2 = 2 \kappa \kappa_{R^3}\,.
\ee

In previous studies \cite{Broedel:2012rc} of the double-copy of YM with higher-derivative operators, the L and R sector couplings were chosen to be the same and that choice results in $g_B=0$. In that case, axion-parity holds, while dilaton-parity is violated even at 3-point by the inclusion of the $F^3$ operator. Note that $F^3$ is not compatible with supersymmetry, so in a supersymmetric context, the amplitudes in \reef{3ptZ} and \reef{3ptphiB} all vanish.

In superstring theory, dilaton-parity is violated by $\alpha'$-corrections. This does not happen at 3-point due to supersymmetry, as discussed above, but it can be seen directly from the $\alpha'^3 e^{-6\varphi}R^4$ effective operator. In type-IIA superstring theory dimensionally reduced to 4d, axion-parity, $B \to -B$, continues to hold to all orders in $\alpha'$. From a double-copy point of view, this is related to the fact that the couplings of the open string in the L and R sectors are identical and the KLT kernel is symmetric. Conversely, as can be seen from (\ref{gdilB}), for a non-L-R symmetric or \textit{heterotic} double-copy, axion- or $B$-parity is generically broken. Such $hhB$ interactions arise in the effective action of the heterotic superstring \cite{Gross:1985rr} from the modification of the $B$ field strength tensor by a Lorentz Chern-Simons term and play an important role in the Green-Schwarz anomaly cancellation mechanism \cite{Green:1984sg}. 

We now turn to 4-point, where the 3-point amplitudes above are needed for identifying the pole terms correctly, both in the L and R sector amplitudes and in the resulting gravitational amplitude.

\subsection{4-Gluon MHV Amplitude and Generalized KKBCJ}
\label{sec:bcj4point}

The starting point is to construct the most general ansatz for the tree-level MHV amplitude in YM + higher derivative operators. We write it as 
\be
\label{eq:fullansatzMHVamps}
\begin{split}
\mathcal{A}_4[1^+ 2^+ 3^- 4^-] 
&=[12]^{2} \langle 34\rangle^{2}
\bigg(\frac{g_{\text{YM}}^{2}}{su}-\frac{g_{F^{3}}^{2}}{\Lambda^{4}}\frac{t}{s}+
  \sum_{k=2}^N \sum_{r=1}^{k-1} \frac{e_{k,r}}{\Lambda^{2k}}  s^{r-1} t^{k-r-1}\bigg)\ ,  \\
 \mathcal{A}_4[1^+ 3^- 2^+ 4^-] 
 &=[12]^{2}\langle 34\rangle^{2} \bigg(
 \frac{g_{\text{YM}}^{2}}{tu} +
  \sum_{k=2}^N \sum_{r=1}^{k-1} \frac{h_{k,r}}{\Lambda^{2k}}  s^{r-1} t^{k-r-1}\bigg) \ .
  \end{split}
\ee
The terms with $g_\text{YM}^2$ arise from the pole terms with two regular YM vertices whereas the $s$-pole term with $g_{F^3}^2$ is from two insertions of $F^3$. There can be no other pole terms in the MHV sector. Contributions from local operators at 4-point are parameterized as polynomial terms in the ansatz with general coefficients $e_{k,r}$ and $h_{k,r}$. 
We have explicitly checked (to all orders we are using) that any local contribution to the MHV amplitudes can be written in the form given in (\ref{eq:fullansatzMHVamps}). The choice of Mandelstam basis in (\ref{eq:fullansatzMHVamps}) is equivalent to choosing a particular basis of higher-dimension operators in the Lagrangian. The local contributions at $O(\Lambda^{-4})$ with coefficients $e_{2,1}$ and $h_{2,1}$ correspond to the combination of  $\textrm{Tr}[F^{4}]$ operators that contribute to the MHV amplitude. The terms at $O(\Lambda^{-6})$ correspond to operators of the schematic form $\textrm{Tr}[D^2F^{4}]$. 

The 4 other arrangements of the external lines are obtained from the above two by cyclic symmetry and momentum relabeling:
\be
  \begin{split}
  \mathcal{A}_4[1^+ 2^+ 4^- 3^-] &= \mathcal{A}_4[1^+ 2^+ 3^- 4^-] \bigg|_{3 \lra 4}\, \\
  \mathcal{A}_4[1^+ 3^- 4^- 2^+] &= 
  \mathcal{A}_4[2^+ 1^+ 3^- 4^-] =
  \mathcal{A}_4[1^+ 2^+ 3^- 4^-] \bigg|_{1 \lra 2}\\
  \mathcal{A}_4[1^+ 4^- 2^+ 3^-] &= \mathcal{A}_4[1^+ 3^- 2^+ 4^-] \bigg|_{3 \lra 4}\, \\
  \mathcal{A}_4[1^+ 4^- 3^- 2^+] &=
  \mathcal{A}_4[2^+ 1^+ 4^- 3^-] =
  \mathcal{A}_4[1^+ 2^+ 3^- 4^-] \bigg|_{1 \lra 2, 3 \lra 4}
  =\mathcal{A}_4[1^+ 2^+ 3^- 4^-]\,.
  \end{split}
\ee

Naturally, one expects the coefficients $e_{k,r}$ and $h_{k,r}$ in \reef{eq:fullansatzMHVamps} to be related since these amplitudes should arise from a Lagrangian with a color-group structure and --- as we shall see --- indeed this is the case: the generalized KKBCJ relations will fix the $h_{k,r}$ completely in terms of the $e_{k,r}$ and the kernel coefficients $a_{k,r}$ and they impose additional constraints on the $e_{k,r}$.

The generalized KKBCJ constraints are imposed using the trace-reversal identities \reef{simpleKK4pt} as well as the L and R sector identities \reef{4ptgenU1L}-\reef{4ptgenBCJL} or  \reef{4ptgenU1R}-\reef{4ptgenBCJR}. The result in the {\bf L sector} is
\be 
 \label{A4result1234L}
 \begin{split}
\mathcal{A}^{\textrm{L}}_4[1^+ 2^+ 3^- 4^-]  =&
[12]^2\<34\>^2
\bigg[ \frac{(g_{\textrm{YM}}^{\textrm{L}})^2}{s u}
- \frac{(g_{F^{3}}^\text{L})^{2}}{\Lambda^4}\frac{t}{s}
+\frac{(g_{\textrm{YM}}^{\textrm{L}})^2}{g^2\Lambda^4} (a_{1,1} - a_{1,0})
+\frac{e^{\textrm{L}}_{3,1}}{\Lambda^6} t  \\
&
\hspace{1.6cm}+ 
\frac{1}{\Lambda^8}\bigg\{
 \frac{(g_{\textrm{YM}}^{\textrm{L}})^2}{g^2} \tilde{a}_{3, 3} \,s^2  + 
 \alpha^\text{L} \,t^2  + 
 e^\text{L}_{4, 2}\,  s t
\bigg\}+\ldots
\bigg]\,, \\
\mathcal{A}^{\textrm{L}}_4[1^+ 3^- 2^+ 4^-]  =&
[12]^2\<34\>^2
\bigg[ \frac{(g_{\textrm{YM}}^{\textrm{L}})^2}{tu}
+\frac{1}{\Lambda^4}\bigg(
\frac{(g_{\textrm{YM}}^{\textrm{L}})^2}{g^2} \big(a_{1,1} - a_{1,0}\big)
- (g_{F^{3}}^\text{L})^{2} \bigg)
+\frac{e^{\textrm{L}}_{3,1}}{\Lambda^6} s  \\
&
\hspace{1.6cm}+
\frac{1}{\Lambda^8}\bigg\{
  -\bigg(\frac{(g_{\textrm{YM}}^{\textrm{L}})^2}{g^2}\tilde{a}_{3, 3}+ \frac{(g_{F^{3}}^{\textrm{L}})^2}{g^2}(a_{1,1}-a_{1,0}) \bigg)\,tu  + \\
  &\hspace{2cm}+ \bigg(\alpha^\text{L}-\frac{(g_\text{YM}^\text{L})^{2}}{g^{2}}\tilde{a}_{3,3}
  - 2\frac{(g_{F^{3}}^{\textrm{L}})^2}{g^2}(a_{1,1}-a_{1,0})+e^\text{L}_{4, 2}
  \bigg)\,s^2
\bigg\}+\ldots
\bigg]\,,
\end{split}
\ee

where 
\be
  \begin{split}
  \tilde{a}_{3,3} &=  a_{3, 3} - a_{3, 2} + a_{3, 1} - a_{3, 0}  \,,
  \\
  \alpha^{\textrm{L}} &= \frac{(g_{\textrm{YM}}^{\textrm{L}})^2}{g^4}\, a_{1, 1}\big(a_{1, 1}  - a_{1, 0}\big) + \frac{(g_{\textrm{YM}}^{\textrm{L}})^2}{g^2}\big( 2a_{3, 3}-a_{3, 2}  \big) 
  + \frac{(g_{F^{3}}^{\textrm{L}})^{2}}{g^2}\big(a_{1, 1}-a_{1, 0}\big) \,.
  \end{split}
\ee
The {\bf R sector} takes the form
\be 
 \label{A4result1234R}
 \begin{split}
\mathcal{A}^{\textrm{R}}_4[1^+ 2^+ 3^- 4^-]  =&
[12]^2\<34\>^2
\bigg[ \frac{(g_{\textrm{YM}}^{\textrm{R}})^2}{s u}
- \frac{(g_{F^{3}}^{R})^{2}}{\Lambda^4}\frac{t}{s}+ \frac{(g_{\textrm{YM}}^{\textrm{R}})^2}{g^2\Lambda^4} a_{1,1}
+\frac{e^{\textrm{R}}_{3,1}}{\Lambda^6} \,t \\
&
\hspace{1.6cm}+ 
\frac{1}{\Lambda^8}\bigg\{
\frac{(g_{\textrm{YM}}^{\textrm{R}})^2}{g^2} a_{3, 3} s^2  + 
 \alpha^\text{R} \,t^2  + 
  e^{\textrm{R}}_{4, 2}  \,s t
\bigg\}+\ldots
\bigg]\,, \\
\mathcal{A}^{\textrm{R}}_4[1^+ 3^- 2^+ 4^-]  =&
[12]^2\<34\>^2
\bigg[ \frac{(g_{\textrm{YM}}^{\textrm{R}})^2}{tu}
+\frac{1}{\Lambda^4}\bigg( \frac{(g_{\textrm{YM}}^{\textrm{R}})^2}{g^2} a_{1,1} 
-(g_{F^{3}}^\text{R})^{2} \bigg)
+\frac{e^{\textrm{R}}_{3,1}}{\Lambda^6} s  \\
&
\hspace{1.6cm}+
\frac{1}{\Lambda^8}\bigg\{  -\bigg( \frac{(g_\text{YM}^\text{R})^{2}}{g^2}\,a_{3,3}
+\frac{(g_{F^{3}}^{\textrm{R}})^{2}}{g^{2}}\,a_{1,1} 
\bigg)\,t u\\
  &\hspace{1.6cm}+ \bigg(\alpha^{\text{R}}-\frac{(g^\text{R}_\text{YM})^{2}}{g^{2}} a_{3,3}
  -2\frac{(g^\text{R}_{F^3})^{2}}{g^{2}} a_{1,1} 
  +e^{\textrm{R}}_{4,2} \bigg)s^2
\bigg\}+\ldots
\bigg]\,,
\end{split}
\ee
where 
\be
  \alpha^R = \frac{(g_{\textrm{YM}}^{\textrm{R}})^2}{g^4}  a_{1, 1}\big( a_{1, 1} - a_{1, 0}\big) 
  + \frac{(g_{\textrm{YM}}^{\textrm{R}})^2}{g^2} \big(2 a_{3, 3} - a_{3, 2} +a_{3, 0}\big) 
  + \frac{(g_{F^{3}}^{\textrm{R}})^{2}}{g^2}a_{1, 1}  \,.
\ee
\begin{table}
    \begin{center}
	\begin{tabular}{|c|c|c|c|c|}
		\hline
		Schematic Operator & Total & Generalized & String & Cubic BAS\\
		\hline
		$ \textrm{Tr}[F^{4}]$& 1& 1 & 0 & $\times$ \\
		\hline
		$\textrm{Tr}[D^{2}F^{4}]$& 2& 1 & 1 & 1 \\
		\hline
		$\textrm{Tr}[D^{4}F^{4}]$& 3& 3 & 1 & 1 \\
		\hline
		$\textrm{Tr}[D^{6}F^{4}]$& 4& 3 & 2 & 2 \\
		\hline
	\end{tabular}
\end{center}
    \caption{Number of tunable parameters in the operator coefficients contributing to the MHV amplitude $\mathcal{A}_4[1^+ 2^+ 3^- 4^-]$ subject to the L or R sector KKBCJ relations (generalized, string, or pure field theory BAS, respectively). 
    The total number of independent MHV operators at that dimension is also listed.  
    The $\times$ indicates that $\textrm{Tr}[F^{4}]$ is  disallowed by the field theory KKBCJ relations.
    }
    \label{fig:tableoperatorsMHV}
\end{table}

Each of the local (i.e.~non-pole) terms in \reef{A4result1234L} or \reef{A4result1234R} correspond to a local operator in the YM + h.d.~effective action. For example, 
\be
 \begin{split}
  \mathcal{A}^{\textrm{L}}_4[1^+ 2^+ 3^- 4^-] &\supset
[12]^2\<34\>^2 \frac{1}{\Lambda^4}
\frac{(g_{\textrm{YM}}^{\textrm{L}})^2}{g^2} \big(a_{1,1} - a_{1,0}\big)\,,
\\
\mathcal{A}^{\textrm{R}}_4[1^+ 2^+ 3^- 4^-] &\supset
[12]^2\<34\>^2 \frac{1}{\Lambda^4}
\frac{(g_{\textrm{YM}}^{\textrm{R}})^2}{g^2} \,a_{1,1},
\end{split}
\ee 
comes from the matrix element of (a particular contraction of) $\Tr{}[F^4]$. As this shows, the generalized KKBCJ allows $\Tr{}[F^4]$ with independent tunable coefficients, $a_\text{L}$ and $a_\text{R}$ given in \reef{aLaR}, for the L and R sectors respectively, from the $\partial^2 \phi^4$ operators \reef{lagrangianoper} in the zeroth-copy model. 
In contrast, this operator is not allowed by the uncorrected field theory KKBCJ relations (which have $a_{k,r}=0$). And while $\Tr{}[F^4]$ does arise the $\alpha'$-expansion of the open (super)string gluon amplitudes, it does so with a fixed untunable coefficient because $a_{1,1}=-1/(6\pi \alpha')$ and $a_{1,0}=0$.   

In Table \ref{fig:tableoperatorsMHV}, the row labeled $\Tr{}[F^4]$   summarizes the above discussion. Similarly, the subsequent rows in the table compare the number of allowed operators at higher order contributing to the  YM + h.d.~MHV tree amplitude when subject to the respective KKBCJ constraints. In the count of tunable couplings, we consider  $g_{\text{YM}}^{\textrm{L}/\textrm{R}}$, $g$ and $g_{F^3}^{\textrm{L}/\textrm{R}}$  fixed by the 3-point amplitudes and we avoid double-counting by taking into account that lower-point parameters often feed into higher-point contributions. The lesson from the Table \ref{fig:tableoperatorsMHV} is that our generalized KKBCJ relations allow a broader range of EFT operators in  single-copy EFTs but still not all operators are allowed. 

For a closer comparison with string theory, the $\alpha'$-expansion of the  4-gluon MHV tree amplitude of type-I open string theory is 
\begin{equation}
\mathcal{A}_{4}^{\text{open}}[1^{+}2^{+}3^{-}4^{-}]
=g_{\text{YM}}^{2}[12]^{2}\langle 34\rangle^{2}\left [ \frac{1}{su}-\frac{\pi^{2}\alpha'^{2}}{6}+\alpha'^{3}\zeta(3)t+O(\alpha'{}^{4})\right ]  .
\end{equation}
Comparing the leading orders, this corresponds to the limit 
\be
{a_{k,l},g,\Lambda\rightarrow \textrm{String Kernel},
~~~~\ g_{F^{3}}^{2}\rightarrow 0, 
~~~~ e_{3,1}^{\textrm{L}} \rightarrow \frac{\zeta_{3}}{\pi^{3}}, \ \ldots  }
\ee
of the L or R sector MHV amplitudes \reef{A4result1234L} and \reef{A4result1234R}, where \textit{String Kernel} refers to equation \reef{stringsas}.

%%%%%%%%%%%%%%%%%%%%%%%%%%%%	
\subsection{4-Graviton MHV from  of YM + h.d.}
\label{sec:dc4point}
%%%%%%%%%%%%%%%%%%%%%%%%%%%%%	

The generalized KLT double-copy formula can be carried out in any choice of KLT basis (and we have explicitly checked the basis-independence) thanks to the generalized KKBCJ relations. For example, we can  compute $\mathcal{M}_4$ as
\be
  \label{M1234}
  \mathcal{M}_4(1234) 
  = \mathcal{A}^\text{L}_4(1^+ 2^+ 3^- 4^-)
  \frac{1}{f_1(s,t)}
  \mathcal{A}^\text{R}_4(1^+ 2^+ 3^- 4^-)\,,
\ee
with $f_1$ given by \reef{f1res}. Using \eqref{A4result1234L} and \eqref{A4result1234R}, we then find
\begin{align}\label{eq:doublecopygravitymhv}
 \mathcal{M}_4(1^+ 2^+ 3^- 4^-) =  [12]^4 \<34\>^4 
 &\bigg[ -\frac{\kappa^2}{stu}+\frac{\kappa_{\varphi}^2+\kappa_{B}^2}{\Lambda^4}
 \frac{1}{s} 
 \nonumber\\
 &-\frac{1}{\Lambda^6} 
 \bigg( 
 \frac{(g^{\textrm{R}}_\text{YM})^2}{g^2}\,e^{\textrm{L}}_{3, 1}
 +\frac{(g^{\textrm{L}}_\text{YM})^2}{g^2}\,e^{\textrm{R}}_{3, 1} 
 - \frac{\kappa^2}{g^2}\,a_{2, 0}\bigg)
 \nonumber\\
 &+\frac{1}{\Lambda^8} 
  \bigg(
     \kappa_{R^3}^2\frac{t u}{s}
     + \bigg( 
      \frac{(g^{\textrm{R}}_\text{YM})^2}{g^2}\,e^{\textrm{L}}_{4, 2}
      +\frac{(g^{\textrm{L}}_\text{YM})^2}{g^2}\,e^{\textrm{R}}_{4, 2}  
      + \tilde{a}_{\Lambda^8}
     \bigg ) s \bigg )
 \nonumber\\
 &+
O\left( \frac{1}{\Lambda^{10}}\right)\bigg]\, ,
\end{align}
where we have used the identification of the gravitational coupling $\kappa$ and the $R^3$-coupling $\kappa_{R^3}$ in \eqref{kappa} and the non-canonical 4-derivative couplings between the dilaton/axion and the graviton in \reef{gdilB}. The constant $\tilde{a}_{\Lambda^8}$ is a linear combination of kernel coefficient $a_{k,r}$. 

Performing a term-by-term analysis of (\ref{eq:doublecopygravitymhv}), we find
\begin{itemize}
    \item $O(\Lambda^{0})$: the leading term  is the tree-level graviton amplitude.
    \item $O(\Lambda^{-4})$: the axion and dilaton can be exchanged only in the $s$-channel since they couple to same-helicity gravitons \reef{3ptphiB}.
    \item $O(\Lambda^{-6})$: this local contribution is the matrix element of $R^4$. It is generated irrespective of the higher-dimension corrections in KKBCJ and the KLT kernel so long as the coefficient $e_{3, 1} \ne 0$ of $D^2F^4$ is non-zero in the L or R sector. Thus, even in the usual field theory double-copy or in the $\alpha'$-expansion of stringy KLT, $R^4$ arises from the double-copy of $D^2F^4$ with the usual leading YM 4-point amplitude.\footnote{While it is true that the MHV matrix element of $R^4$ mathematically is directly the square of the MHV matrix element of $\Tr{}F^{4}$, it is {\em not} the case that $R^4$ arises from $\Tr{}F^{4}$ in the actual double-copy.} 
 The new feature in the generalized double-copy is that $R^4$ can be generated from the $a_{2,0}$-controlled 4-derivative correction to the BAS model even in the absence of $D^2F^4$ in both the L and R sectors of (YM + h.d.).

    \item $O(\Lambda^{-8})$: the pole term arises from the factorization into two $R^3$-vertices. Since they are only $+++$ and $---$, the exchange happens only in the $s$-channel. In addition there is a local term from $\nabla^2R^4$ at this order and its coefficient is determined by the double-copy of the matrix element $\Tr{}D^4F^4$ and the usual YM amplitude along with various kernel coefficients $a_{k,r}$ suppressed into the constant $\tilde{a}_{\lambda^8}$ in \reef{eq:doublecopygravitymhv}.
    \end{itemize}

There are no other poles in 
 $\mathcal{M}_4(1^+ 2^+ 3^- 4^-)$ than those shown in \reef{eq:doublecopygravitymhv}, hence higher-order contributions all arise from local operators. 
 It is perhaps curious to mention that the full residue of the $s$-channel of the MHV amplitude \reef{eq:doublecopygravitymhv} factorizes into the palatable form
\begin{equation}
\lim_{s\rightarrow 0}s\ \mathcal{M}_{4}[1^{+} 2^{+} 3^{-} 4^{-}]=\frac{[12]^{4}\langle 34\rangle^{4}}{g^2}\left ( \frac{(g_{\textrm{YM}}^{\textrm{L}})^{2}}{t}+\frac{(g_{F^{3}}^{\textrm{L}})^{2}t}{\Lambda^{4}}\right )\left ( \frac{(g_{\textrm{YM}}^{\textrm{R}})^{2}}{t}+\frac{(g_{F^{3}}^{\textrm{R}})^{2}t}{\Lambda^{4}}\right ) \,.
\end{equation}
\noindent Upon expanding the above expression, the $t^{-2}$ term corresponds to the exchange of the graviton with the regular Einstein-Hilbert 3-point interactions, the $t^{2}$ term corresponds to the exchange with a graviton with two $R^{3}$ interactions, and $t^{0}$ term corresponds to dilaton-axion exchange.

It is quite interesting to note that while the generalized KKBCJ relations allow for a wider range of higher-derivative operators in the L and R MHV amplitudes, there are (for generic input) no new operators appearing in the 4-point MHV gravity amplitude (see Table \ref{fig:tableoperatorsMHV}); rather, its Wilson coefficients are merely shifted by the $a_{k,r}$ parameters. For example, in (\ref{eq:doublecopygravitymhv}), we can absorb the nonzero $a_{2,0}$ into $e_{3,1}^{\textrm{R}}$ as
\begin{equation}
\label{shifte31}
e_{3,1}^{\textrm{R}}\rightarrow e_{3,1}^{\textrm{R}}+ \frac{(g_{\text{YM}}^{\textrm{R}})^{2}}{g^{2}}a_{2,0} \, .
\end{equation}

\begin{table}
    \begin{center}
	\begin{tabular}{|c|c|c|c|c|c|c|c|}
		\hline
	    Schematic Operator & MHV & NSD & SD  \\
		\hline
		$ R^{4}$ & \text{1 of 1} & \text{0 of 0}  & \text{0 of 1}  \\
		\hline
		$ \nabla^{2}R^{4}$ & \text{1 of 1} & \text{0 of 0} & \text{1 of 1} \\
		\hline
		$ \nabla^{4}R^{4}$ & \text{2 of 2} & \text{1 of 1} & \text{1 of 2} \\
		\hline	
		$ \nabla^{6}R^{4}$ & \text{2 of 2} & \text{0 of 0} & \text{1 of 1} \\
		\hline	
		$ \nabla^{8}R^{4}$ & \text{3 of 3} & \text{1 of 1} & \text{1 of 2} \\
		\hline
		$ \nabla^{10}R^{4}$ & \text{3 of 3} & \text{1 of 1} & \text{2 of 2} \\
		\hline	
	\end{tabular}
\end{center}
    \caption{Number of independent local operators out of the total possible contributing to the 4-graviton amplitude: we list how many operators are generated by any version of the KLT double copy (generalized / field theory BAS / string kernel).      
    }
    \label{fig:tableoperatorsMHVGR}
\end{table}

The pattern continues to higher orders; we have explicitly checked up to and including $O(p^{18})$, i.e.~$D^{10}R^4$. Table \ref{fig:tableoperatorsMHVGR} summarizes the number of independent local operators that contribute to the MHV amplitude and we find that each one of them is produced in any of the KLT double-copies. In particular, we find at 4-point that the entire effect of the higher-derivative corrections to the double-copy kernel can be absorbed into the R (or L) sector Wilson coefficients of the YM+h.d.~amplitudes. Of course, there are also other combinations of the double-copy spectrum. For example, we can double-copy $\mathcal{A}^\text{L}_4(1^+ 2^+ 3^- 4^-)$ with $\mathcal{A}^\text{R}_4(1^- 2^- 3^+ 4^+)$
to get the $\mathcal{M}_4(ZZ\bar{Z}\bar{Z})$ amplitude. It is noteworthy (but somewhat trivial at 4-point) that the same shift of Wilson coefficients that absorbs the effect of the $a_{k,r}$-coefficients of the double-copy kernel for the 4-graviton MHV amplitude also does the job for these other helicity combinations obtained from double-copies of the 4-gluon MHV amplitudes. 

This phenomenon is reminiscent of the single-valued projection that occurs in the string double-copy \cite{Stieberger:2014hba}, but is much more general. While the single-valued projection that occurs in string theory is tied to a particular expansion of higher dimensional operators, our result seems to hold for a large class of EFTs without a clear high energy completion. 

%%%%%%%%%%%%%%%%%%%
\subsection{SD and NSD Sectors}
%%%%%%%%%%%%%%%%%%%
\label{sec:SDNSD4}

One can similar perform an analysis of YM with higher-derivative corrections in the Self-Dual (SD, or all-plus) and Next-To-Self-Dual Sector (NSD, or one-minus) at 4-point. Up to the order $D^{10} F^4$ 
we have checked that the matrix elements of local operators can be written with a common spinor-helicity prefactor times a Mandelstam polynomial:
\be 
  \label{SDandNSD1}
  \begin{split}
  \text{SD ($++++$):}~~~\frac{1}{\<12\>\<23\>\<34\>\<41\>}\times P_\text{SD}(s,t)\,,\\
  \text{NSD ($+++-$):}~~~\frac{\<4|1.3|4\>^2}{\<12\>\<23\>\<34\>\<41\>}\times P_\text{NSD}(s,t)\,,
  \end{split}
\ee
where $\<4|1.3|4\> = \<41\>[13]\<34\>$ and $P_\text{SD}$ and $P_\text{NSD}$ are local Mandelstam polynomials. We impose cyclic symmetry for the ansatz of the SD local terms. Including the appropriate pole terms and subjecting the ansatz to the generalized KKBCJ relations (\reef{simpleKK4pt} and the L or R sector identities \reef{4ptgenU1L}-\reef{4ptgenBCJL} and  \reef{4ptgenU1R}-\reef{4ptgenBCJR}), we find the result for the R-sector SD amplitude is
\be
  \label{ARSD}
  \begin{split}
  \mathcal{A}^{\textrm{R}}_4[1^+ 2^+ 3^+ 4^+]  =&
   \frac{s t u}{\<12\>\<23\>\<34\>\<41\>}
   \bigg(
    \frac{2 g_\text{YM}^\text{R} g_{F^3}^\text{R}}{\Lambda^2} 
        + \frac{2g_\text{YM}^\text{R} g_{F^3}^\text{R}}{g^2 \Lambda^6} a_{1,1}\,t^2 
    \\
    & \hspace{3cm}
    -
    \frac{d^\text{R}_1}{2\Lambda^6}(s^2 + t^2 + u^2)
    + 
    \frac{d^\text{R}_2}{\Lambda^8}\, s t u
    + \ldots
   \bigg)\,, 
   \\
  \end{split}
\ee
and the R-sector NSD amplitude  is
\be
  \label{ARNSD}
  \begin{split}
  \mathcal{A}^{\textrm{R}}_4[1^+ 2^+ 3^+ 4^-]  =&
   \frac{\<4|1.3|4\>^2}{\<12\>\<23\>\<34\>\<41\>}
   \bigg(
    \frac{g_\text{YM}^\text{R} g_{F^3}^\text{R}}{\Lambda^2} 
    + \frac{g_\text{YM}^\text{R} g_{F^3}^\text{R}}{g^2 \Lambda^6} a_{1,1}\,s u 
    -\frac{c^\text{R}_1}{\Lambda^8} s t u
    + \ldots\bigg)\,.
  \end{split}
\ee
The leading term in each case arises from the pole diagram with a 3-point vertex of YM and one from $F^3$ and they have been given previously (see \cite{Broedel:2012rc} and references therein). The rest of the terms are local and can be rewritten in a manifestly polynomial form in terms of spinor-helicity variables that is less compact. The L-sector SD and NSD amplitudes are similar and, to the orders shown, they are found by simply taking $a_{1,1} \to a_{1,1}-a_{1,0}$ and changing the superscripts R to L. 

One can compute the double-copy of various combinations of the MHV, SD, and NSD amplitudes to get all the possible gravity-dilaton-axion amplitudes. In particular, amplitudes with an odd number of external axions are nonvanishing whenever the L and R sector couplings are distinct, as noticed already at 3-point. 

Let us simply do one illustrative example here, the SD graviton amplitude. Using \reef{M1234} we find from \reef{ARSD} that
\be
  \label{M4SD}
 \begin{split}
 \mathcal{M}_4(1^+ 2^+ 3^+ 4^+) 
 = &
 4 \kappa \kappa_{R^3} \,s t u\, \frac{[12][23][34][41]}{\<12\>\<23\>\<34\>\<41\>}
 \\
 &+
 \frac{1}{5 g^2 \Lambda^8}
 \Big( g_{F^3}^\text{L} g_\text{YM}^\text{L} d^\text{R}_1+g_{F^3}^\text{R} g_\text{YM}^\text{R} d^\text{L}_1
 -2 \kappa \kappa_{R^3}(2a_{1,1}-a_{1,0})
 \Big)\\
 &~~\hspace{1.5cm}\times
 \Big([12]^4[34]^4 s+[13]^4[24]^4 t+[14]^4[23]^4 u\Big)
 +\ldots
 \end{split}
\ee
Let us first comment on the leading term. This is a pole term with two sets of contributions: one from the graviton exchange with a regular vertex from $R$ and another from $R^3$, contributing 
$2 \kappa \kappa_{R^3}$ to the overall factor. Next, there are scalar exchanges which kinematically take the same form as the graviton exchange: the dilaton exchange comes with coupling $\kappa_\varphi^2$ while the axion exchange contributes $(i\kappa_B)^2$. Thanks to the relation \reef{kapparel}, the total scalar exchange therefore contributes a coupling 
$\kappa_\varphi^2 - \kappa_B^2 = 2 \kappa \kappa_{R^3}$. Thus, the combined graviton+dilaton+axion exchange accounts precisely for the overall factor $4 \kappa \kappa_{R^3}$ in the first line of \reef{M4SD}.

Next, we notice that there is no contribution corresponding to $R^4$. There does exist a unique operator $R^4$ with a nonvanishing matrix element in the SD sector, but it is not produced in the double copy. For $D^2 R^4$, there is likewise a unique SD matrix element and as shown above it \textit{is} produced by the double-copy. 

Table \ref{fig:tableoperatorsMHVGR} summarizes the results for higher-derivative operators produced by the double-copy of the 4-point SD and NSD graviton amplitudes. We have computed 4-point amplitudes too with external scalars too, but will not clutter the presentation by presenting them here.

%%%%%%%%%%%%%%%%%%%%%%%%%%%%%%%%%
%%%%%%%%%%%%%%%%%%%%%%%%%%%%%%%%%
%%%%%%%%%%%%%%%%%%%%%%%%%%%%%%%%%

\section{Generalized KLT at 5-Point}
\label{sec:5pt}

We now analyse the KLT bootstrap equations at 5-point: we set up the problem and present the conditions on the zeroth copy that ensures that the matrix of amplitudes has rank 2. We then solve the equations perturbatively and show that --- very importantly --- this does not place restrictions on the coefficients $a_{k,r}$ for the 4-point KLT kernel. 
As an example, we apply the generalized KLT double-copy to YM theory with higher derivative corrections in the SD (all-plus) sector. 

\subsection{5-Point Bootstrap Equations}
\label{s:boot5pt}
We use cyclic symmetry and momentum relabeling to write the $(n-1)! = 4! = 24$ doubly color-ordered amplitudes in terms of 8 functions:
\begin{equation}
\begin{array}{lllll}
m_{5}[12345|12345] &= g_1[12345] \,,
&~~~~~&
m_{5}[12345|13254] &= g_5[12345]\,,
\\
m_{5}[12345|12354] &= g_2[12345] \,,
&~~~~~&
m_{5}[12345|13524] &= g_6[12345]\,, 
\\
m_{5}[12345|12453] &= g_3[12345] \,,
&~~~~~&
m_{5}[12345|14253] &= g_7[12345]\,, 
\\
m_{5}[12345|12543] &= g_4[12345] \,,
&~~~~~&
m_{5}[12345|15432] &= g_8[12345] \,.
\end{array}    
\label{basisindepentamplitudesss}
\end{equation}
Cyclic symmetry requires $g_1[12345] = g_1[51234]$ and likewise for $g_{6}$, $g_{7}$, and $g_8$. We do not assume reversal symmetry. 

Imposing the minimal rank condition is equivalent to requiring all 3$\times$3 minors to vanish, thus reducing the 24$\times$24 matrix to rank $(n-3)!=2$. These conditions allow us to analytically solve for $g_4$, $g_5$, $g_6$, $g_7$, and $g_8$ in terms of $g_1$, $g_2$, and $g_3$. For example, 
\be
  \begin{split}
  g_4[12345] =& 
  \Big(g_1[12345] g_1[12354] g_2[12435] 
  - g_2[12345] g_2[12354] g_2[12435] 
     \\
     &~~~~-g_1[12354] g_2[51243]g_3[12345] + 
        g_2[12354] g_3[12345] g_3[12435]\Big)
        \\&
        \Big(g_1[12345] g_3[12435]-g_2[12345] g_2[51243] 
     \Big)^{-1}
       \end{split}
\ee
The remaining analytic solutions are given in Appendix \ref{app:5pt}. In addition, there are bootstrap equations that relate $g_1$, $g_2$, and $g_3$. One can explicitly check that the BAS amplitudes and the string kernel amplitudes of \cite{Mizera:2016jhj} solve these equations. 

Let us now outline how we set up the ansatz for solving the functions $g_i$ perturbatively in the momentum expansion. The leading order terms are the usual BAS amplitudes and the subleading terms have both local contributions as well as pole terms from factorization into BAS 3-point amplitudes and 4-point local contributions parameterized by the coefficients $a_{k,r}$ in Section \ref{s:pert4pt}. 
For example, the $s_{12}$-factorization channel of $g_1$ is 
\be
  \begin{split}
 s_{12}\, g_1[12345] \bigg|_{s_{12} = 0}=& 
 s_{12}\,  m[12345|12345]  \bigg|_{s_{12} = 0}\\
 =&
 g^3 \bigg( \frac{1}{s_{34}} +\frac{1}{s_{45}} \bigg)
 + m_3[12P|12P] \, \tilde{m}_4[345P|345P]
 \\
 =&
  g^3 \bigg( \frac{1}{s_{34}} +\frac{1}{s_{45}} \bigg)
  +
    \frac{g(a_{1, 0} - 2 a_{1, 1})}{\Lambda^{4}} s_{35} - 
 \frac{g \,a_{2, 0}}{\Lambda^6}  s_{35}^2 + \ldots,
  \end{split}
\ee
where $P=P_{12}$ is on-shell ($P^2=0$), $m_3[12P|12P] = g$ and $\tilde{m}_4[345P|345P] = f_1(s_{34},s_{35})\big|_\text{local}$ 
are the higher-derivative corrections to the 4-point bi-adjoint that are encoded in the function  $f_1$ given in \reef{f1res}.

Including all five factorization channels and all possible local counterterms compatible with cyclic symmetry gives the following ansatz for $g_1$
\be\label{eq:g1ansatzsolved}
  \begin{split}
 g_1[12345] =& g^3 \bigg( \frac{1}{s_{12}s_{34}} + \frac{1}{s_{23}s_{45}}+\frac{1}{s_{34}s_{51}} +\frac{1}{s_{45}s_{12}} +\frac{1}{s_{51}s_{23}}\bigg)\\
  &+ \frac{g}{\Lambda^{4}} (a_{1, 0} - 2 a_{1, 1}) \bigg(
  \frac{s_{35}}{s_{12}} + \frac{s_{41}}{s_{23}} 
  + \frac{s_{13}}{s_{45}}+ \frac{s_{24}}{s_{51}} 
  + \frac{s_{52}}{s_{34}}\bigg)
  + w_{1} \\
  &- \frac{g a_{2, 0} }{\Lambda^6} 
  \bigg( \frac{s_{35}^2}{s_{12}} + \frac{s_{41}^2}{s_{23}} 
  +\frac{s_{13}^2}{s_{45}} + \frac{s_{24}^2}{s_{51}} 
  + \frac{s_{52}^2}{s_{34}}\bigg) 
      \\
      &+ w_1'\big(s_{12} + s_{23} + s_{34} + s_{45} + s_{51}\big)+\ldots 
 \end{split}
\ee
In the ansatz, the local terms have coefficients $w_i$. 

Remarkably, the 5-point rank 2 bootstrap equations fix the couplings 
$w_1$ and $w_1'$ of the 5-point local contact terms completely in terms of the 4-point bi-adjoint, namely
\be
 \label{w1w1p}
 w_1 = 2\frac{g}{\Lambda^4} \big(2 a_{1, 1}-a_{1,0}\big)
 ~~~~\text{and}~~~~w_1'  = -\frac{g}{\Lambda^6}a_{2,0}\,.
\ee
One has to go to cubic order in the Mandelstams in order to find local 5-point operators whose coefficients are not fixed by 4-point coefficients. Starting at $O(p^4)$ (not shown above) we include also parity odd terms, but we have found that none of them are allowed by the  bootstrap. We have solved the bootstrap equations up to and including $O(p^6)$.

It is very important that the 5-point bootstrap equations do not place any restrictions on the $a_{k,r}$-parameters that were free in the 4-point solution. This is a very relevant consistency check: if 5-point has constrained 4-point, then we would have needed to go to 6-point to understand if that gave even further constraints. As it is, the lack of such constraints is an important clue that our generalization of the KLT kernel is based on sound principles.

\subsection{Comparison with String Theory at 5-point}
The string KLT amplitudes in \cite{Mizera:2016jhj} are  dimensionless, for example 
\begin{equation}\label{eq:stringresult5pg1}
\begin{split}
g_{1}^{\text{string}}[12345]&=\cot \left(\pi  \alpha'  s_{12}\right) \cot \left(\pi  \alpha'  s_{34}\right)+\cot \left(\pi  \alpha'  s_{51}\right) \cot \left(\pi  \alpha'  s_{34}\right) \\
&+\cot
   \left(\pi  \alpha'  s_{12}\right) \cot \left(\pi  \alpha'  s_{45}\right)+\cot \left(\pi  \alpha'  s_{23}\right) \cot \left(\pi  \alpha'  s_{45}\right)\\
   &+\cot
   \left(\pi  \alpha'  s_{23}\right) \cot \left(\pi  \alpha'  s_{51}\right)+1  \\
   &=\frac{1}{\alpha'^{2}\pi ^2}\left (\frac{1}{ s_{12} s_{45}}+\frac{1}{ s_{23} s_{45}}+\frac{1}{s_{23} s_{51}}+\frac{1}{ s_{12} s_{34}}+\frac{1}{s_{51}
   s_{34}}\right )\\
   &+\frac{1}{3}\bigg(
  \frac{s_{35}}{s_{12}} + \frac{s_{41}}{s_{23}} 
  + \frac{s_{13}}{s_{45}}+ \frac{s_{24}}{s_{51}} 
  + \frac{s_{52}}{s_{34}}-2\bigg)+O(\alpha'^{2}) \ .
   \end{split}
\end{equation}
To compare \reef{eq:g1ansatzsolved}-\reef{w1w1p} with \reef{eq:stringresult5pg1}, we therefore multiply $g_1$ by $\Lambda$ and then find a match when the $a_{k,r}$ take the values \reef{stringsas} up to and including order $O(p^6)$. At $O(p^6)$ there are four coefficients of local operators that are unrelated to the 4-point  $a_{k,r}$'s and the comparison to \reef{eq:stringresult5pg1} shows that all four coefficients are zero in the string kernel. In the generalized kernel they can take any value. 

\subsection{Example: Higher-Derivative YM to Gravity at 5-Point}

Due to the tower of local corrections to the 4-point YM amplitudes in Sections \ref{sec:bcj4point} and \ref{sec:SDNSD4}, the analysis of the 5-point factorization channels takes a bit more effort. We assume a general ansatz for the pole contribution takes the form
\begin{equation}
    \mathcal{A}_5^{\text{YM}}[1^+ 2^+ 3^+ 4^+ 5^+] = \frac{1}{\langle 12 \rangle \langle 23 \rangle \langle 34 \rangle \langle 45 \rangle \langle 51 \rangle}\Big[P_+(s_{ij}) + P_-(s_{ij})\epsilon(1,2,3,4)\Big],
\end{equation}
where $P_+$ and $P_-$ are cyclically invariant polynomials in a basis of independent Mandelstam invariants $\{s_{12},s_{23},s_{34},s_{45},s_{51}\}$ and 
\begin{equation}
\epsilon(1,2,3,4) = \epsilon_{\mu\nu\rho\sigma}p_1^\mu p_2^\nu p_3^\rho p_4^\sigma 
=\frac{i}{4}\Big(\langle 12\rangle \langle 34\rangle [23][14]-\langle 14\rangle \langle 23\rangle [12][34]\Big)\,.
\end{equation}
Since this ansatz is cyclically invariant, we only need to match a single factorization channel to the known 3- and 4-point amplitudes. For the 45 channel we match to 
\begin{align}
    \mathcal{A}_4^{\text{YM}}[1^+ 2^+ 3^+ (P_{45})^+]\mathcal{A}_3^{\text{YM}}[(-P_{45})^- 4^+ 5^+] &= \frac{[45]}{\langle 12 \rangle \langle 23 \rangle \langle 34 \rangle \langle 51 \rangle}P_{\text{SD}}(s_{12},s_{13}) \nonumber \ , \\
    \mathcal{A}_4^{\text{YM}}[1^+ 2^+ 3^+ (P_{45})^-]\mathcal{A}_4^{\text{YM}}[(-P_{45})^+ 4^+ 5^+] &\\
    = \frac{[45]}{\langle 12 \rangle \langle 23 \rangle \langle 34 \rangle \langle 51 \rangle}(s_{12}(s_{51}-s_{23})+& s_{23}s_{34})^2P_{\text{NSD}}(s_{12},s_{13})\,,
\end{align}
where $P_{\text{SD}}$ and $P_{\text{NSD}}$ were defined in \reef{SDandNSD1} and their explicit form can be inferred from the L and R sector results for the 4-point SD and NSD  amplitudes presented in Section \ref{sec:SDNSD4}
where $\<45\>=0$ for the equality to hold. Combining this with the 5-particle ansatz gives the following residue matching condition:
\begin{equation}
\begin{split}
    P_+(s_{45}=0)&-\frac{i}{4}(s_{12}(s_{51}-s_{23})+s_{23}s_{34})P_-(s_{45}=0) \\
    &= P_{\text{SD}}(s_{12},s_{13}) + (s_{12}(s_{51}-s_{23})+s_{23}s_{34})^2P_{\text{NSD}}(s_{12},s_{13}).
\end{split}
\end{equation}

Both sides of this equation are polynomials that can be written in terms of the independent Mandelstam invariants $\{s_{12},s_{23},s_{34},s_{51}\}$, so we can match terms on both sides and fix all coefficients in the ansatz. 

Next, we have to add contributions of local terms to the ansatz. For example, the two independent SD contributions from $\Tr[F^5]$ are
\begin{equation}
\begin{split}
&\mathcal{A}_{5}^{\text{YM}}[1^{+} 2^{+} 3^{+} 4^{+} 5^{+}]\supset \ \frac{c_{5,1}}{\Lambda^6 }[12][23][34][45][51] \\
&\hspace{40mm}+\frac{c_{5,2}}{\Lambda^6 }\Big(
 [12]^2 [34] [35] [45] + \text{cyclic perms}\Big) \ .   
 \end{split}
\end{equation}

The ansatz of pole terms (fixed by 3- and 4-point input) and all possible local terms is then subject to the L and R generalized KKBCJ relations at 5-point. This is efficiently done in the form of setting all $3 \times 3$ matrices
\begin{equation}\label{eq:generalizedBCJrelationsss5p}
\det \begin{bmatrix}
m_5[12345|12345] & m_5[12345|12354] & \mathcal{A}^{\textrm{R}}_{5}[12345] \\ 
m_5[12354|12345] & m_5[12354|12354] & \mathcal{A}^{\textrm{R}}_{5}[12354]  \\ 
m_5[\alpha_{i}|12345]  & m_5[\alpha_{i}|12354] & \mathcal{A}^{\textrm{R}}_{5}[\alpha_{i}] 
\end{bmatrix},
\end{equation}
to zero. 
No constraints are placed by these constraints on the lower-point coefficients, but some coefficients of the 5-point local operators are fixed; for example we find
$c_{5,1} = 0$  and $c_{5,2} = 0$ 
for both the L and R sectors. This is summarized in the row labeled $\Tr[F^5]$ in Table \ref{fig:tableoperatorsSD}.

\begin{table}
    \begin{center}
	\begin{tabular}{|c|c|c|c|c|}
		\hline
		Schematic Operator & Total & Bootstrapped & String & FT\\
		\hline
		$ \textrm{Tr}[F^{5}]$& 2 & $\times$ & $\times$ & $\times$ \\
		\hline
		$\textrm{Tr}[D^{2}F^{5}]$& 5 & $\times$ & $\times$ & $\times$ \\
		\hline
		$\textrm{Tr}[D^{4}F^{5}]$& 14 & 1 & 1 & 1 \\
		\hline
		$\textrm{Tr}[D^{6}F^{5}]$& 28 & 4 & 2 &  2\\
		\hline
	\end{tabular}
\end{center}
    \caption{Number of operators contributing to  $\mathcal{A}^{\textrm{R}}_5[1^+ 2^+ 3^+ 4^+ 5^+]$ after imposing the KKBCJ relations in the generalized, string, and BAS form. We also list the total number of independent operators. The $\times$ indicates that no operator at that order is allowed.}
    \label{fig:tableoperatorsSD}
\end{table}

The double-copy of $\mathcal{A}[1^+ 2^+ 3^+ 4^+ 5^+]$ with itself gives the graviton amplitude  expression for $\mathcal{M}(1^+ 2^+ 3^+ 4^+ 5^+)$, which we have computed up to $O(\Lambda^{-14})$. 
As at 4-point, we find that the same operators are produced by the generalized double-copy kernel as with BAS or strings, however, their coefficients are now in certain cases shifted but the parameters of the kernel.  

\begin{table}
    \begin{center}
	\begin{tabular}{|c|c|c|c|}
		\hline
		Schematic Operator & SD    \\
		\hline
		$R^{5}$ & \text{0 of 1} \\
		\hline
		$\nabla^{2}R^{5}$ & \text{0 of 1}  \\
		\hline
		$\nabla^{4}R^{5}$ & \text{1 of 3}  \\
		\hline
		$\nabla^{6}R^{5}$ & \text{1 of 3}  \\
		\hline
	\end{tabular}
\end{center}
    \caption{Number of tunable parameters in $\mathcal{M}_5[1^+ 2^+ 3^+ 4^+ 5^+]$ of the given form and number of tuneable parameters compatible with the double copy. Note that the generalized, string and cubic BAS double copy all generate the same set of higher dimension operators. The $\times$ indicates that the higher dimension operator of the given form are not just fixed, but disallowed by the double copy.} 
    \label{fig:tableGRoperatorsSD}
\end{table}

%%%%%%%%%%%%%%%%%%%%%%%%%%%%%%%%%
%%%%%%%%%%%%%%%%%%%%%%%%%%%%%%%%%
%%%%%%%%%%%%%%%%%%%%%%%%%%%%%%%%% 
\section{Alternative Double-Copy Constructions}
\label{sec:QBAS}

We showed in the Introduction how the zeros of the determinant of the rank $R_n$ submatrix of zeroth-copy amplitudes may provide ``missing poles'' needed for the double copy to work, but  that they can also give rise to potentially dangerous spurious poles, such as in the example of Section \ref{sec:nonlocal}. No such spurious poles arose in the higher-derivative generalizations of the double-copy kernel at 4- and 5-point studied in Sections \ref{sec:bootstrap}-\ref{sec:5pt}.

In this section we initiate the study of whether there can be other versions of the double-copy which are not anchored on the leading BAS model. The central property we examine is whether the rank $R_n$ determinant has zeroes in unphysical locations that could lead to spurious singularities in the double-copy amplitudes {\em unless} additional cancellations take place. 

\subsection{Modification of KLT at 3-Point}
\label{s:altKLT3pt}

In Section \ref{sec:bootstrap3pt} we classified the most general 3-point bi-adjoint scalar amplitudes 
\begin{align}
	m_3[123|123] = g+\lambda_3\,,&& m_3[123|132] = -g+\lambda_3\,,
\end{align}
where $g$ is the cubic BAS coupling and $\lambda_3$ is the coupling associated with the cubic interaction with $d^{abc}$-contractions; see \reef{fanddat3pt}. We found that rank 1 at 3-point required one of these two couplings to be zero. Let us now examine what happens at 4-point.

It follows from the 3-particle input that the 
 three independent doubly-ordered amplitudes at  4-point are
\begin{align}
\label{mdabc1}
m_4[1234|1234] =& (g+\lambda_3)^2\left(\frac1{s}+\frac1{u}\right)\,,\\
\label{mdabc2}
m_4[1234|1243] =& (-g^2+\lambda_3^2)\ \frac1{s}\,,\\
\label{mdabc3}
m_4[1234|1432] =& (-g+\lambda_3)^2\left(\frac1{s}+\frac1{u}\right)\,.
\end{align}
For generic non-zero values of $g$ and $\lambda_3$, the $6\times 6$ matrix of amplitudes \reef{mdabc1} in this model has rank 6; i.e.~it is full rank. The determinant is 
\be
	\frac{2048 g^3 \lambda_3 ^7 \left(3 g^2-\lambda_3 ^2\right)}{s^2 t^2 (s+t)^2}\,,
\ee
and it has no kinematic zeros, so no spurious poles arise at 4-point. However, it does vanish for $g=0$, $\lambda_3=0$, and $\lambda_3 = \pm \sqrt{3} g$ which means that those cases have lower ranks. We find
\begin{center}
    4-point~~~~
	\begin{tabular}{|c|c|c|}
		\hline
		Couplings & Matrix Rank & Spurious Singularities?\\
		\hline
		$g\ne 0$, $\lambda_3\ne 0$ & 6 & No\\
		\hline
		$g\ne 0$, $\lambda_3= \sqrt{3}g$ & 5 & No\\
		\hline
		$g=0$, $\lambda_3\ne 0$ & 3 & No\\
		\hline
		$g\ne 0$, $\lambda_3= 0$ & 1 & No\\
		\hline
	\end{tabular}
\end{center}
The 3rd column asks if there are spurious zeroes in the determinant at the given rank, and in each case at 4-point the answer is no. We proceed to 5-point. 

As explained in Section \ref{sec:5pt}, the 5-point amplitudes are determined by 8 functions $g_1$-$g_8$ and they determine the full $24 \times 24$ matrix of zeroth copy amplitudes $m_5$. We calculate them from the known 3-particle vertices with general $g$ and $\lambda_3$, compute the rank, and examine the associated determinants for zeros. The results are summarized as
\begin{center}
    5-point~~~~
	\begin{tabular}{|c|c|c|}
		\hline
		Couplings & Matrix Rank & Spurious Singularities?\\
		\hline
		$g\ne 0$, $\lambda_3\ne 0$ & 24 & Yes\\
		\hline
		$g\ne 0$, $\lambda_3= \sqrt{3}g$ & 21 & Yes\\
		\hline
		$g=0$, $\lambda_3\ne 0$ & 11 & Yes\\
		\hline
		$g\ne 0$, $\lambda_3= 0$ & 2 & No\\
		\hline
	\end{tabular}
\end{center}
Thus, despite first appearances, only the model with $\lambda_3=0$, i.e.~the BAS model, leads to a double-copy without spurious singularities appearing in the kernel. This is a serious potential obstacle for the double-copy since it means that additional cancellation has to take place among the terms in the KLT sum to avoid spurious poles in the double-copy 5-point amplitude. We discuss such cancellations briefly for the case with $g=0$ and $\lambda_3\ne 0$ in Section \ref{s:spurcancel}.

%%%%%%%%%%%%% 

\subsection{Modification of KLT at 4-Point}
\label{s:altKLT4pt}

Having classified the possible generalizations of the double copy at 3-point, let us now set $g=\lambda_3 = 0$ and examine the possibility for zeroth-copy models with constant 4-point interactions $\phi^4$.  

There are different ways four bi-adjoint scalars $\phi^{aa'}$ can be contracted with group-invariant tensors. This is simplest to classify using the amplitudes. We know from Section \ref{s:boot4pt} that three amplitudes determine the full $6\times6$ matrix and we simply parameterize them with three constants,
\be
	m_4[1234|1234] = \alpha_{1}\,,
	~~~~~
	m_4[1234|1243] = \alpha_{2}\,,
	~~~~~
	m_4[1234|1432] = \alpha_{3}\,.
\ee
The resulting matrix ranks at 4-point are summarized below
\begin{center}
	\begin{tabular}{|c|c|}
		\hline
		Couplings & Matrix Rank\\
		\hline
		$\alpha_{1}\ne 0$, $\alpha_{2}\ne 0$, $\alpha_{3}\ne 0$ & 6\\
		\hline
		$\alpha_{1}\ne 0$, $\alpha_{2}\ne 0$, $\alpha_{3}= -4 \alpha_{2} - \alpha_{1}$ & 5\\
		\hline
		$\alpha_{1}\ne 0$, $\alpha_{2}\ne 0$, $\alpha_{3}= 2 \alpha_{2} - \alpha_{1}$ & 4\\
		\hline
		$\alpha_{1}\ne 0$, $\alpha_{2}= 0$, $\alpha_{3}= - \alpha_{1}$ & 3\\
		\hline
		$\alpha_{1}=-2\alpha_{2}$, $\alpha_{2}\ne 0$, $\alpha_{3}= -2 \alpha_{2}$ & 2\\
		\hline
		$\alpha_{1}=\alpha_{2}=\alpha_{3}$ & 1\\
		\hline
	\end{tabular}
\end{center}
No kinematically spurious zeros arise in the determinants at 4-point since the amplitudes are  constants, so we need to go to 6-point to assess the model further. 

At 6-point we find that all cases have spurious singularities in the kernel from zeros in the determinant, {\em except} the model with $\alpha_{1}=\alpha_{2}=\alpha_{3} \equiv \lambda$. This model is\footnote{This was the modification we added to the BAS model in Section \ref{sec:nonlocal} and found in that case at 4-point that the rank 2 system had spurious poles. Note that this problem disappears at 4-point when the cubic coupling $g$ of the BAS model is set to zero, because then the rank is reduced from 2 to 1.}
\begin{align}
\label{ddL}
\mathcal{L} = -\frac{1}{2}\left(\partial_\mu \phi^{aa'}\right)^2+ \lambda\, d^{abcd}\tilde{d}^{a'b'c'd'}\phi^{aa'}\phi^{bb'}\phi^{cc'}\phi^{dd'}\,,
\end{align}
and the $120 \times 120$ matrix of its tree amplitudes has rank 10. 

An appealing feature of a kernel based on the zeroth copy \reef{ddL} is that it double-copies $F^4$ with itself to $R^4$; this is not true of the  BAS or stringy double-copy. However, by now, we have learned the lesson that higher-point calculations may change our outlook. And this is the case here too: the $5040 \times 5040$ matrix of 8-point amplitudes has rank 273 and there are spurious zeros in the $273 \times 273$ determinants. Therefore, unless there are additional cancellations at 8-point, this model does not lead to a healthy double-copy. 

\subsection{Cancellation Spurious Singularities}
\label{s:spurcancel}
In the previous two subsections, we have encountered examples with spurious poles in the kernel. While they represent a potential problem, it is worth noting that the L and R sector amplitudes must obey the KKBCJ relations defined by the kernel. Does this allow one to cancel the spurious poles in the generalized KLT sum to recover a sensible local tree amplitude as a result of the double-copy? We briefly examine this question here. 

As an example where such cancellations do happen, consider the kernel based on the zeroth copy with cubic interactions only with $g=0$ and $\lambda_3 \ne 0$. A model whose tree amplitudes solve the corresponding rank 3 generalized KKBCJ relations is
\be
  \mathcal{L}_{\text{int}} = d^{abc}\phi^a\phi^b\phi^c \,,
\ee
and it is easy to see at 5-point that the spurious poles in the kernel do get canceled. The double-copy of this theory with itself is the abelian $\phi^3$ model. 

Thus encouraged, we tried the same with a less trivial model
\be
  \mathcal{L}_{\text{int}} = d^{abc}Z^a F^b_{\mu\nu} F^{c\mu\nu} + \text{h.c.}\,,
\ee
for a complex adjoint scalar field $Z$, 
but in this case the spurious poles were not cancelled at 5-point. 

Based on these examples, we note that the issues with spurious poles appear to arise for kernels with non-minimal rank, i.e.~with rank $>(n-3)!$. One may therefore speculate that minimal rank $(n-3)!$ is a key ingredient for a physical double-copy multiplication kernel. 

\section{Discussion and Outlook}
\label{sec:discussion}

The double-copy is a very efficient method for computing gravity amplitudes that is playing an increasingly important role in modern gravity computations; for example, one of the recent exciting applications is to gravitional-wave physics 
\cite{Goldberger:2016iau,Shen:2018ebu,Cheung:2018wkq,Kosower:2018adc,Bern:2019nnu,Antonelli:2019ytb,Bern:2019crd,Bjerrum-Bohr:2020syg,Almeida:2020mrg}. 
A fully systematic understanding of the double-copy is still an open question. This is true in many contexts and of particular relevance for our work is the precise role of higher dimension operators in the double-copy.  Although a number of procedures for double-copying higher dimension operators have been developed \cite{Broedel:2012rc,Elvang:2020kuj,Low:2020ubn,Carrasco:2019yyn}, no pattern has emerged for which local counterterms can and cannot be derived from the double-copy. These problems are not simply a formality but have far-reaching implications. Infinite counterterms are necessary to regulate UV divergences and finite counterterms are important for determining the (existence of) regularization schemes that preserve certain symmetries. In gravitational-wave calculations, higher-dimension operators account for finite-size corrections \cite{Goldberger:2004jt,Bern:2019crd}.

In this paper we introduced a novel bottom-up approach to the double-copy in the KLT formulation. It was based on the KLT algebra and how it links the identity element (``zeroth copy'') to the kernel that determines the multiplication rule. We showed how this gives a KLT bootstrap formalism for the zeroth copy model whose tree amplitudes determine the kernel. At 4- and 5-point we solved the KLT bootstrap equations and found a generalized double-copy kernel based on a low-energy expansion that generalizes the $\alpha'$-expansion of the string theory KLT kernel. We applied the generalized double-copy to YM theory and $\chi$PT. Many interesting questions remain to be studied and we now discuss some of them. \\

%%%%%%
{\bf Similarity Transformations.} \\
In our examples with the generalized KLT kernel based on BAS+h.d., we found that the double-copy contains the same operators as the standard field theory double copy, but with shifts in some of their Wilson coefficients. It is tempting to ask if this shift can be encoded more systematically. To examine this, consider at 4-point performing a {\em similarity transformation} on the BAS solutions with superscripts ${}^{(0)}$:
\be
  \begin{split}
  \label{simlar}
  m_4[\beta|\alpha] 
  &= Z^\text{L}[\alpha]\,Z^\text{R}[\beta] m_4^{(0)}[\beta|\alpha]  \,,\\
  \mathcal{A}_4^\text{L}[\alpha] 
  &= Z^\text{L}[\alpha] \mathcal{A}_4^{\text{L}(0)}[\alpha]\,,\\
  \mathcal{A}_4^\text{R}[\beta] 
  &= Z^\text{R}[\beta] \mathcal{A}_4^{\text{R}(0)}[\beta]\,.
  \end{split}  
\ee
Note that the amplitudes $\mathcal{A}_4^{\text{L/R}(0)}[\alpha]$ may include whatever higher-derivative corrections are compatible with the BAS KKBCJ relations. In general, there are fewer operators in $\mathcal{A}_4^{\text{L/R}(0)}[\alpha]$ than in $\mathcal{A}_4^{\text{L/R}}[\alpha]$.

It follows from the definition \reef{simlar} that if $m_4^{(0)}[\beta|\alpha]$ solves the rank $(n-3)!$ bootstrap equations, then so does $m_4[\beta|\alpha]$ (the rank of a matrix does not change when the rows and columns of a matrix are rescaled).
Similarly, the single-copy amplitudes $\mathcal{A}_4$ solve the generalized KKBCJ relations whenever $\mathcal{A}_4^{(0)}$ is compatible with the field theory ones, as can be seen from 
\bea
\label{RBCJmatrixnot3}
   \mathds{1} \otimes \text{R} = \text{R}\!:~~~~~~~~~
    m_4[\delta|\alpha]\frac{1}{m_4[\beta|\alpha]}\mathcal{A}_n^\text{R}[\beta] = \mathcal{A}_n^\text{R}[\delta]\,,
\eea
and similarly for the L sector. 

When applied to the double-copy, we see that 
\be
  \mathcal{M}_4 
  = \mathcal{A}_n^\text{L}[\alpha]\frac{1}{m_4[\beta|\alpha]}\mathcal{A}_n^\text{R}[\beta] 
  = \mathcal{A}_n^{\text{L}(0)}[\alpha]\frac{1}{m_4^{(0)}[\beta|\alpha]}\mathcal{A}_n^{\text{R}(0)}[\beta]\,.
\ee

It may now seem plausible that all double-copies can be equivalently obtained from the field theory BAS kernel. However, one has to be more careful: 
\begin{enumerate}
\item First of all, one must ensure that the LHS of \reef{simlar} is local; no spurious poles are allowed to arise from the product with the similarity factors. 

\item The cyclicity properties of the amplitudes $m_4[\beta|\alpha]$ are ensured if $Z^\text{L/R}$ are both cyclic. 

\item It is not \emph{a priori} clear that all solutions to the 4-point KLT bootstrap equation are related to the BAS amplitude via a similarity transformation such as the one in the first line of \reef{simlar}.

\item With a cyclic choice for $Z^\text{L/R}$, one may not be able to produce all possible solutions $\mathcal{A}_n^{\text{L/R}}$ to the generalized KKBCJ relations. 

\end{enumerate}
To simultaneously enforce all these properties is non-trivial. In the perturbative context of BAS+h.d., one can solve items 1 and 2 by choosing 
$Z^\text{L/R} = 1 + s u\,  P^\text{L/R}$, where $P^\text{L/R}$ is a Mandelstam polynomial symmetric in $s$ and $u$. We can then test whether our generalized solution for $m_4$ given in Section \ref{s:pert4pt} for BAS+h.d.~can be reproduced and fix most of the constants in the ansatz $P^\text{L/R}$.\footnote{We have done this to order $O(p^6)$ as a preliminary test.} So this resolves item 3 above. The resulting similarity functions $Z^\text{L/R}$ do indeed produce the most general solution to the generalized KKBCJ equations given in Section \ref{sec:bcj4point} for the MHV YM+h.d.~amplitudes (to the orders we have checked), so that addresses item 4. This requires shifts in the YM Wilson coefficients of  $\mathcal{A}_n^{\text{L/R}(0)}$ such as the one given in \reef{shifte31}. This then explains (to the orders checked) why we found no new operators in the double-copy at 4-point and why their coefficients could be understood as shifts of the Wilson coefficients of the L and R copies. Note though that this assumes that the L and R Wilson coefficients are sufficiently generic. 

At 5-point and higher, it becomes much more challenging to overcome the potential obstructions from the constraints listed above. Another issue arises if one studies double-copies outside the regime of the low-energy expansion. Then it becomes much harder to ensure the absense of spurious poles. 
Future explorations may shed light on these questions. \\
%%%%%%

{\bf Positivity Constraints.} \\
If a similarity transformation exists for the double-copy of EFTs, they move the higher-derivative corrections from the generalized kernel into shifts of the Wilson coefficients of the L and R sector input amplitudes. It would be interesting to examine if such shifts are in conflict with Swampland positivity constraints.\footnote{The string kernel itself does not obey any obvious positivity conditions like those in \cite{Arkani-Hamed:2020blm}, which makes sense as the $\phi^{3}$ theory has no ground state and is not well-defined non-perturbatively.} Potentially relevant for this research direction is the observation that imposing the string KKBCJ relations along with the positivity constraints of \cite{Arkani-Hamed:2020blm} is enough to uniquely fix the Wilson coefficients of YM to those in Type-I superstring theory \cite{Huang:2020nqy}. Understanding the interplay of generalized KKBCJ relations, similarity transforms, and positivity restrictions on Wilson coefficients could give some bottom-up understanding of why the strings kernel takes such a particular form. \\

%%%%%%

{\bf Connection to BCJ Double-Copy} \\
While we have focused on the KLT formula, there is in the field theory ($\alpha'\rightarrow 0$) limit an alternative formulation of the double-copy, first introduced by Bern, Carrasco and Johansson (BCJ) \cite{Bern:2008qj}, based on a trivalent graphical expansion and the principle of \textit{color-kinematics duality} (see the review \cite{Bern:2019prr}). One of the primary advantages of this approach is that it has been proposed --- and tested in numerous cases --- to generalize to loop integrands, making possible otherwise prohibitively difficult high-loop-order calculations in maximal supergravity \cite{Bern:2018jmv}. Focusing on tree-level, it is natural to ask if the generalized double-copy presented in this paper has a BCJ-like formulation. 

The recent papers \cite{Carrasco:2019yyn,Carrasco:2021ptp}  incorporate higher-derivative corrections and generalized color-tensors in the BCJ double-copy. This approach makes use of generalized \textit{color weights} $\hat{c}_{s/t/u}$ (see (18) of \cite{Carrasco:2019yyn} for a precise definition) that depend on both color tensors and Mandelstam invariants in such a way that the usual \textit{adjoint-type} color identities remain true. Taking such objects, we can construct a natural zeroth-copy by making the usual BCJ replacement of kinematic numerators with color-factors; at 4-point 
\begin{equation}
    \label{BCJgenzerothcopy}
    m_4 = \frac{\hat{c}^{(L)}_s \hat{c}^{(R)}_s}{s} + \frac{\hat{c}^{(L)}_t \hat{c}^{(R)}_t}{t}+\frac{\hat{c}^{(L)}_u \hat{c}^{(R)}_u}{u}\,.
\end{equation}
Similar to the output of the KLT bootstrap described in this paper, this can be identified as a scattering amplitude of a higher-derivative corrected BAS model, though \textit{a priori} these two approaches are not obviously related. Curiously, we find that when (\ref{BCJgenzerothcopy}) is expressed as a matrix in L and R color-orderings, the result has rank 1 for all choices of parameters in the generalized color-weights defined in \cite{Carrasco:2019yyn}. Moreover, we have found that up to $O\big(p^8\big)$, the parameters in (\ref{BCJgenzerothcopy}) can always be chosen to reproduce the general solution to the KLT bootstrap equations (\ref{f2res}). In this sense (\ref{BCJgenzerothcopy}) provides a closed form \textit{solution} to the 4-point bootstrap equations. It remains unknown whether the higher-multiplicity generalized color-factors described in \cite{Carrasco:2021ptp} likewise provide a solution to the bootstrap conditions and whether a BCJ-like double-copy procedure can be devised to reproduce the results of the generalized double-copy (\ref{M1234}) presented in this paper. We leave these and related important questions to future work. \\

%%%%%%%
{\bf Exact Solutions to the Bootstrap Equations: Trunctions} \\
The field theory and string zeroth copies are exact solutions to the rank $(n-3)!$ bootstrap equations. We have found generalizations of these that solve the bootstrap equations as an order-by-order low-energy expansion corresponding to adding higher-derivative terms to the BAS model. A natural question is if there are new solutions that solve the bootstrap equations exactly? 

To address this, we take the 4-point solution \reef{f2res}-\reef{f1res} as the starting point and examine if the low-energy expansion truncates for certain choices of coefficients $a_{i,j}$, i.e.~if there are choices of a finite set of non-zero coefficients such that the rank of the 6$\times$6 matrix is exactly 1, rather than solving this constraint order by order in the low-energy expansion. Interestingly, such solutions do exist! 

For example, setting $a_{1,1}=0$ or $a_{1,1}=a_{1,0}$ (equivalently, $a_\text{L}=0$ or $a_\text{R}=0$) while taking all other $a_{i,j}$ = 0 is an exact solution to the $f_2$ condition \reef{f2cond}, moreover, they give local solutions for $f_1$: specifically for $a_\text{L}=0$, we have
\be  
  \label{trunc}
  f_1(s,t) = - \frac{g^2t }{s u}   - 4 \frac{a_\text{R}}{\Lambda^4} \,t\,, 
  ~~~~~
  f_2(s,t) =- \frac{g^2}{s} - 4 \frac{a_\text{R}}{\Lambda^4} \,u\,.
\ee
When this is used as input at 5-point, one finds that the $24 \times 24$ matrix indeed has rank 2, so the solution truncates consistently; this is true for both solutions $a_\text{L}=0$ and $a_\text{R}=0$.\footnote{Similarly, one can include the contribution with coefficient $a_{2,0}$ with no further restrictions in the 4- and 5-point bootstrap. At higher-orders, one finds that certain choices of the $a_{i,j}$'s admit finite truncations.}

When the contribution from $a_\text{R}$ in \reef{trunc} is regarded as a perturbation of the BAS model and the kernel is expanded in small $s_{ij}/\Lambda^2$, no spurious poles arise, because this is just like the general perturbative solution in Section \ref{s:pert4pt}. However, if we attempt to regard the solution \reef{trunc} as an {\em exact} solution with no expansion in $s_{ij}/\Lambda^2$, we have to beware of potential spurious poles in the kernel (e.g.~in $1/f_1(s,t)$). It is clear that $f_1$ in \reef{trunc} in addition to the zero at $t=0$ (which provides the missing $t$-channel pole in the double-copy) also has a zero that cannot be a physical pole (it is not even a massive pole). This means that amplitudes $\mathcal{A}^\text{L/R}_4$ that are double-copied with this kernel must have zeros that cancel the spurious poles; this is similar to the discussion of potential cancellation of spurious poles in Section \ref{s:spurcancel}. Something nice can indeed happen to cancel these poles. To see this, consider the L and R amplitudes for YM+h.d.~in \reef{A4result1234L} and \reef{A4result1234R}. Setting $a_\text{L} = 0$ (i.e.~$a_{1,1}=a_{1,0}$), $g_{F^3} = 0$ and all other higher-derivative contributions to zero, one finds
\be
\mathcal{A}^{\textrm{L}}_4[1^+ 2^+ 3^- 4^-]  =
[12]^2\<34\>^2
\frac{(g_{\textrm{YM}}^{\textrm{L}})^2}{s u} \,,
~~~~~
\mathcal{A}^{\textrm{R}}_4[1^+ 2^+ 3^- 4^-]  =
(g_{\textrm{YM}}^{\textrm{R}})^2 [12]^2\<34\>^2
\bigg[ \frac{1}{s u}
+ \frac{4 a_\text{R}}{g^2\Lambda^4} 
\bigg]
\ee
When these amplitudes are double-copied using \reef{M1234} with $f_1$ given by \reef{trunc}, the entire $a_\text{R}$ dependence cancels and the result is simply the pure Einstein gravity amplitude
\be  
  \mathcal{M}_4(1^+ 2^+ 3^- 4^-) = \kappa^2 \frac{[12]^4\<34\>^4}{stu} \,.
\ee
Here \reef{kappa} was used to identify $\kappa$. In a sense this is a version of the similarity transformations \reef{simlar} at work for a finite  (i.e.~non-perturbative)  modification of the zeroth and single-copy models. In particular, this example shows that for the choice of kernel given by \reef{trunc}, it possible at 4-point to double-copy YM with YM+$F^4$ to give Einstein gravity $\sqrt{-g}R$ without higher-derivative terms!

Finally, let us note that the solution \reef{trunc} can be written in the form \reef{BCJgenzerothcopy} with manifestly local generalized color-factors $\hat{c}_{s/t/u}$. The exact solutions to the KLT bootstrap equations, their relation to the BCJ-like formulation \cite{Carrasco:2019yyn,Carrasco:2021ptp}, and the issues of spurious poles deserve further investigation. \\

%%%%%%%
{\bf Exact Solutions to the Bootstrap Equations: Z-theory} \\
Any function of the form
\begin{equation}\label{mostgeneralansatz}
f_{2}^\text{ansatz}(s,t)=\frac{1}{s}\frac{G_{1}(s)G_{2}(t)}{G_{3}(s+t)} \,,
\end{equation}
for general $G_1$, $G_2$, and $G_3$ solves the 4-point KLT bootstrap equation \reef{f2cond}. Equation (\ref{mostgeneralansatz}) is not the most general ansatz to the bootstrap equations, but is curious nonetheless. For example, the string solution is of the form (\ref{mostgeneralansatz}) with $G_{2}(t)=G_{3}(s+t)=1$ and $G_{1}(s)=s/\sin(\alpha' s)$. Furthermore, there is another solution to (\ref{mostgeneralansatz}) motivated by string theory; the double partial amplitudes of non-abelian Z-theory take the form of (\ref{mostgeneralansatz}) \cite{Broedel:2013tta,Mafra:2011nw,Carrasco:2016ygv}: 
\begin{equation}\label{ztheoryintegral}
Z_{1234}[1243]=\frac{1}{s}\frac{\Gamma(1+\alpha's)\Gamma(1+\alpha'u)}{\Gamma(1+\alpha's+\alpha'u)}    \ .
\end{equation}
\noindent The double partial amplitudes of non-abelian Z-theory can be identified with the disk integrals that appear at tree-level in open string integrands and encode all non-trivial $\alpha'$-dependence for type-I open-string amplitudes at tree-level. This $\alpha'$-dependence can be extracted using generalized double copy procedures, where the double copy of Z-theory amplitudes with SYM yields type-I open string amplitudes: ``type-I=Z $\otimes$ SYM''. In addition to providing another ``stringy'' solution to the bootstrap equations, one might wonder what physical meaning the minimal rank condition could have for Z-theory amplitudes. \\

%%%%%% 
{\bf Other Directions} \\
We have formulated a general framework for generalizations of the double copy and explored it in examples. There are several future directions. For example, it would be interesting to generalize our construction to theories with massive and/or (anti-)fundamental states \cite{Johnson:2020pny,Chiodaroli:2015rdg,Naculich:2014naa,Johansson:2015oia,Brown:2018wss,Johansson:2019dnu}. The (anti-)fundamental KLT kernel could be a particularly interesting target as it retains many desirable features of the bi-adjoint kernel, such as local poles in $\mathbf{S}_{n}(B_{\textrm{L}},B_{\textrm{R}})$ to at least 8-point \cite{Naculich:2014naa,Johansson:2019dnu}, a positive geometry interpretation \cite{Arkani-Hamed:2017mur,Herderschee:2019wtl,Herderschee:2020lgb}, and even a loop generalization \cite{Carrasco:2020ywq}. More exotically, a KLT kernel for quartic kernels was recently constructed using intersection theory \cite{Kalyanapuram:2020axt}. It would also be interesting to understand our formalism in the context of the CHY formalism \cite{He:2016iqi} and homotopy algebras \cite{Borsten:2021hua}.

\section*{Acknowledgements}

We would like to thank Alan (Shih-Kuan) Chen, Lance Dixon, Sebastian Mizera, Radu Roiban, and Fei Teng for useful questions, comments, and discussions. HC, HE, and AH were supported in part by Department of Energy grant DE-SC0007859. HC, AH, CRTJ, and SP were supported in part by Leinweber Fellowships. SP was supported in part by a Barbour Scholarship from Rackham at the University of Michigan. As a special pandemic acknowledgement, HE would like to express immense gratitude to her childrens' wonderful daycare teachers: thank you Amanda, Amber, Angelique, Courtney, Darius, Derek, Karen, Kellie, Kim, Mary, and Megan.

\appendix

\section{Generalized KKBCJ Relations as Null Vectors}
\label{sec:nullvectors}

We now show that the $\text{BAS}\times \text{BAS}$ bootstrap equation and deformed BCJ relation yield that vectors orthogonal to $\mathcal{A}^{\textrm{R}}[\alpha]$ span the kernel of the column space of $m[\alpha|\beta]$, thereby proving that there are only $(n-1)!-R_{n}$ linearly independent generalized KKBCJ relations. A similar argument holds for $\mathcal{A}^{\textrm{L}}[\alpha]$, whose orthogonal vectors span the nullspace, or the kernel of the row space, of $m[\alpha|\beta]$.

We can choose the $(n-1)!$ color orderings as
\begin{equation}
    \label{all_color_order}
    \{\text{all } (n-1)! \text{ color-orderings}\} = B_R \bigcup \bar{B}_R  \,,
\end{equation}
where $B_R$ is some BCJ basis and  $\bar{B}_R$ is the complement of $B_R$ in the $(n-1)!$ color orderings. And $B_R \bigcup \bar{B}_R$ means $B_R$ occupies the first $R_n$ slots, while $\bar{B}_R$ takes the remaining $(n-1)! - R_n$ slots, of the $(n-1)!$ color-orderings.

Then, an explicit basis for these $(n-1)! - R_n$ different null vectors take the form 
\begin{eqnarray}
    \vec{n}_{i}^{\textrm{R}}
    &=&\begin{pmatrix}
        \mathbf{m}_n [\alpha_i,B_{\textrm{L}} ] \mathbf{S}_n[B_L,B_R],  
        &
        0,                                    &                                   
        \cdots,                                &                                  
        -1,                                    &                                  
        \cdots                                &                                   
    \end{pmatrix} \nonumber \\
    &\equiv&\begin{pmatrix}
        \mathbf{m}_n [\alpha_i,B_{\textrm{L}} ] \mathbf{S}_n[B_L,B_R],  
        &                                
        \mathbf{-1}_{\alpha_i}                                                                       
    \end{pmatrix},
\end{eqnarray}
\noindent where $\alpha_{i}$ corresponds to some color-ordering in $\bar{B}_R$ and the $-1$ entry is  at the corresponding position of $\alpha_i$ in Eq.(\ref{all_color_order}). So the defined vector $\mathbf{-1}_{\alpha_i}$ takes value $-1$ at the corresponding position of $\alpha_i$ and 0 elsewhere. And different choices of the dummy BCJ basis $B_L$ will give the same null vectors.

Taking the product of $\vec{n}_{i}^{\textrm{R}}$ with $\mathbf{m}_{n}$ and $\mathcal{A}^{\textrm{R}}$, one finds

\begin{eqnarray}
    \label{product_zero}
    &&\vec{n}_{i}^{\textrm{R}} \cdot \mathbf{m}_n\left[B_R \bigcup \bar{B}_R, B_L' \bigcup \bar{B}_L' \right] \nonumber \\
    &=& \begin{pmatrix}
        \mathbf{m}_n [\alpha_i,B_{\textrm{L}} ] \mathbf{S}_n[B_L,B_R],  
        &                                
        \mathbf{-1}_{\alpha_i}                                                                       
    \end{pmatrix} 
    \cdot 
    \left(
    \begin{array}{c|c}
            \mathbf{m}_n[B_R, B_L']    & \mathbf{m}_{n}[B_R, \bar{B}_L']  \\
            \hline
            \mathbf{m}_n[\bar{B}_R, B_L'] & \mathbf{m}_n[\bar{B}_R, \bar{B}_L']
        \end{array}
    \right)  \nonumber \\
    &=& 
    \begin{pmatrix}
        \mathbf{m}_n [\alpha_i,B_{\textrm{L}} ] \mathbf{S}_n[B_L,B_R] \mathbf{m}_n[B_R, B_L'] - \mathbf{m}_n[\alpha_i, B_L'],  
        &                                
        \mathbf{m}_n [\alpha_i,B_{\textrm{L}} ] \mathbf{S}_n[B_L,B_R]  \mathbf{m}_{n}[B_R, \bar{B}_L'] - \mathbf{m}_n[\alpha_i, \bar{B}_L']                                                                     
    \end{pmatrix} \nonumber \\
    &=& \vec{\mathbf{0}}, \nonumber \\
    && \nonumber \\
    &&\vec{n}_{i}^{\textrm{R}} \cdot \mathbf{A}^{\textrm{R}}\left[B_R \bigcup \bar{B}_R \right] \nonumber \\
    &=& \begin{pmatrix}
        \mathbf{m}_n [\alpha_i,B_{\textrm{L}} ] \mathbf{S}_n[B_L,B_R],  
        &                                
        \mathbf{-1}_{\alpha_i}                                                                       
    \end{pmatrix} 
    \cdot 
    \begin{pmatrix}
        \mathbf{A}^{\textrm{R}} [B_R ],  
        &                                
        \mathbf{A}^{\textrm{R}} [\bar{B}_R ]                                                            
    \end{pmatrix} \nonumber \\
    &=& \mathbf{m}_n [\alpha_i,B_{\textrm{L}} ] \mathbf{S}_n[B_L,B_R] \mathbf{A}^{\textrm{R}} [B_R ] - \mathbf{A}^{\textrm{R}} [\alpha_i ] \nonumber \\
    &=& 0,
\end{eqnarray}

where we used the bootstrap equation and the deformed KKBCJ relations, and $B_L'$ is some BCJ basis and $\bar{B}_L'$ is its complement in the $(n-1)!$ color orderings. A different choice of $B_L'$ corresponds to a trivial rearrangement of column vectors of $\mathbf{m}_n\left[B_R \bigcup \bar{B}_R, B_L' \bigcup \bar{B}_L' \right]$. Eq.(\ref{product_zero}) explicitly shows that $\vec{n}_{i}^{\textrm{R}}$ is orthogonal to column vectors of the $(n-1)! \times (n-1)!$ $\mathbf{m}_n$ matrix, and to $\mathbf{A}^{\textrm{R}}$ as well.

\noindent The $\vec{n}_{i}^{\textrm{R}}$'s are manifestly linearly independent due to the different locations of -1 entry in each vector, so the $\vec{n}_{i}^{\textrm{R}}$'s correspond to a complete basis for the kernel of the column space of $\mathbf{m}_n$, and also the space of the generalized KKBCJ relations of $\mathbf{A}^{\textrm{R}}$. Since there are only $(n-1)!-R_{n}$ linearly independent null vectors of $\mathbf{m}_{n}$, there are olny $(n-1)!-R_{n}$ linearly independent generalized KKBCJ relations of $\mathbf{A}^{\textrm{R}}$.\footnote{If we have reversal identity for the BAS theory, $m_n[\alpha | \beta^T] = (-1)^{x_1} m_n[\alpha | \beta] $ and $m_n[\alpha^T | \beta] = (-1)^{x_2} m_n[\alpha | \beta]$, for signs determined by possibly $n$-dependent integers $x_1$ and $x_2$, we also have corresponding reversal identity for the L and R sector of the single copy theory.}

\section{Pions, Special Galileons and Born-Infeld Photons}
\label{app:NLSM}

In Section \ref{sec:HDYM} we discuss in detail which higher-derivative Yang-Mills operators that can be double-copied and what corrections they map to in the resulting theory of gravity. This is of course not the only effective field theory that can be double-copied via the formalism introduced in Section \ref{sec:introklt}. A far simpler theory is a non-linear sigma model of pions, also known as chiral perturbation theory, whose 4-point amplitude at leading order is
\begin{align}
\label{eq:NLSM4pt}
	\mathcal{A}^0_4[1234]=\frac{t}{f_\pi^2},
\end{align}
where $f_\pi$ is the pion decay constant. 

To find double-copy-compatible corrections, we start with an ansatz\footnote{While it is \emph{a priori} possible that the amplitude begins with a constant contribution at order $\Lambda^0$, such a term is forbidden by the leading order BCJ relations.},
\begin{align}
\label{eq:NLSMansatz}
\mathcal{A}_4[1234]=&\frac{t}{f_\pi^2}+\frac{b_{1,1} s^2+b_{1,2} s t+b_{1,3} t^2}{\Lambda ^4}+\frac{b_{2,1} s^3+b_{2,2} s^2 t+b_{2,3} s t^2+b_{2,4} t^3}{\Lambda ^6}\nonumber\\
&+\frac{b_{3,1} s^4+b_{3,2} s^3t+b_{3,3} s^2 t^2+b_{3,4} s t^3+b_{3,5} t^4}{\Lambda ^8}+O\left(\frac{1}{\Lambda^{10}}\right)\,,
\end{align}
with appropriate superscripts on the parameters $b_{i,j}$ for the left and right KKBCJ-compatible amplitudes. 

Since we are dealing with a scalar theory, every other color-ordering is a simple relabeling of \eqref{eq:NLSMansatz}. As a result the (right) KKBCJ relations of Section \ref{sec:bootstrap} can be rewritten as consistency conditions on the (right) amplitude such as,
\begin{align}
	\mathcal{A}_4[1234]&=m[1234|1432]\,m[1243|1432]^{-1} \mathcal{A}_4[1243]\,,
\end{align}
where $\mathcal{A}_4[1243]$ is given by a $3\leftrightarrow 4$ relabelling of $\mathcal{A}_4[1234]$. 

Solving such consistency conditions using $m[\alpha|\beta]$ given in Section \ref{sec:bootstrap}, gives us the final NLSM amplitudes,
\begin{align}
\label{eq:NLSMfinal4}
	\mathcal{A}^{\textrm{R}}_4[1234]=&\frac{t}{\left(f^{\textrm{R}}_\pi\right)^2}+\frac{a_{1,1} s t u}{\left(f^{\textrm{R}}_\pi\right)^2 g^2 \Lambda^4}+\frac{b^{\textrm{R}}_{2,4}\,t(s^2+t^2+u^2)}{2\Lambda ^6}-\frac{b^{\textrm{R}}_{3,3}\, st^2u}{\Lambda ^{8}}+O\left(\frac{1}{\Lambda^{10}}\right)\,,\\
	\mathcal{A}^{\textrm{L}}_4[1234]=&\frac{t}{\left(f^{\textrm{L}}_\pi\right)^2}+\frac{(a_{1,1}-a_{1,0}) s t u}{\left(f^{\textrm{L}}_\pi\right)^2 g^2 \Lambda^4}+\frac{b^{\textrm{L}}_{2,4}\,t(s^2+t^2+u^2)}{2\Lambda ^6}-\frac{b^{\textrm{L}}_{3,3}\, st^2u}{\Lambda ^{8}}+O\left(\frac{1}{\Lambda^{10}}\right)\,.
\end{align}
Note that all corrections that are quadratic in Mandelstam variables are disallowed. This is important in Born-Infeld theory as we discuss below.

To construct Born-Infeld amplitudes in all helicity sectors, we will also need the self-dual, next-to-self-dual and MHV Yang-Mills amplitudes that are compatible with the KLT kernel developed in Section \ref{sec:bootstrap}. These are given in \eqref{ARSD}, \eqref{ARNSD} and \eqref{A4result1234R}.

We now construct higher derivative corrections to the special Galileon and Born-Infeld amplitudes,
\begingroup
\allowdisplaybreaks
\begin{align}
	\label{eq:sGal4}
	\mathcal{M}^\text{sGal}_4(1234)=&-\frac{stu}{\left(f^{\textrm{L}}_\pi\right)^2 \left(f^{\textrm{R}}_\pi\right)^2\,g}-\frac{(b^{\textrm{L}}_{2,4}\left(f^{\textrm{L}}_\pi\right)^2+b^{\textrm{R}}_{2,4}\left(f^{\textrm{R}}_\pi\right)^2)s t u (s^2+ t^2 + u^2)}{2\left(f^{\textrm{L}}_\pi\right)^2 \left(f^{\textrm{R}}_\pi\right)^2\,g^2\,\Lambda^{6}}\nonumber\\
	&+\frac{((b^{\textrm{L}}_{3,3}\left(f^{\textrm{L}}_\pi\right)^2+b^{\textrm{R}}_{3,3}\left(f^{\textrm{R}}_\pi\right)^2)g^2+a_{2,0}\Lambda^4)\, s^2 t^2 u^2}{\left(f^{\textrm{L}}_\pi\right)^2 \left(f^{\textrm{R}}_\pi\right)^2\, g^4\,\Lambda^{8}}+O\left(\frac{1}{\Lambda^{10}}\right)\,,\\
	\label{eq:BIMHV4}
	\mathcal{M}^\text{BI}_4(1^+2^+3^-4^-)=&[12]^2\langle 34\rangle^2\left[-\frac{\left(g^{\textrm{R}}_{\text{YM}}\right)^2}{\left(f^{\textrm{L}}_\pi\right)^2 g^2}+\frac{\left(g^\textrm{R}_{F^3}\right)^2 t u}{\left(f_\pi^\textrm{L}\right)^2g^2\Lambda^4}-\frac{\left(g^\textrm{R}_\text{YM}\right)^2 b^\textrm{L}_{2,4} \left(s^2 +t^2+ u^2\right)}{2g^2\Lambda^{6}} \right.\nonumber\\
	&\left.-\frac{\left(e^R_{3,1}\,g^2-a_{2,0}\,\left(g^\textrm{R}_\text{YM}\right)^2\right) stu}{\left(f^\textrm{L}_\pi\right)^2 g^4\Lambda^{6}}+O\left(\frac{1}{\Lambda^{8}}\right)\right]\,,\\
	\label{eq:BINSD4}
	\mathcal{M}^\text{BI}_4(1^+2^+3^+4^-)=&[12]^2[3|1|4\rangle^2\left[-\frac{ g^\text{R}_{F^3}g^\text{R}_\text{YM} }{\left(f^\textrm{L}_\pi\right)^2 g^2 \Lambda^2}-\frac{g^\text{R}_{F^3}\,g^\textrm{R}_\text{YM}\, b^\textrm{L}_{2,4} \left(s^2 +t^2+ u^2\right)}{2g^2\Lambda^{8}}\nonumber\right.\\
	&\left.-\frac{(c^R_{1}g^2-a_{2,0}g^\text{R}_{F^3} g^\text{R}_\text{YM})stu}{\left(f^\textrm{L}_\pi\right)^2\,g^4\,\Lambda^8}
	+O\left(\frac{1}{\Lambda^{10}}\right)\right]\,,\\
	\label{eq:BISD4}
	\mathcal{M}^\text{BI}_4(1^+2^+3^+4^+)=&\frac{s^2 t u^2}{\ang{12}\ang{23}\ang{34}\ang{41}}\left[-\frac{2 g^\text{R}_{F^3}g^\text{R}_\text{YM} }{\left(f^\textrm{L}_\pi\right)^2 g^2 \Lambda^2}+\frac{\left(d^\text{R}_{1} g^2 - 4 a_{1,1} g^\text{R}_{F^3}g^\text{R}_\text{YM}\right) (s^2 + t^2+ u^2)}{4\left(f^\textrm{L}_\pi\right)^2 g^4 \Lambda^6}\right]\nonumber\\
	&+
	O\left(\frac{1}{\Lambda^{8}}\right)\,.
\end{align}
\endgroup

One of the applications of this analysis is determining whether or not duality symmetry is anomalous. Born-Infeld theory is known to have an electromagnetic duality at tree-level which manifests on the scattering amplitudes as an optical helicity conservation rule. In order for the symmetry to be non-anomalous at 1-loop, duality-violating amplitudes should be removable by the addition of local counterterms for e.g. a $\Lambda^{-6}$ counterterm at 4-point. This was found to be true \cite{Elvang:2019twd}, though the 1-loop regularization scheme was incompatible with the double-copy, i.e. the local counterterms necessary to restore the symmetry were not produced by the double-copy \cite{Elvang:2020kuj}. It is interesting that the more nuanced approach to higher-derivative corrections in the double-copy explored in this paper results in \eqref{eq:BINSD4} and \eqref{eq:BISD4} which also lack $\Lambda^{-6}$ corrections, in keeping with the previous analysis.

At 5-point, higher-derivative corrections added to the kernel do not change a previously noted \cite{Elvang:2018dco} result that the first BCJ-compatible correction to $\chi$PT occurs at 14-derivative order. In particular, this means that the WZW term is
incompatible with the generalized KKBCJ constraints. 

\section{Analytic Expressions for the 5-Point Bootstrap}
\label{app:5pt}
We arrange the $24 \times 24$ matrix $m[a|b]$ according to the following ordering of the permutations of momenta $2,3,4,5$:
\be
  \label{24basisperms}
  \begin{array}{llllllll}
 &\{12345, 12354, 
 12435, 12453, 
 12534, 12543, 
 13245, 13254, \\
 &\phantom{\{}13425, 13452, 
 13524, 13542, 
 14235, 14253, 
 14325, 14352,\\
 &\phantom{\{}14523, 14532, 
 15234, 15243, 
 15324, 15342, 
 15423, 15432\}\,.
  \end{array}
\ee
As described in Section \ref{s:boot5pt}, the 5-point bootstrap equations can be solved for $g_4$-$g_8$ in terms of $g_1$, $g_3$, and $g_3$. The following are the results, we indicate which minors are set to set to zero to obtain each relation using the notation that 
\be
  \text{Minor}_3[r_1,r_ 2,r_3 ; c_1, c_2, c_3],
\ee
denotes the $3 \times 3$ minor with rows $r_1$, $r_ 2$, $r_3$ and columns $c_1$, $c_ 2$, $c_3$ with labels refering to the basis \reef{24basisperms}. We use cyclic symmetry of $g_1$ to simplify the results as well as momentum relabeling.

From $\text{Minor}_3[1,2,3; 1,2,4]$ we get 
\be
\begin{split}
	\label{eq:g4bootstrap}
	g_4[12345] =& 
	\Big(g_1[12345] g_1[12354] g_2[12534] 
	- g_2[12345] g_2[12354] g_2[12534] 
	\\
	&~~~~-g_1[12345] g_2[41253]g_3[12354] + 
	g_2[12345] g_3[12345] g_3[12534]\Big)
	\\&
	\Big(g_1[12354] g_3[12534]-g_2[12354] g_2[41253] 
	\Big)^{-1}.
\end{split}
\ee
From $\text{Minor}_3[1,2,7; 1,2,3]$ we get 
\be
\begin{split}
\label{eq:g5bootstrap}
	g_5[12345] =& 
	\Big(g_1[13245] g_1[13254] g_3[51234] 
	- g_1[13254] g_2[45123] g_2[51324] 
	\\
	&~~~~-g_2[13245] g_2[13254]g_3[51234] + 
	g_2[13245] g_2[45123] g_3[25413]\Big)
	\\&
	\Big(g_1[13245] g_3[25413]-g_2[13254] g_2[51324] 
	\Big)^{-1}.
\end{split}
\ee
From $\text{Minor}_3[1,2,3; 1,2,11]$ we get 
\be
\begin{split}
	g_6[12345] =& 
	\Big(g_1[12345] g_1[12354] g_3[35124] 
	- g_1[12345] g_3[12435] g_3[41235] 
	\\
	&~~~~-g_2[12345] g_2[12354]g_3[35124] + 
	g_2[12345] g_2[51243] g_3[41235]\Big)
	\\&
	\Big(g_1[12354] g_2[51243]-g_2[12354] g_3[12435] 
	\Big)^{-1}.
\end{split}
\ee
From $\text{Minor}_3[1,2,3; 1,2,14]$ we get 
\begingroup
\allowdisplaybreaks
\begin{align}
	g_7[12345] =& 
	\Big(g_1[12345] g_1[12354] g_5[12435] 
	- g_1[12345] g_3[12435] g_5[23541] 
	\nonumber \\
	&~~~~-g_2[12345] g_2[12354]g_5[12435] + 
	g_2[12345] g_2[51243] g_5[23541]\Big)
	\\&
	\Big(g_1[12354] g_2[51243]-g_2[12354] g_3[12435] 
	\Big)^{-1}. \nonumber
\end{align}
\endgroup
where $g_5$ is as given in \eqref{eq:g5bootstrap}.
From $\text{Minor}_3[1,2,3; 1,2,24]$ we get 
\be
\begin{split}
	g_8[12345] =& 
	\Big(g_1[12345] g_1[12354] g_4[43512] 
	- g_1[12345] g_3[12435] g_4[54123] 
	\\
	&~~~~-g_2[12345] g_2[12354]g_4[43512] + 
	g_2[12345] g_2[51243] g_4[54123]\Big)
	\\&
	\Big(g_1[12354] g_2[51243]-g_2[12354] g_3[12435] 
	\Big)^{-1}.
\end{split}
\ee
where $g_4$ is as given in \eqref{eq:g4bootstrap}.

The perturbative solutions for the $g$ functions are
\begingroup
\allowdisplaybreaks
\begin{align}
g_1[12345]=&g^3 \bigg( \frac{1}{s_{12}s_{34}} + \frac{1}{s_{23}s_{45}}+\frac{1}{s_{34}s_{51}} +\frac{1}{s_{45}s_{12}} +\frac{1}{s_{51}s_{23}}\bigg)\nonumber\\
&+ \frac{g}{\Lambda^{4}} (a_{1, 0} - 2 a_{1, 1}) \bigg(
\frac{s_{35}}{s_{12}} + \frac{s_{41}}{s_{23}} 
+ \frac{s_{13}}{s_{45}}+ \frac{s_{24}}{s_{51}} 
+ \frac{s_{52}}{s_{34}}+1\bigg)\nonumber\\
&- \frac{g a_{2, 0} }{\Lambda^6} 
\bigg( \frac{s_{35}^2}{s_{12}} + \frac{s_{41}^2}{s_{23}} 
+\frac{s_{13}^2}{s_{45}} + \frac{s_{24}^2}{s_{51}} 
+ \frac{s_{52}^2}{s_{34}}\bigg) +O\left(\frac{1}{\Lambda^5}\right),\\
g_2[12345]=&g^3 \bigg(-\frac{1}{s_{12}s_{45}} - \frac{1}{s_{23}s_{45}}\bigg)\nonumber\\
&+ \frac{g}{\Lambda^{4}} \bigg(
(2 a_{1, 1}-a_{1, 0}) \frac{s_{13}}{s_{45}} +a_{1,0} \left(\frac{s_{35}}{s_{12}} 
+ \frac{s_{14}}{s_{23}}\right)+a_{1,1} \left(\frac{s_{45}}{s_{12}} 
+ \frac{s_{45}}{s_{23}}-2\right)\bigg)\nonumber\\
&+\frac{g a_{2, 0} }{\Lambda^6} 
\bigg( \frac{s_{13}^2}{s_{45}} - \frac{s_{14}s_{15}}{s_{23}} 
-\frac{s_{34}s_{35}}{s_{12}} -2s_{13}\bigg) +O\left(\frac{1}{\Lambda^5}\right),\\
g_3[12345]&=g^3 \bigg(-\frac{1}{s_{12}s_{45}}\bigg)\nonumber\\
&+ \frac{g}{\Lambda^{4}} \bigg(a_{1,0} \left(\frac{s_{13}}{s_{45}} 
+ \frac{s_{35}}{s_{12}}-1\right)+a_{1,1} \left(\frac{s_{45}}{s_{12}} 
+ \frac{s_{12}}{s_{45}}\right)\bigg)\nonumber\\
&+\frac{g a_{2, 0} }{\Lambda^6} 
\bigg( -\frac{s_{13}s_{23}}{s_{45}} - \frac{s_{34}s_{35}}{s_{12}} +s_{14}+s_{25}\bigg) +O\left(\frac{1}{\Lambda^5}\right),\\
g_4[12345]&=g^3 \bigg(\frac{1}{s_{12}s_{45}}+\frac{1}{s_{12}s_{34}}\bigg)\nonumber\\
&+ \frac{g}{\Lambda^{4}} \bigg((a_{1, 0}-2 a_{1, 1}) \frac{s_{35}}{s_{12}} +a_{1,0} \left( 
- \frac{s_{13}}{s_{45}}- \frac{s_{25}}{s_{34}}\right)+a_{1,1} \left(
-\frac{s_{12}}{s_{45}}-\frac{s_{12}}{s_{34}}+2\right)\bigg)\nonumber\\
&+\frac{g a_{2, 0} }{\Lambda^6} 
\bigg(\frac{s_{15}s_{25}}{s_{34}} + \frac{s_{13}s_{23}}{s_{45}}-\frac{s_{35}^2}{s_{12}} +2s_{35}\bigg) +O\left(\frac{1}{\Lambda^5}\right),\\
g_5[12345]&=g^3 \bigg(\frac{1}{s_{23}s_{45}}\bigg)\nonumber\\
&+ \frac{g}{\Lambda^{4}} \bigg(a_{1,0} \left( 
1- \frac{s_{14}}{s_{23}}- \frac{s_{13}}{s_{45}}\right)+a_{1,1} \left(
-\frac{s_{23}}{s_{45}}-\frac{s_{45}}{s_{23}}\right)\bigg)\nonumber\\
&+\frac{g a_{2, 0} }{\Lambda^6} 
\bigg(\frac{s_{14}s_{45}}{s_{23}} - \frac{s_{13}s_{12}}{s_{45}}+\frac{s_{14}^2}{s_{23}} +s_{12}+2s_{24}\bigg) +O\left(\frac{1}{\Lambda^5}\right),\\
g_6[12345]&=\frac{g}{\Lambda^{4}} a_{1,0}+\frac{g a_{2, 0} }{\Lambda^6} \left(s_{13}+2s_{14}-s_{23}+s_{24}\right)+ O\left(\frac{1}{\Lambda^5}\right),\\
g_7[12345]&=-\frac{g}{\Lambda^{4}} a_{1,0}-\frac{g a_{2, 0} }{\Lambda^6} \left(s_{13}+2s_{14}-s_{23}+s_{24}\right)+ O\left(\frac{1}{\Lambda^5}\right),\\
g_8[12345]&=g^3 \bigg( \frac{1}{s_{12}s_{34}} + \frac{1}{s_{23}s_{45}}+\frac{1}{s_{34}s_{51}} +\frac{1}{s_{45}s_{12}} +\frac{1}{s_{51}s_{23}}\bigg)\nonumber\\
&+ \frac{g}{\Lambda^{4}} (a_{1, 0} - 2 a_{1, 1}) \bigg(
\frac{s_{35}}{s_{12}} + \frac{s_{41}}{s_{23}} 
+ \frac{s_{13}}{s_{45}}+ \frac{s_{24}}{s_{51}} 
+ \frac{s_{52}}{s_{34}}+1\bigg)\nonumber\\
&- \frac{g a_{2, 0} }{\Lambda^6} 
\bigg( \frac{s_{35}^2}{s_{12}} + \frac{s_{41}^2}{s_{23}} 
+\frac{s_{13}^2}{s_{45}} + \frac{s_{24}^2}{s_{51}} 
+ \frac{s_{52}^2}{s_{34}}\bigg) +O\left(\frac{1}{\Lambda^5}\right).
\end{align}
\endgroup

%%%%%%%%%%%%%%%%%%%%%%%%%%%%
\bibliographystyle{JHEP}
\bibliography{Draft.bib}

\providecommand{\href}[2]{#2}\begingroup\raggedright\begin{thebibliography}{10}

\bibitem{Kawai:1985xq}
H.~Kawai, D.~Lewellen and S.~Tye, \emph{{A Relation Between Tree Amplitudes of
  Closed and Open Strings}},
  \href{https://doi.org/10.1016/0550-3213(86)90362-7}{\emph{Nucl. Phys. B}
  {\bfseries 269} (1986) 1}.

\bibitem{Bern:2019prr}
Z.~Bern, J.~J. Carrasco, M.~Chiodaroli, H.~Johansson and R.~Roiban, \emph{{The
  Duality Between Color and Kinematics and its Applications}},
  \href{https://arxiv.org/abs/1909.01358}{{\ttfamily 1909.01358}}.

\bibitem{Monteiro:2014cda}
R.~Monteiro, D.~O'Connell and C.~D. White, \emph{{Black holes and the double
  copy}}, \href{https://doi.org/10.1007/JHEP12(2014)056}{\emph{JHEP} {\bfseries
  12} (2014) 056} [\href{https://arxiv.org/abs/1410.0239}{{\ttfamily
  1410.0239}}].

\bibitem{Kleiss:1988ne}
R.~Kleiss and H.~Kuijf, \emph{{Multi - Gluon Cross-sections and Five Jet
  Production at Hadron Colliders}},
  \href{https://doi.org/10.1016/0550-3213(89)90574-9}{\emph{Nucl. Phys. B}
  {\bfseries 312} (1989) 616}.

\bibitem{Bern:2008qj}
Z.~Bern, J.~Carrasco and H.~Johansson, \emph{{New Relations for Gauge-Theory
  Amplitudes}}, \href{https://doi.org/10.1103/PhysRevD.78.085011}{\emph{Phys.
  Rev. D} {\bfseries 78} (2008) 085011}
  [\href{https://arxiv.org/abs/0805.3993}{{\ttfamily 0805.3993}}].

\bibitem{Mizera:2016jhj}
S.~Mizera, \emph{{Inverse of the String Theory KLT Kernel}},
  \href{https://doi.org/10.1007/JHEP06(2017)084}{\emph{JHEP} {\bfseries 06}
  (2017) 084} [\href{https://arxiv.org/abs/1610.04230}{{\ttfamily
  1610.04230}}].

\bibitem{Tseytlin:1997csa}
A.~A. Tseytlin, \emph{{On nonAbelian generalization of Born-Infeld action in
  string theory}},
  \href{https://doi.org/10.1016/S0550-3213(97)00354-4}{\emph{Nucl. Phys. B}
  {\bfseries 501} (1997) 41}
  [\href{https://arxiv.org/abs/hep-th/9701125}{{\ttfamily hep-th/9701125}}].

\bibitem{Broedel:2012rc}
J.~Broedel and L.~J. Dixon, \emph{{Color-kinematics duality and double-copy
  construction for amplitudes from higher-dimension operators}},
  \href{https://doi.org/10.1007/JHEP10(2012)091}{\emph{JHEP} {\bfseries 10}
  (2012) 091} [\href{https://arxiv.org/abs/1208.0876}{{\ttfamily 1208.0876}}].

\bibitem{Carrasco:2019yyn}
J.~J.~M. Carrasco, L.~Rodina, Z.~Yin and S.~Zekioglu, \emph{{Simple encoding of
  higher derivative gauge and gravity counterterms}},
  \href{https://doi.org/10.1103/PhysRevLett.125.251602}{\emph{Phys. Rev. Lett.}
  {\bfseries 125} (2020) 251602}
  [\href{https://arxiv.org/abs/1910.12850}{{\ttfamily 1910.12850}}].

\bibitem{Carrasco:2021ptp}
J.~J.~M. Carrasco, L.~Rodina and S.~Zekioglu, \emph{{Composing Effective
  Prediction at Five Points}},
  \href{https://arxiv.org/abs/2104.08370}{{\ttfamily 2104.08370}}.

\bibitem{Cachazo:2013iea}
F.~Cachazo, S.~He and E.~Y. Yuan, \emph{{Scattering of Massless Particles:
  Scalars, Gluons and Gravitons}},
  \href{https://doi.org/10.1007/JHEP07(2014)033}{\emph{JHEP} {\bfseries 07}
  (2014) 033} [\href{https://arxiv.org/abs/1309.0885}{{\ttfamily 1309.0885}}].

\bibitem{BjerrumBohr:2009rd}
N.~Bjerrum-Bohr, P.~H. Damgaard and P.~Vanhove, \emph{{Minimal Basis for Gauge
  Theory Amplitudes}},
  \href{https://doi.org/10.1103/PhysRevLett.103.161602}{\emph{Phys. Rev. Lett.}
  {\bfseries 103} (2009) 161602}
  [\href{https://arxiv.org/abs/0907.1425}{{\ttfamily 0907.1425}}].

\bibitem{Plahte:1970wy}
E.~Plahte, \emph{{Symmetry properties of dual tree-graph n-point amplitudes}},
  \href{https://doi.org/10.1007/BF02824716}{\emph{Nuovo Cim. A} {\bfseries 66}
  (1970) 713}.

\bibitem{Feng:2010my}
B.~Feng, R.~Huang and Y.~Jia, \emph{{Gauge Amplitude Identities by On-shell
  Recursion Relation in S-matrix Program}},
  \href{https://doi.org/10.1016/j.physletb.2010.11.011}{\emph{Phys. Lett. B}
  {\bfseries 695} (2011) 350}
  [\href{https://arxiv.org/abs/1004.3417}{{\ttfamily 1004.3417}}].

\bibitem{Stieberger:2009hq}
S.~Stieberger, \emph{{Open \textbackslash{}\& Closed vs. Pure Open String Disk
  Amplitudes}},  \href{https://arxiv.org/abs/0907.2211}{{\ttfamily 0907.2211}}.

\bibitem{BjerrumBohr:2010zs}
N.~E.~J. Bjerrum-Bohr, P.~H. Damgaard, T.~Sondergaard and P.~Vanhove,
  \emph{{Monodromy and Jacobi-like Relations for Color-Ordered Amplitudes}},
  \href{https://doi.org/10.1007/JHEP06(2010)003}{\emph{JHEP} {\bfseries 06}
  (2010) 003} [\href{https://arxiv.org/abs/1003.2403}{{\ttfamily 1003.2403}}].

\bibitem{Boels:2014dka}
R.~H. Boels and T.~Hansen, \emph{{String theory in target space}},
  \href{https://doi.org/10.1007/JHEP06(2014)054}{\emph{JHEP} {\bfseries 06}
  (2014) 054} [\href{https://arxiv.org/abs/1402.6356}{{\ttfamily 1402.6356}}].

\bibitem{Elvang:2018dco}
H.~Elvang, M.~Hadjiantonis, C.~R. Jones and S.~Paranjape, \emph{{Soft Bootstrap
  and Supersymmetry}},
  \href{https://doi.org/10.1007/JHEP01(2019)195}{\emph{JHEP} {\bfseries 01}
  (2019) 195} [\href{https://arxiv.org/abs/1806.06079}{{\ttfamily
  1806.06079}}].

\bibitem{Elvang:2020kuj}
H.~Elvang, M.~Hadjiantonis, C.~R.~T. Jones and S.~Paranjape,
  \emph{{Electromagnetic Duality and D3-Brane Scattering Amplitudes Beyond
  Leading Order}}, \href{https://doi.org/10.1007/JHEP04(2021)173}{\emph{JHEP}
  {\bfseries 04} (2021) 173}
  [\href{https://arxiv.org/abs/2006.08928}{{\ttfamily 2006.08928}}].

\bibitem{Bern:2017tuc}
Z.~Bern, A.~Edison, D.~Kosower and J.~Parra-Martinez, \emph{{Curvature-squared
  multiplets, evanescent effects, and the U(1) anomaly in $N=4$ supergravity}},
  \href{https://doi.org/10.1103/PhysRevD.96.066004}{\emph{Phys. Rev. D}
  {\bfseries 96} (2017) 066004}
  [\href{https://arxiv.org/abs/1706.01486}{{\ttfamily 1706.01486}}].

\bibitem{Bern:2017rjw}
Z.~Bern, J.~Parra-Martinez and R.~Roiban, \emph{{Canceling the U(1) Anomaly in
  the $S$ Matrix of $N$=4 Supergravity}},
  \href{https://doi.org/10.1103/PhysRevLett.121.101604}{\emph{Phys. Rev. Lett.}
  {\bfseries 121} (2018) 101604}
  [\href{https://arxiv.org/abs/1712.03928}{{\ttfamily 1712.03928}}].

\bibitem{Carrillo-Gonzalez:2019aao}
M.~Carrillo~Gonz\'alez, R.~Penco and M.~Trodden, \emph{{Shift symmetries, soft
  limits, and the double copy beyond leading order}},
  \href{https://doi.org/10.1103/PhysRevD.102.105011}{\emph{Phys. Rev. D}
  {\bfseries 102} (2020) 105011}
  [\href{https://arxiv.org/abs/1908.07531}{{\ttfamily 1908.07531}}].

\bibitem{Gomis:2021hxa}
J.~Gomis, Z.~Yan and M.~Yu, \emph{{KLT factorization of winding string
  amplitudes}}, \href{https://doi.org/10.1007/JHEP06(2021)057}{\emph{JHEP}
  {\bfseries 06} (2021) 057}
  [\href{https://arxiv.org/abs/2103.05013}{{\ttfamily 2103.05013}}].

\bibitem{Guillen:2021mwp}
M.~Guillen, H.~Johansson, R.~L. Jusinskas and O.~Schlotterer, \emph{{Scattering
  Massive String Resonances through Field-Theory Methods}},
  \href{https://arxiv.org/abs/2104.03314}{{\ttfamily 2104.03314}}.

\bibitem{Chiodaroli:2015rdg}
M.~Chiodaroli, M.~Gunaydin, H.~Johansson and R.~Roiban, \emph{{Spontaneously
  Broken Yang-Mills-Einstein Supergravities as Double Copies}},
  \href{https://doi.org/10.1007/JHEP06(2017)064}{\emph{JHEP} {\bfseries 06}
  (2017) 064} [\href{https://arxiv.org/abs/1511.01740}{{\ttfamily
  1511.01740}}].

\bibitem{Chiodaroli:2017ehv}
M.~Chiodaroli, M.~Gunaydin, H.~Johansson and R.~Roiban, \emph{{Gauged
  Supergravities and Spontaneous Supersymmetry Breaking from the Double Copy
  Construction}},
  \href{https://doi.org/10.1103/PhysRevLett.120.171601}{\emph{Phys. Rev. Lett.}
  {\bfseries 120} (2018) 171601}
  [\href{https://arxiv.org/abs/1710.08796}{{\ttfamily 1710.08796}}].

\bibitem{Chiodaroli:2018dbu}
M.~Chiodaroli, M.~Günaydin, H.~Johansson and R.~Roiban, \emph{{Non-Abelian
  gauged supergravities as double copies}},
  \href{https://doi.org/10.1007/JHEP06(2019)099}{\emph{JHEP} {\bfseries 06}
  (2019) 099} [\href{https://arxiv.org/abs/1812.10434}{{\ttfamily
  1812.10434}}].

\bibitem{Bautista:2019evw}
Y.~F. Bautista and A.~Guevara, \emph{{On the Double Copy for Spinning Matter}},
   \href{https://arxiv.org/abs/1908.11349}{{\ttfamily 1908.11349}}.

\bibitem{Bern:2019nnu}
Z.~Bern, C.~Cheung, R.~Roiban, C.-H. Shen, M.~P. Solon and M.~Zeng,
  \emph{{Scattering Amplitudes and the Conservative Hamiltonian for Binary
  Systems at Third Post-Minkowskian Order}},
  \href{https://doi.org/10.1103/PhysRevLett.122.201603}{\emph{Phys. Rev. Lett.}
  {\bfseries 122} (2019) 201603}
  [\href{https://arxiv.org/abs/1901.04424}{{\ttfamily 1901.04424}}].

\bibitem{Brandhuber:2021kpo}
A.~Brandhuber, G.~Chen, G.~Travaglini and C.~Wen, \emph{{A new gauge-invariant
  double copy for heavy-mass effective theory}},
  \href{https://arxiv.org/abs/2104.11206}{{\ttfamily 2104.11206}}.

\bibitem{Momeni:2020hmc}
A.~Momeni, J.~Rumbutis and A.~J. Tolley, \emph{{Kaluza-Klein from
  Colour-Kinematics Duality for Massive Fields}},
  \href{https://arxiv.org/abs/2012.09711}{{\ttfamily 2012.09711}}.

\bibitem{Momeni:2020vvr}
A.~Momeni, J.~Rumbutis and A.~J. Tolley, \emph{{Massive Gravity from Double
  Copy}}, \href{https://doi.org/10.1007/JHEP12(2020)030}{\emph{JHEP} {\bfseries
  12} (2020) 030} [\href{https://arxiv.org/abs/2004.07853}{{\ttfamily
  2004.07853}}].

\bibitem{Johnson:2020pny}
L.~A. Johnson, C.~R.~T. Jones and S.~Paranjape, \emph{{Constraints on a Massive
  Double-Copy and Applications to Massive Gravity}},
  \href{https://doi.org/10.1007/JHEP02(2021)148}{\emph{JHEP} {\bfseries 02}
  (2021) 148} [\href{https://arxiv.org/abs/2004.12948}{{\ttfamily
  2004.12948}}].

\bibitem{Chiodaroli:2015wal}
M.~Chiodaroli, M.~Gunaydin, H.~Johansson and R.~Roiban, \emph{{Complete
  construction of magical, symmetric and homogeneous N=2 supergravities as
  double copies of gauge theories}},
  \href{https://doi.org/10.1103/PhysRevLett.117.011603}{\emph{Phys. Rev. Lett.}
  {\bfseries 117} (2016) 011603}
  [\href{https://arxiv.org/abs/1512.09130}{{\ttfamily 1512.09130}}].

\bibitem{Johansson:2014zca}
H.~Johansson and A.~Ochirov, \emph{{Pure Gravities via Color-Kinematics Duality
  for Fundamental Matter}},
  \href{https://doi.org/10.1007/JHEP11(2015)046}{\emph{JHEP} {\bfseries 11}
  (2015) 046} [\href{https://arxiv.org/abs/1407.4772}{{\ttfamily 1407.4772}}].

\bibitem{Johansson:2015oia}
H.~Johansson and A.~Ochirov, \emph{{Color-Kinematics Duality for QCD
  Amplitudes}}, \href{https://doi.org/10.1007/JHEP01(2016)170}{\emph{JHEP}
  {\bfseries 01} (2016) 170}
  [\href{https://arxiv.org/abs/1507.00332}{{\ttfamily 1507.00332}}].

\bibitem{delaCruz:2015dpa}
L.~de~la Cruz, A.~Kniss and S.~Weinzierl, \emph{{Proof of the fundamental BCJ
  relations for QCD amplitudes}},
  \href{https://doi.org/10.1007/JHEP09(2015)197}{\emph{JHEP} {\bfseries 09}
  (2015) 197} [\href{https://arxiv.org/abs/1508.01432}{{\ttfamily
  1508.01432}}].

\bibitem{Brown:2018wss}
R.~W. Brown and S.~G. Naculich, \emph{{KLT-type relations for QCD and bicolor
  amplitudes from color-factor symmetry}},
  \href{https://doi.org/10.1007/JHEP03(2018)057}{\emph{JHEP} {\bfseries 03}
  (2018) 057} [\href{https://arxiv.org/abs/1802.01620}{{\ttfamily
  1802.01620}}].

\bibitem{Brown:2016hck}
R.~W. Brown and S.~G. Naculich, \emph{{Color-factor symmetry and BCJ relations
  for QCD amplitudes}},
  \href{https://doi.org/10.1007/JHEP11(2016)060}{\emph{JHEP} {\bfseries 11}
  (2016) 060} [\href{https://arxiv.org/abs/1608.05291}{{\ttfamily
  1608.05291}}].

\bibitem{Naculich:2015coa}
S.~G. Naculich, \emph{{Amplitudes for massive vector and scalar bosons in
  spontaneously-broken gauge theory from the CHY representation}},
  \href{https://doi.org/10.1007/JHEP09(2015)122}{\emph{JHEP} {\bfseries 09}
  (2015) 122} [\href{https://arxiv.org/abs/1506.06134}{{\ttfamily
  1506.06134}}].

\bibitem{Anastasiou:2016csv}
A.~Anastasiou, L.~Borsten, M.~J. Duff, M.~J. Hughes, A.~Marrani, S.~Nagy
  et~al., \emph{{Twin supergravities from Yang-Mills theory squared}},
  \href{https://doi.org/10.1103/PhysRevD.96.026013}{\emph{Phys. Rev. D}
  {\bfseries 96} (2017) 026013}
  [\href{https://arxiv.org/abs/1610.07192}{{\ttfamily 1610.07192}}].

\bibitem{delaCruz:2016wbr}
L.~de~la Cruz, A.~Kniss and S.~Weinzierl, \emph{{Double Copies of Fermions as
  Matter that Interacts Only Gravitationally}},
  \href{https://doi.org/10.1103/PhysRevLett.116.201601}{\emph{Phys. Rev. Lett.}
  {\bfseries 116} (2016) 201601}
  [\href{https://arxiv.org/abs/1601.04523}{{\ttfamily 1601.04523}}].

\bibitem{Chiodaroli:2013upa}
M.~Chiodaroli, Q.~Jin and R.~Roiban, \emph{{Color/kinematics duality for
  general abelian orbifolds of N=4 super Yang-Mills theory}},
  \href{https://doi.org/10.1007/JHEP01(2014)152}{\emph{JHEP} {\bfseries 01}
  (2014) 152} [\href{https://arxiv.org/abs/1311.3600}{{\ttfamily 1311.3600}}].

\bibitem{Huang:2012wr}
Y.-t. Huang and H.~Johansson, \emph{{Equivalent D=3 Supergravity Amplitudes
  from Double Copies of Three-Algebra and Two-Algebra Gauge Theories}},
  \href{https://doi.org/10.1103/PhysRevLett.110.171601}{\emph{Phys. Rev. Lett.}
  {\bfseries 110} (2013) 171601}
  [\href{https://arxiv.org/abs/1210.2255}{{\ttfamily 1210.2255}}].

\bibitem{Huang:2013kca}
Y.-t. Huang, H.~Johansson and S.~Lee, \emph{{On Three-Algebra and
  Bi-Fundamental Matter Amplitudes and Integrability of Supergravity}},
  \href{https://doi.org/10.1007/JHEP11(2013)050}{\emph{JHEP} {\bfseries 11}
  (2013) 050} [\href{https://arxiv.org/abs/1307.2222}{{\ttfamily 1307.2222}}].

\bibitem{Gross:1985rr}
D.~J. Gross, J.~A. Harvey, E.~J. Martinec and R.~Rohm, \emph{{Heterotic String
  Theory. 2. The Interacting Heterotic String}},
  \href{https://doi.org/10.1016/0550-3213(86)90146-X}{\emph{Nucl. Phys. B}
  {\bfseries 267} (1986) 75}.

\bibitem{Green:1984sg}
M.~B. Green and J.~H. Schwarz, \emph{{Anomaly Cancellation in Supersymmetric
  D=10 Gauge Theory and Superstring Theory}},
  \href{https://doi.org/10.1016/0370-2693(84)91565-X}{\emph{Phys. Lett. B}
  {\bfseries 149} (1984) 117}.

\bibitem{Stieberger:2014hba}
S.~Stieberger and T.~R. Taylor, \emph{{Closed String Amplitudes as
  Single-Valued Open String Amplitudes}},
  \href{https://doi.org/10.1016/j.nuclphysb.2014.02.005}{\emph{Nucl. Phys. B}
  {\bfseries 881} (2014) 269}
  [\href{https://arxiv.org/abs/1401.1218}{{\ttfamily 1401.1218}}].

\bibitem{Goldberger:2016iau}
W.~D. Goldberger and A.~K. Ridgway, \emph{{Radiation and the classical double
  copy for color charges}},
  \href{https://doi.org/10.1103/PhysRevD.95.125010}{\emph{Phys. Rev. D}
  {\bfseries 95} (2017) 125010}
  [\href{https://arxiv.org/abs/1611.03493}{{\ttfamily 1611.03493}}].

\bibitem{Shen:2018ebu}
C.-H. Shen, \emph{{Gravitational Radiation from Color-Kinematics Duality}},
  \href{https://doi.org/10.1007/JHEP11(2018)162}{\emph{JHEP} {\bfseries 11}
  (2018) 162} [\href{https://arxiv.org/abs/1806.07388}{{\ttfamily
  1806.07388}}].

\bibitem{Cheung:2018wkq}
C.~Cheung, I.~Z. Rothstein and M.~P. Solon, \emph{{From Scattering Amplitudes
  to Classical Potentials in the Post-Minkowskian Expansion}},
  \href{https://doi.org/10.1103/PhysRevLett.121.251101}{\emph{Phys. Rev. Lett.}
  {\bfseries 121} (2018) 251101}
  [\href{https://arxiv.org/abs/1808.02489}{{\ttfamily 1808.02489}}].

\bibitem{Kosower:2018adc}
D.~A. Kosower, B.~Maybee and D.~O'Connell, \emph{{Amplitudes, Observables, and
  Classical Scattering}},
  \href{https://doi.org/10.1007/JHEP02(2019)137}{\emph{JHEP} {\bfseries 02}
  (2019) 137} [\href{https://arxiv.org/abs/1811.10950}{{\ttfamily
  1811.10950}}].

\bibitem{Antonelli:2019ytb}
A.~Antonelli, A.~Buonanno, J.~Steinhoff, M.~van~de Meent and J.~Vines,
  \emph{{Energetics of two-body Hamiltonians in post-Minkowskian gravity}},
  \href{https://doi.org/10.1103/PhysRevD.99.104004}{\emph{Phys. Rev. D}
  {\bfseries 99} (2019) 104004}
  [\href{https://arxiv.org/abs/1901.07102}{{\ttfamily 1901.07102}}].

\bibitem{Bern:2019crd}
Z.~Bern, C.~Cheung, R.~Roiban, C.-H. Shen, M.~P. Solon and M.~Zeng,
  \emph{{Black Hole Binary Dynamics from the Double Copy and Effective
  Theory}}, \href{https://doi.org/10.1007/JHEP10(2019)206}{\emph{JHEP}
  {\bfseries 10} (2019) 206}
  [\href{https://arxiv.org/abs/1908.01493}{{\ttfamily 1908.01493}}].

\bibitem{Bjerrum-Bohr:2020syg}
N.~E.~J. Bjerrum-Bohr, T.~V. Brown and H.~Gomez, \emph{{Scattering of Gravitons
  and Spinning Massive States from Compact Numerators}},
  \href{https://doi.org/10.1007/JHEP04(2021)234}{\emph{JHEP} {\bfseries 04}
  (2021) 234} [\href{https://arxiv.org/abs/2011.10556}{{\ttfamily
  2011.10556}}].

\bibitem{Almeida:2020mrg}
G.~L. Almeida, S.~Foffa and R.~Sturani, \emph{{Classical Gravitational
  Self-Energy from Double Copy}},
  \href{https://doi.org/10.1007/JHEP11(2020)165}{\emph{JHEP} {\bfseries 11}
  (2020) 165} [\href{https://arxiv.org/abs/2008.06195}{{\ttfamily
  2008.06195}}].

\bibitem{Low:2020ubn}
I.~Low, L.~Rodina and Z.~Yin, \emph{{Double Copy in Higher Derivative Operators
  of Nambu-Goldstone Bosons}},
  \href{https://doi.org/10.1103/PhysRevD.103.025004}{\emph{Phys. Rev. D}
  {\bfseries 103} (2021) 025004}
  [\href{https://arxiv.org/abs/2009.00008}{{\ttfamily 2009.00008}}].

\bibitem{Goldberger:2004jt}
W.~D. Goldberger and I.~Z. Rothstein, \emph{{An Effective field theory of
  gravity for extended objects}},
  \href{https://doi.org/10.1103/PhysRevD.73.104029}{\emph{Phys. Rev. D}
  {\bfseries 73} (2006) 104029}
  [\href{https://arxiv.org/abs/hep-th/0409156}{{\ttfamily hep-th/0409156}}].

\bibitem{Arkani-Hamed:2020blm}
N.~Arkani-Hamed, T.-C. Huang and Y.-T. Huang, \emph{{The EFT-Hedron}},
  \href{https://doi.org/10.1007/JHEP05(2021)259}{\emph{JHEP} {\bfseries 05}
  (2021) 259} [\href{https://arxiv.org/abs/2012.15849}{{\ttfamily
  2012.15849}}].

\bibitem{Huang:2020nqy}
Y.-t. Huang, J.-Y. Liu, L.~Rodina and Y.~Wang, \emph{{Carving out the Space of
  Open-String S-matrix}},
  \href{https://doi.org/10.1007/JHEP04(2021)195}{\emph{JHEP} {\bfseries 04}
  (2021) 195} [\href{https://arxiv.org/abs/2008.02293}{{\ttfamily
  2008.02293}}].

\bibitem{Bern:2018jmv}
Z.~Bern, J.~J. Carrasco, W.-M. Chen, A.~Edison, H.~Johansson, J.~Parra-Martinez
  et~al., \emph{{Ultraviolet Properties of $\mathcal N = 8$ Supergravity at
  Five Loops}}, \href{https://doi.org/10.1103/PhysRevD.98.086021}{\emph{Phys.
  Rev. D} {\bfseries 98} (2018) 086021}
  [\href{https://arxiv.org/abs/1804.09311}{{\ttfamily 1804.09311}}].

\bibitem{Broedel:2013tta}
J.~Broedel, O.~Schlotterer and S.~Stieberger, \emph{{Polylogarithms, Multiple
  Zeta Values and Superstring Amplitudes}},
  \href{https://doi.org/10.1002/prop.201300019}{\emph{Fortsch. Phys.}
  {\bfseries 61} (2013) 812} [\href{https://arxiv.org/abs/1304.7267}{{\ttfamily
  1304.7267}}].

\bibitem{Mafra:2011nw}
C.~R. Mafra, O.~Schlotterer and S.~Stieberger, \emph{{Complete N-Point
  Superstring Disk Amplitude II. Amplitude and Hypergeometric Function
  Structure}},
  \href{https://doi.org/10.1016/j.nuclphysb.2013.04.022}{\emph{Nucl. Phys. B}
  {\bfseries 873} (2013) 461}
  [\href{https://arxiv.org/abs/1106.2646}{{\ttfamily 1106.2646}}].

\bibitem{Carrasco:2016ygv}
J.~J.~M. Carrasco, C.~R. Mafra and O.~Schlotterer, \emph{{Semi-abelian
  Z-theory: NLSM$+\phi^{3}$ from the open string}},
  \href{https://doi.org/10.1007/JHEP08(2017)135}{\emph{JHEP} {\bfseries 08}
  (2017) 135} [\href{https://arxiv.org/abs/1612.06446}{{\ttfamily
  1612.06446}}].

\bibitem{Naculich:2014naa}
S.~G. Naculich, \emph{{Scattering equations and BCJ relations for gauge and
  gravitational amplitudes with massive scalar particles}},
  \href{https://doi.org/10.1007/JHEP09(2014)029}{\emph{JHEP} {\bfseries 09}
  (2014) 029} [\href{https://arxiv.org/abs/1407.7836}{{\ttfamily 1407.7836}}].

\bibitem{Johansson:2019dnu}
H.~Johansson and A.~Ochirov, \emph{{Double copy for massive quantum particles
  with spin}}, \href{https://doi.org/10.1007/JHEP09(2019)040}{\emph{JHEP}
  {\bfseries 09} (2019) 040}
  [\href{https://arxiv.org/abs/1906.12292}{{\ttfamily 1906.12292}}].

\bibitem{Arkani-Hamed:2017mur}
N.~Arkani-Hamed, Y.~Bai, S.~He and G.~Yan, \emph{{Scattering Forms and the
  Positive Geometry of Kinematics, Color and the Worldsheet}},
  \href{https://doi.org/10.1007/JHEP05(2018)096}{\emph{JHEP} {\bfseries 05}
  (2018) 096} [\href{https://arxiv.org/abs/1711.09102}{{\ttfamily
  1711.09102}}].

\bibitem{Herderschee:2019wtl}
A.~Herderschee, S.~He, F.~Teng and Y.~Zhang, \emph{{On Positive Geometry and
  Scattering Forms for Matter Particles}},
  \href{https://doi.org/10.1007/JHEP06(2020)030}{\emph{JHEP} {\bfseries 06}
  (2020) 030} [\href{https://arxiv.org/abs/1912.08307}{{\ttfamily
  1912.08307}}].

\bibitem{Herderschee:2020lgb}
A.~Herderschee and F.~Teng, \emph{{Open associahedra and scattering forms}},
  \href{https://doi.org/10.1007/JHEP12(2020)134}{\emph{JHEP} {\bfseries 12}
  (2020) 134} [\href{https://arxiv.org/abs/2008.06418}{{\ttfamily
  2008.06418}}].

\bibitem{Carrasco:2020ywq}
J.~J.~M. Carrasco and I.~A. Vazquez-Holm, \emph{{Loop-Level Double-Copy for
  Massive Quantum Particles}},
  \href{https://doi.org/10.1103/PhysRevD.103.045002}{\emph{Phys. Rev. D}
  {\bfseries 103} (2021) 045002}
  [\href{https://arxiv.org/abs/2010.13435}{{\ttfamily 2010.13435}}].

\bibitem{Kalyanapuram:2020axt}
N.~Kalyanapuram, \emph{{On Polytopes and Generalizations of the KLT
  Relations}}, \href{https://doi.org/10.1007/JHEP12(2020)057}{\emph{JHEP}
  {\bfseries 12} (2020) 057}
  [\href{https://arxiv.org/abs/2009.10114}{{\ttfamily 2009.10114}}].

\bibitem{He:2016iqi}
S.~He and Y.~Zhang, \emph{{New Formulas for Amplitudes from Higher-Dimensional
  Operators}}, \href{https://doi.org/10.1007/JHEP02(2017)019}{\emph{JHEP}
  {\bfseries 02} (2017) 019}
  [\href{https://arxiv.org/abs/1608.08448}{{\ttfamily 1608.08448}}].

\bibitem{Borsten:2021hua}
L.~Borsten, B.~Jurco, H.~Kim, T.~Macrelli, C.~Saemann and M.~Wolf,
  \emph{{Double Copy from Homotopy Algebras}},
  \href{https://arxiv.org/abs/2102.11390}{{\ttfamily 2102.11390}}.

\bibitem{Elvang:2019twd}
H.~Elvang, M.~Hadjiantonis, C.~R.~T. Jones and S.~Paranjape,
  \emph{{All-Multiplicity One-Loop Amplitudes in Born-Infeld Electrodynamics
  from Generalized Unitarity}},
  \href{https://doi.org/10.1007/JHEP03(2020)009}{\emph{JHEP} {\bfseries 03}
  (2020) 009} [\href{https://arxiv.org/abs/1906.05321}{{\ttfamily
  1906.05321}}].

\end{thebibliography}\endgroup
%%%%%%%%%%%%%%%%%%%%%%%%%%%%
%%%%%%%%%%%%%%%%%%%%%%%%%%%%
%%%%%%%%%%%%%%%%%%%%%%%%%%%%
%%%%%%%%%%%%%%%%%%%%%%%%%%%%
%%%%%%%%%%%%%%%%%%%%%%%%%%%%

\end{document}